\pdfoutput=1
\documentclass[fleqn,usenatbib,usedcolumn]{mnras}
\usepackage[british]{babel}             
\usepackage{txfonts}                    
\usepackage[T1]{fontenc}                
\usepackage{graphicx}                   
\usepackage{color}

\usepackage{amsmath}

\def \LGalaxies{\texttt{L-Galaxies}\,}
\hypersetup{pdfauthor={D. Izquierdo-Villalba}, pdftitle={Building up pseudobulges in SAMs}, pdfkeywords={galaxies:theoretical -- galaxies:evolution -- methods:numerical}}
  
\volume{{\rm submitted}}


\title[Pseudobulges in a $\Lambda$CDM universe]{The build-up of pseudobulges in a hierarchical universe}

\author[Izquierdo-Villalba D. et al.]{
David Izquierdo-Villalba$^{1}$\thanks{E-mail: dizquierdo@cefca.es},
Silvia Bonoli$^{1,3,4}$,
Daniele Spinoso$^{1}$,
Yetli Rosas-Guevara$^{1,3}$, \newauthor \space
Bruno M. B. Henriques$^{5}$,
Carlos Hern\'{a}ndez-Monteagudo$^{2}$
\\
\\
$^{1}$ Centro de Estudios de F\'{\i}sica del Cosmos de Arag\'{o}n (CEFCA), Plaza San Juan, 1, E-44001, Teruel, Spain\\
$^{2}$ Centro de Estudios de F\'{\i}sica del Cosmos de Arag\'{o}n (CEFCA) - Unidad Asociada al CSIC, Plaza San Juan, 1, E-44001, Teruel, Spain\\
$^{3}$ Donostia International Physics Centre (DIPC), Paseo Manuel de Lardizabal 4, 20018 Donostia-San Sebastian, Spain\\
$^{4}$ IKERBASQUE, Basque Foundation for Science, E-48013, Bilbao, Spain\\
$^{5}$ Department of Physics, ETH Zurich, CH-8093 Zurich, Switzerland
}

\begin{document}
\maketitle
\begin{abstract}
We study the cosmological build-up of pseudobulges using the \LGalaxies semi-analytical model for galaxy formation with a new approach for following separately the assembly of classical bulges and pseudobulges. Classical bulges are assumed to be the result of violent processes (i.e., mergers and starbursts), while the formation of pseudobulges is connected to the secular growth of disks. We apply the model to both the \texttt{Millennium} and the \texttt{Millennium II} simulations, in order to study our results across a wide range of stellar masses ($\rm 10^{7}\,{-}\,10^{11.5} M_{\odot}$). We find that $z\,{=}\,0$ pseudobulges  mainly reside in galaxies of $\rm M_{stellar} \,{\sim}\, 10^{10}\,{-}\,10^{10.5} M_{\odot}$ ($ \rm M_{halo} \,{\sim}\, 10^{11.5}\,{-}\,10^{12} M_{\odot}$) and we recover structural properties of these objects  (e.g., sizes and bulge-to-total ratios) that are in good agreement with observational results. Tracing their formation history, we find that pseudobulges assembled in galaxies with a very quiet merger history, as opposed to the host galaxies of classical bulges. Regarding the bulge structure, we find that  $\,{\sim}\, 30\%$ of the galaxies with a predominant pseudobulge feature a composite structure, hosting both a pseudo and a classical bulge component. The classical component typically constitutes ${\sim}\, 10\%$ of the total bulge galaxy mass. When looking at the properties of the host galaxies, we find that $z\,{=}\,0$ pseudobulges are hosted by main sequence galaxies, characterized by  a stellar population which is generally younger compared to the one of the hosts of classical bulges. 



\end{abstract}


\section{INTRODUCTION}
According to the current picture of galaxy formation and evolution, the collapse of primordial and diffuse gas into condensed structures follows the aggregation of dark matter halos  \citep{WhiteandRees1978,ForcadaandWhite1997,WhiteFrenk1991,BirnboimandDekel2003}. During this process, the hot gas cools down and settles into rotationally-supported disks which act as birthplaces for galaxies. At later times, protogalaxies grow and evolve via a combination of \textit{ex-situ} and \textit{in-situ} processes which gradually shape their morphology, giving rise to the diverse population of galaxies in the mature universe, characterized by different proportions of bulge and disk components and complex features such as spiral arms and bar structures. While \textit{ex-situ} mechanisms can be dynamically fast and violent phenomena which take place, for instance, during galaxy mergers, the \textit{in-situ} processes comprise phenomena such as cooling of gas, internal star formation activity and instabilities in the galactic structure \citep[see][]{KormendyandKennicutt2004,Kormendy2013}. These processes can be long compared to the dynamical time of the galaxy,  in which case they are referred to with the term \textit{secular}.

It is broadly accepted that elliptical galaxies and classical bulges are formed via galaxy encounters during their hierarchical growth \citep{KauffmannWhiteGuiderdoni1993,BaughFrenk1996,vanDokkum2005,Benson2002,Menci2004,Moorthy2006,Eliche2006,Ryan2008,Carpineti2012,Kormendy2013}. Despite sharing common properties, a slightly different formation scenario has been proposed for each of them. Elliptical structures are expected to be the result of collisions between galaxies with similar baryonic mass, during which any memory of previous structural features, such as bulge morphology or disk component, is lost and the final galaxy is transformed in a pure-bulge \citep{Eliche2006,Cote1998,Barnes1999}. Classical bulges, on the other hand, are formed in galaxy encounters with small satellites where the nuclear region of the central galaxy experiences a significant growth as a consequence of the satellite mass incorporation \citep{Doyon1994,Aguerri2001,Tacconi2002,vanDokkum2005,Hammer2005,Bournaud2005,Dasyra2006A,Dasyra2007,Hopkins_and_Somerville2009,Rahimi2010}. This seems to be a simplified scenario, as pointed out by e.g. \cite{Hopkins2009,Ueda2014}, who found that some remnants of a equal-mass galaxy merger can still host a small disk component.\\

On the other hand, bulge structures developed in isolated galaxies
are thought to follow a different formation pathway than ellipticals and classical bulges \citep[see e.g.][]{KormendyandKennicutt2004,Athanassoula2005}. Within this evolutionary channel, morphological modifications are mostly governed by the self-gravity of the galactic disk: spatially extended and massive disks are susceptible to undergo a wide range of dynamical instabilities, characterized by the formation of non-axisymmetric and/or spiral structures usually referred to as \textit{bars} or \textit{spiral arms}, respectively \citep{Kalnajs1972,Ostriker1973,Combes&Sanders1981,Toomre1981,Efstathio1982,Pfenniger1990,MoMaoWhite1997,Athanassoula2005,Sellwood2016}. In particular, bar instabilities can have an in important role in shaping galaxy morphology by acting on the disk via angular momentum redistribution and gravitational torques \citep{Athanassoula2012A}. One of the net effects of these complex dynamical processes is the formation of a nuclear structure known as \textit{pseudobulge} shortly after bar formation, as consequence for instance of the buckling of the nuclear stellar orbits \citep{Pfenniger1990,Bureau1999,Combes2009,Athanassoula2012A,Kormendy2013}.\\

The basic elements of this picture are supported by several studies based on observational data. \cite{Doyon1994,Papovich2005,Tamburri2014}, for instance, showed that classical bulges are usually characterized by \textit{elliptical-like} properties, such as high Sersic indexes ($\rm n > 2$), old stellar populations, lack of star-forming activity and stellar kinematics dominated by velocity dispersion. On the other hand, pseudobulges display properties more related to disk-like structures, such as lower Sersic indexes ($\rm n<2$) or ongoing star formation \citep[see e.g.][]{DroryandFisher2007,FisherandDrory2008,Fisher2009}. Nevertheless, deviations from this \textit{archetypal} behaviour for pseudobulges/classical bulges have been found, for instance, by \cite{Ribeiro2016}. To complicate this picture further, some works have argued that bulge formation could not be a consequence of just mergers and bar instabilities \citep{Noguchi1998,Noguchi1999,Obreja2013,Laurikainen2016A}. In order to shed light on the possible formation mechanisms of pseudobulges, classical bulges and ellipticals new efforts have been pursued from an observational perspective by, e.g., \cite{Gadotti2009}. This work supported the idea of classical bulges and pseudobulges being formed via different evolutionary pathways, which would leave their respective imprint in the bulge structural properties. According to this work, Sersic indexes and \textit{bulge-to-total} ratios in classical bulges follow an elliptical-like correlation, suggesting a structural similarity between these two classes of objects. However, at the same time, classical bulges appear to be offset in the mass-to-size relation, as to confirm that classical bulges are not just ellipticals surrounded by disks. Finally, \cite{Gadotti2009} showed that classical- and pseudo-bulges overlap when their host structural parameters (such as bulge or disk scale lengths) are taken into account. These findings suggest that bulge formation is an extremely complex phenomenon, which might be shaped by both mergers and secular processes during the complex  cosmological evolution of galaxies \citep[see e.g.][]{Bournaud2002,Obreja2013,Erwin2015,Laurikainen2016A}.\\

Bulge formation has been extensively studied also via numerical approaches. \citet{Noguchi1998,Noguchi1999}, for instance, used simulations of isolated galaxies to introduce the \textit{clumpy-origin} bulge formation mechanism. This scenario is based on the radial migration and aggregation of several stellar clumpy structures during the high-redshift assembly of galaxy disks. Similar results have been obtained by \cite{Dekel2009} in a theoretical work. \cite{spinoso2017} analyzed the bar-induced formation of a pseudobulge structure within a Milky Way-like galaxy produced by the ErisBH cosmological zoom-in simulation \citep{Bonoli2016}. According to their analysis, a combination of the central black hole feedback at high redshift and the galaxy quiet merger history at lower one could have delayed the growth of the galaxy bulge, producing a disk more prone to bar instabilities at $z<0.5$. Nevertheless, all these numerical works could only focus on the analysis of few specific objects, suffering low statistics issues. Semi-analytical models (SAMs) have shown to be an useful tool to shed light to this complicated bulge formation paradigm under a statistical point of view \citep[see e.g.][]{Gargiulo2015,Guo2011,Lacey2016,Lagos2018}, despite some intrinsic limitations in  modelling galaxy  evolution processes. For instance, by using the \LGalaxies SAM, \citet{Shankar2012} could reproduce some observed properties of early type galaxies, such as effective radii or black hole- bulge mass relation. Other recent works  used a simple approach to model the bulge growth and were able to \textit{naturally} obtain the observed fraction of bulge galaxies and the galaxy size - stellar mass relation \citep{Tonini2016,Lagos2018}.\\


In this work, we use an updated version of the \LGalaxies semi-analytical model \citep{Henriques2015} to study the evolution of bulges, following separately classical and pseudo-bulge components. The code is run on both the \texttt{Millennium} and \texttt{Millennium II} merger tress \citep{Springel2005,Boylan-Kolchin2009}, enabling us to study a wide range in stellar mass ($\rm 10^{7}\,{-}\,10^{11.5} M_{\odot}$). The main novelty of our approach is that we differentiate between \textit{merger}-driven and \textit{secularly}-driven disk instabilities, linking the former to the growth of classical bulges, and the latter to the formation of bars and pseudobulges.  The outline of this work is as follow: In Section~\ref{sec:SAM_MILL} we describe the main characteristics of \LGalaxies and \texttt{Millennium} simulations. We present updates in the model that lead to a better description of galaxy morphology, and our approach in following the formation and evolution of bulge structures. In Section~\ref{sec:results} we present our results, focusing on the properties of galaxies that host pseudobulges across cosmic time and on the structural properties of the simulated pseudobulges.  Finally, in Section~\ref{sec:Conclusions} we summarize our main findings.

\section{\texttt{L-Galaxies} Semi-analytical model} \label{sec:SAM_MILL}

In this section we first briefly describe the \LGalaxies semi-analytical model, from the dark matter simulations to the prescriptions adopted to describe baryonic processes (extensively detailed in \cite{Henriques2015}). We then focus on the modification introduced in this work to better describe galaxy morphology and the pseudobulges build-up.

\subsection{The semi-analytical model: Framework}

\subsubsection{Dark matter simulations}
The backbones of \LGalaxies are the catalogues of merger-trees obtained by the \texttt{Millennium} \citep[hereafter MS,][]{Springel2005} and \texttt{Millennium II} \citep[MSII,][]{Boylan-Kolchin2009} $N$-body simulations. The first one follows the cosmological evolution of $\rm N\,{=}\,2160^3$ dark matter (DM) particles ($ \mathrm{m_p}\,{=}\,8.6 \times 10^8\, \mathrm {M_{\odot}}/h$) inside a periodic box of 500 ${\rm Mpc}/h$ on a side, from $z\,{=}\, $127 to the present. The latter can be thought as a high-resolution version of the MS, as it follows the same number of particles with a 125 times higher mass resolution ($ \mathrm{m_p = 6.885\,{\times}\,10^6\,M_{\odot}}/h$) in a smaller box ($\rm 100\,Mpc/h$ on a side). Both simulations were originally run with $\Omega_{\rm m}\,{=}\,0.25$, $\Omega_b\,{=}\,0.045$, $\Omega_{\rm \Lambda}\,{=}\,0.75$, $h\,{=}\,\rm 0.73 \,km\,s^{-1}\,Mpc^{-1} $, $n\,{=}\,1$, $\sigma_{8}\,{=}\,0.9$ \citep{Colless2001}.\\

Data from the MS and MSII simulations were stored respectively at 63 and 68 epochs (snapshots), spaced approximately logarithmically in time at $z\,{>}\,0.7$ and linearly at $z\,{<}\,0.7$ (where $\rm \Delta t \,{\sim}\, 300\,Myr$).  DM halos and subhalos were identified within the snapshots by using a \textit{friend-of-friend} (FOF) group-finder and an extended version of the $\rm SUBFIND$ algorithm \citep{Springel2001}. Halo catalogs were built by considering only bound structures with at least 20 particles, which translates into a minimum halo-mass of $\rm M_{\rm halo}^{\rm min} \,{=}\, 1.72 \times 10^{10} \, M_{\odot}/\textit{h}$ and $\rm M_{\rm halo}^{\rm min} \,{=}\, 1.38 \times 10^{8} \, M_{\odot}/\textit{h}$ for MS and MSII, respectively. Halos and subhalos were finally arranged in merger trees structures, thus allowing to follow the evolutionary path of any DM halo in the simulations. These merger trees are the skeleton of our SAM, but, because of the finite number of outputs of the DM simulations, the time resolution they offer is not enough to properly trace the baryonic physics. Therefore, to accurately follow the galaxy evolution between two consecutive DM snapshots ($\rm \Delta t\,{\sim}\,300\,Myr$), the SAM does an internal time discretization between them with approximately $\rm{\sim}\, 5{-}20 \,Myr$ of time resolution. These extra temporal subdivisions used by the SAM are called \textit{sub-steps}.\\

We \textit{want to stress} that the accuracy of the results presented in this work at $\rm M_{stellar}\,{<}\,10^9 M_{\odot}$ for MS are limited by halo mass resolution issues. In these cases, we will rely in the MSII predictions whose limitation is at $\rm M_{stellar}\,{\sim}\,10^{8}M_{\odot}$.\\

The latest \LGalaxies version was tuned on a re-scaled versions of MS and MSII simulations \citep{Henriques2015}. The re-scaling procedure \citep{AnguloandWhite2010} allows the two simulations to match the cosmological parameters provided by Planck first-year data \citep{PlanckCollaboration2014}) $\Omega_{\rm m} \,{=}\,0.315$, $\Omega_{\rm \Lambda}\,{=}\,0.685$, $\Omega_{\rm b}\,{=}\,0.045$, $\sigma_{8}\,{=}\,0.9$ and $h\,{=}\,0.673\, \rm km\,s^{-1}\,Mpc^{-1}$. After re-scaling, the particle mass corresponds to $\mathrm{m_p}\,{=}\,\rm  1.43 \,{\times}\,10^9\, M_{\odot}/\textit{h}$ and $\mathrm{m_p}\,{=}\,7.68\,{\times}\,10^{6}\, \mathrm{M_{\odot}}/\textit{h}$ for MS and MSII respectively.

\subsubsection{Baryonic physics}
\label{barPhys}
The starting point of the galaxy evolution model is the infall of baryonic  matter onto every newly-resolved DM halo \citep[see e.g.][]{WhiteFrenk1991}. This process is modeled by associating an amount of matter $\rm M_{bar}$ to each halo, proportionally\footnote{ $\rm M_{bar} \,{=}\, f_b \cdot M_{halo}\,{=}\,\Omega_b / \Omega_m\cdot M_{halo} \,{=}\,0.155 \cdot M_{halo}$ \citep{PlanckCollaboration2014}} to its DM mass $\rm M_{halo}$. This process is repeated at each snapshot, in order to keep the baryonic fraction of each halo fixed in time to the value $\rm f_b\,{=}\,0.155$. The baryonic component initially assumes the form of a diffuse, spherical, quasi-static, hot atmosphere of pristine (i.e. zero-metallicity) gas, with radius equal to the halo \textit{virial radius} $\rm R_{200c}$. A fraction $\rm M_{cool}$ of this atmosphere is then allowed to gradually condensate and migrate towards the DM halo center. In particular, following \cite{WhiteandRees1978}, the gas \textit{cooling rate} $\rm \dot{M}_{cool}$ is determined by the amount of hot gas enclosed within the halo \textit{cooling radius} $\rm r_{cool}$ \citep[defined as the radius at which
$\rm t_{cool}(r)\,{=}\,t_{dyn,h}$, as in][]{DeLucia2004}. This implies the presence of two different cooling regimes: the \textit{rapid infall} ($\rm r_{cool}\,{>}\,R_{200c}$) which leads to the fast condensation of the whole hot atmosphere, and the slower \textit{hot phase} ($\rm r_{cool}\,{<}\,R_{200c}$), in which only a fraction of the hot gas is allowed to cool down. The cold gas then settles into a disc-like structure by inheriting specific angular momentum from its host DM halo \citep[see][]{Guo2011} and constitutes the mass reservoir which fuels star formation (SF) processes. After each cooling episode, the mass $\rm M_{cold}$ and dynamical time $\rm t_{dyn,disk}$ of each galaxy cold-gas disk uniquely define the instantaneous star formation rate (SF), by which galaxies build-up their stellar disk in time. As it is widely accepted, feedback from supernovae (SNe) can have a severe impact on star formation within galactic disks. To model this crucial phenomenon, SNe in \LGalaxies inject energy in the cold-gas disk, helping to re-heat a fraction of it to the hot atmosphere and eventually ejecting a fraction of the hot gas beyond $\rm R_{200c}$. The reincorporation of these ejecta at later times helps regulating  the low-$z$ star formation, especially in low-mass satellites galaxies \citep{Henriques2015}.\\

Regarding the bulge component, the galaxies are allowed to develop/grow a dense pack of stars in the nuclear region via mergers and disk instabilities (DI). While the former is a natural consequence of the hierarchical growth of the DM halos, the latter plays a crucial role in galaxies in isolation and closely related with star formation. In the following sections we discuss about these two different channels and in \hyperref[Appendix:Radius]{Appendix~\ref{Appendix:Radius}} we present an improvement in the redshift evolution of the effective radius vs. stellar mass plane for early and latte type galaxies by adding energy dissipation during bulge formation in major mergers.\\

Finally, in order to prevent the stellar component of massive galaxies to over-grow, the model introduces feedback from central super-massive black holes (BHs) as an additional mechanism to regulate star formation at low redshifts. The so-called \textit{radio-mode} BH feedback is defined to be proportional to the matter from the galaxy hot atmosphere and the the central BH mass\footnote{A better modeling of black hole growth and its spin evolution in \LGalaxies will be presented in Izquierdo-Villalba et al. in prep} \citep[see e.g.][]{Croton2006}. The hot atmosphere content, in turn, depends indirectly on all the \textit{large scale} effects acting on it. Among these, the most important for the scope of this work are \textit{environmental processes}, such as ram pressure or tidal interactions which can completely remove the hot gas atmosphere around satellite galaxies and eventually destroy their stellar and gas components \citep[see details in][]{Guo2011,Henriques2015}.

\begin{figure*} 
	\centering
    \includegraphics[width=2.0\columnwidth]{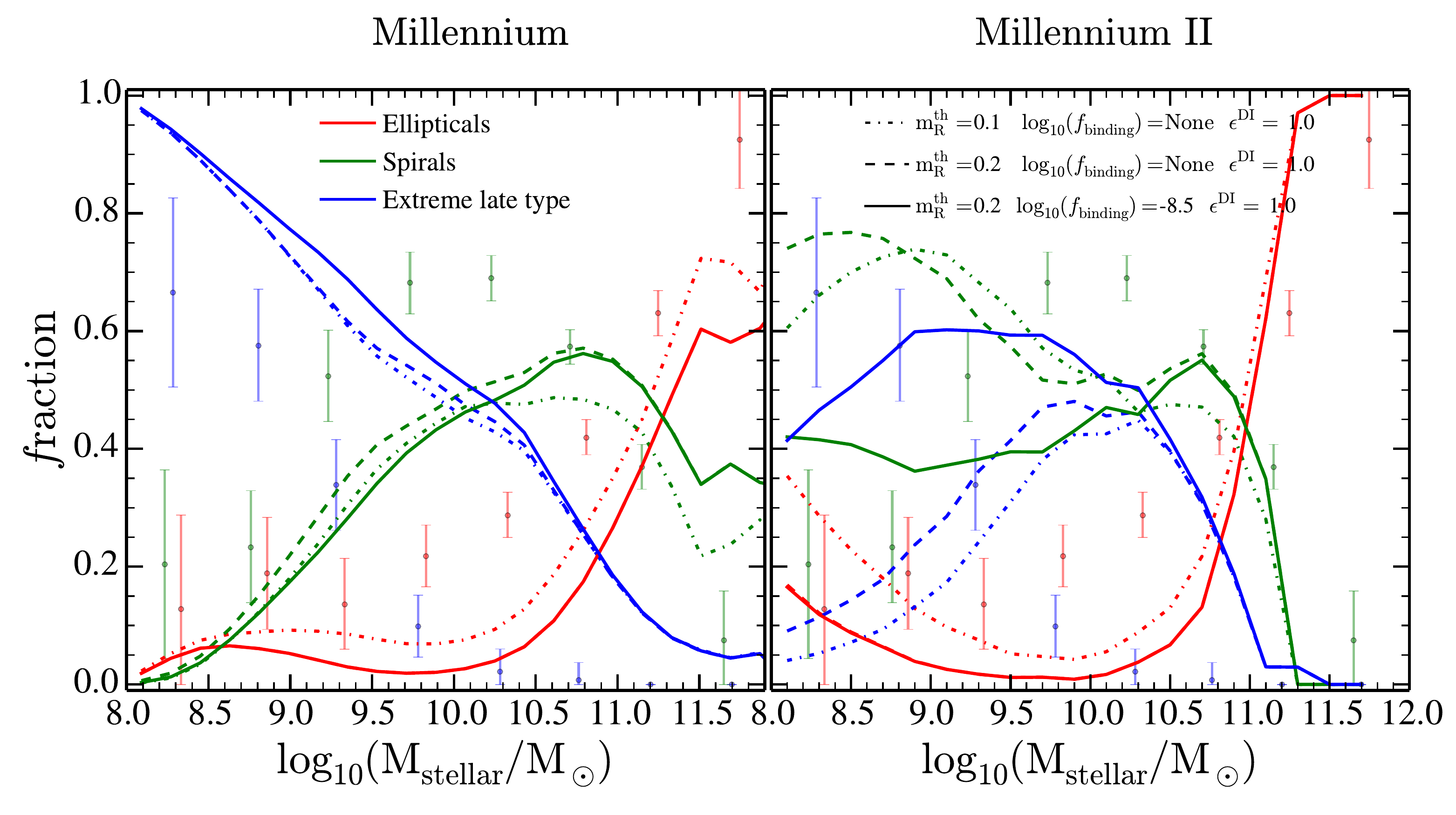}
    \caption{Fraction of different morphological types as a function of stellar mass for the MS (left panel) and MSII (right panel) at $z \,{=}\, 0$. The DI stability parameter is set to $\rm \rm \epsilon^{DI} \,{=}\, 1.0$ (see Section~\ref{sec:DI}). Following  \protect\cite{Henriques2015} we define as early type (red curves), spiral (green curves) and extremely late-type (blue curves) galaxies with respectively bulge-to-total ratio $\rm B/T\,{>}\,0.7$, $0.01\,{<}\,\mathrm{B/T}\,{<}\,0.7$ and $\rm B/T\,{<}\,0.01$. The points of corresponding colours represents the observational constrains presented in \protect\cite{Conselice2006}. Line styles are associated to different set of the parameters $\rm m_R^{th}$ and $\rm \mathit{f}_{binding}^{th}$.}
\label{fig:Morphology_merger_effect}
\end{figure*}

\subsection{The build-up of bulges through mergers} \label{sec:Morpho_and_smooth}

Galaxy morphology in \LGalaxies is mainly driven by  mergers and disk instabilities. Here we describe how the model treats these processes and the modifications we introduced to better describe the abundance of the different morphological types across a wide range of stellar masses and to follow secular-evolution processes.

\subsubsection{Smooth accretion: a new recipe for extreme minor mergers}\label{sec:mergers_and_smooth_accretion}

Galactic encounters are driven by the merger of the parent dark matter haloes. The time-scale of these processes is given by the dynamical friction experienced by the merging galaxies, as presented in \citet{Guo2011}. In the standard picture of \LGalaxies, the ratio $\rm m_R\,{=}\,(M_{cold,1}^{gas} + M_{stellar,1}) / (M_{cold,2}^{gas} + M_{stellar,2})$ between the baryonic masses of the two galaxies is used to differentiate between \textit{major} ($\rm m_R\,{>}\,m_R^{th}$) and \textit{minor} ($\rm m_R\,{<}\,m_R^{th}$) interactions. In the \textit{standard} version of the model, $\rm m_R^{th}$ is set to 0.1. Major mergers are assumed to be able to completely destroy the disks of the two interacting galaxies, leading to a pure spheroidal remnant which suffers a \textit{collisional starburst}. In minor mergers, instead, the disk of the larger galaxy survives and experiences a burst of star formation, while its bulge incorporates the entire stellar mass of the satellite that survived stripping (as modelled by \cite{Guo2011}).\\

In Fig~\ref{fig:Morphology_merger_effect} we show how the standard \LGalaxies model recovers the morphological distribution of galaxies, as a functions of stellar mass, for both the MS (left) and MSII (right) runs. Lines refer to ellipticals (red), spirals (green) and extreme late types (blue), while coloured dots represent a collection of observational data, as in \cite{Conselice2006}\footnote{\cite{Conselice2006} defined ellipticals as galaxies with a morphological type T within $\rm -4 \,{<}\, T \,{<}\, -3$, which would corresponds to bulge-to-total ratios of about $\rm [0.6-0.7]$ \citep[see][for more details]{MoWhite2010,SimienandVaucouleurs1986S}}. Morphological types definition is somewhat arbitrary \citep[see][]{Lagos2008,Guo2011,Gargiulo2015}; in what follows we define extreme late types, spirals and ellipticals as galaxies with \textit{bulge-to-total} ratios (hereafter $\rm B/T$) of, respectively, $\rm B/T\,{<}\,0.01$, $0.01\,{<}\,\mathrm{B/T}\,{<}\,0.7$ and $\rm B/T\,{>}\,0.7$. As we can see, Fig~\ref{fig:Morphology_merger_effect} in dash-dotted lines shows that the \citet{Henriques2015} standard version of \LGalaxies ($\rm m_R^{th}\,{=}\,0.1$, $f_{\rm binding}\,{=}\,\rm None$ and $\rm \epsilon^{DI} \,{=}\,1.0$) reproduces the general trend presented in \cite{Conselice2006} on both MS and MSII. Nevertheless, in both cases the population of (extreme late-type) spiral and elliptical galaxies is (over-) under-predicted in the range $\rm 10^{10}\,{-}\,10^{11} M_{\odot}$ (blue, green and red solid lines, respectively). Besides, the MSII does not converge with the MS, showing a large excess with respect to observations in the spiral population at low stellar masses $\rm\,{<}\,10^{9.5} M_{\odot}$.\\

After a detailed analysis of the impact that the current treatment of merger events has in the definition of galaxy morphology and its dependence on the resolution of the DM simulation used,  we found that an improvement in the morphological distribution of galaxies and a reasonable convergence between MS and MSII can be reached when including the following two modifications: (i) set the threshold between major and minor mergers to the value $\rm m_R^{th}\,{=}\,0.2$ and (ii) introduce a new approach in the treatment of  extreme minor-mergers. The first modification leads to a better convergence between the MS and the MSII in terms of the number density of major merger events (see the details in \hyperref[Appendix:Smooth_accretion]{Appendix~\ref{Appendix:Smooth_accretion}}), and helps increasing the fraction of spirals in galaxies below $\rm{\sim}\,10^{11} M_{\odot}$, as can be seen in Fig~\ref{fig:Morphology_merger_effect} (dashed lines). The second change has a very strong effect on both the convergence of the number density of minor merger events and on the morphological distribution of small galaxies (i.e, $\rm M_{stellar}\,{<}\,10^{9.5} M_{\odot}$). In this mass range star formation in the disk can stall, as the cold-gas content of these low-mass galaxies is typically too low to trigger star formation\footnote{in \LGalaxies the threshold for star formation is $\rm M_{crit} \,{=}\, 2.4{\times}10^{9} M_{\odot}$  (see Eq.S14 of \cite{Henriques2015}). Note that a more accurate description of star formation might come by linking this process with the molecular gas component instead of the total cold gas (see \cite{Lagos2011}), as also discussed in \cite{Henriques2015}}. Therefore, the only events leading to morphological changes for galaxies with $\rm M_{stellar}\,{<}\,10^{9.5} M_{\odot}$ are mergers. In the MSII, in particular, these small galaxies experience a significant number of extreme minor mergers, as the simulation is able to resolve much smaller structures compared to the MS (the most extreme and numerous encounters are with satellite galaxies of the order of $\rm M_{stellar} \,{\sim} 10^5 M_{\odot}$). If such extreme interactions are treated as \textit{normal} minor mergers the bulges of these small galaxies grow by incorporating the stellar mass of the satellites, while their disks are unable to increase in mass, as star formation is stalled (and merger-induced bursts are negligible as less than 0.2\% of the cold gas mass is transformed into stars). This leads to the large fraction of spirals (and lack of extreme disk), as shown in Fig~\ref{fig:Morphology_merger_effect} with dash and dotted lines. 
We thus update the model, introducing a new set of prescriptions to treat these extreme minor mergers, to which we refer with the term \textit{smooth accretions}  \citep[see e.g.][]{Abadi2003,Peniarubia2006,Sales2007,Kazantzidis2008}. In those extreme minor mergers, one might expect that the stellar satellite mass might not be able to reach the bulge of the central galaxies, but gets disrupted by the disk of the central galaxy and get incorporated by it.\\

\begin{figure}
	\centering
\includegraphics[width=1.\columnwidth]{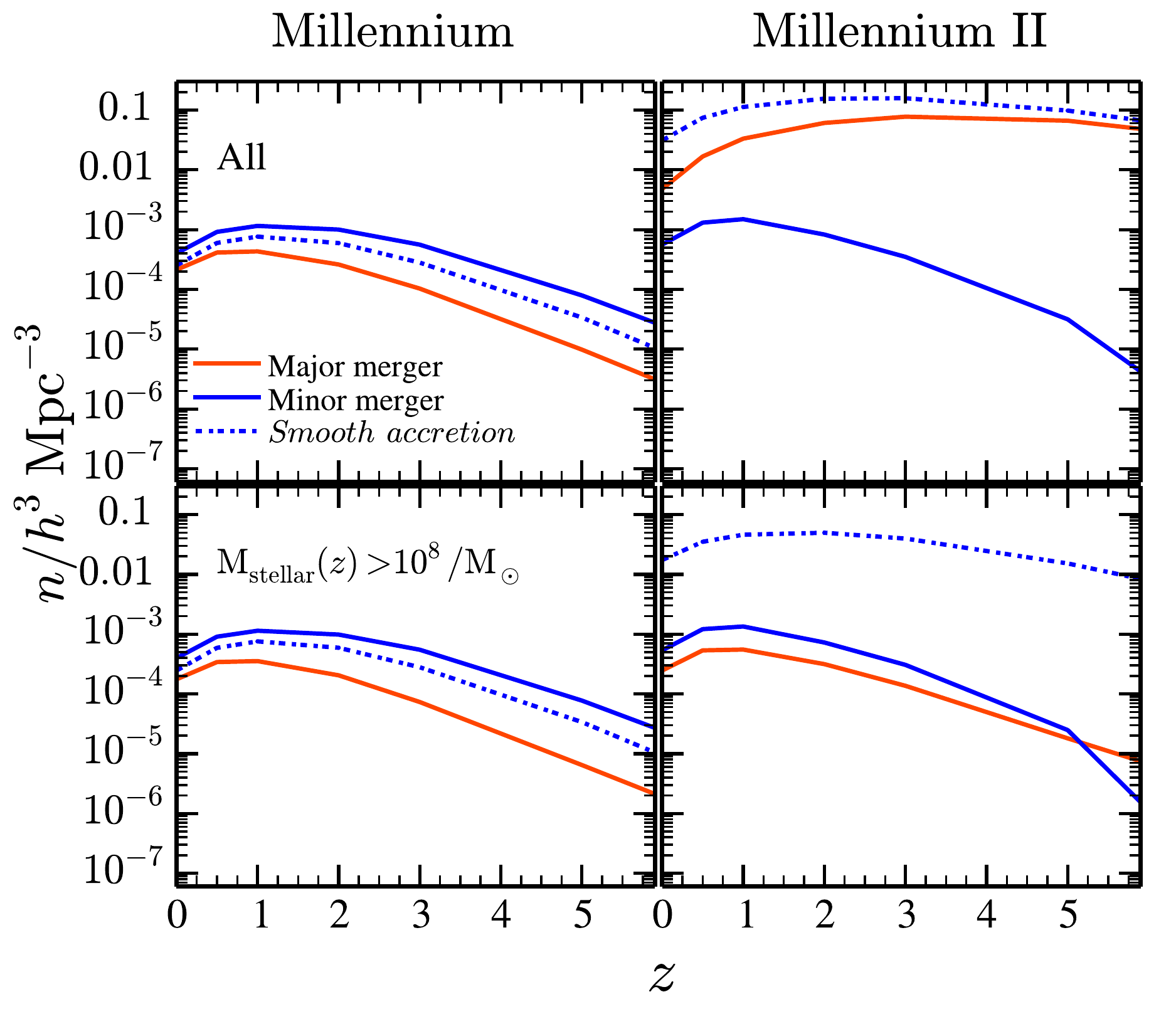}
\caption{Number density of major mergers (red solid line), minor mergers (solid blue line) and \textit{smooth accretion} (dashed blue line) as a function of redshift. The left and right columns display the results for the \texttt{Millennium} and \texttt{Millennium II} simulation, respectively. In the first row we present the results for \textit{all} galaxies in the simulation. The second one represents the same but for galaxies with $\rm M_{stellar}\,{>}\,10^8 M_{\odot}$ at a given redshift.} 
\label{fig:all_n_density}
\end{figure}

\begin{figure*} 
	\centering
    \includegraphics[width=2.0\columnwidth]{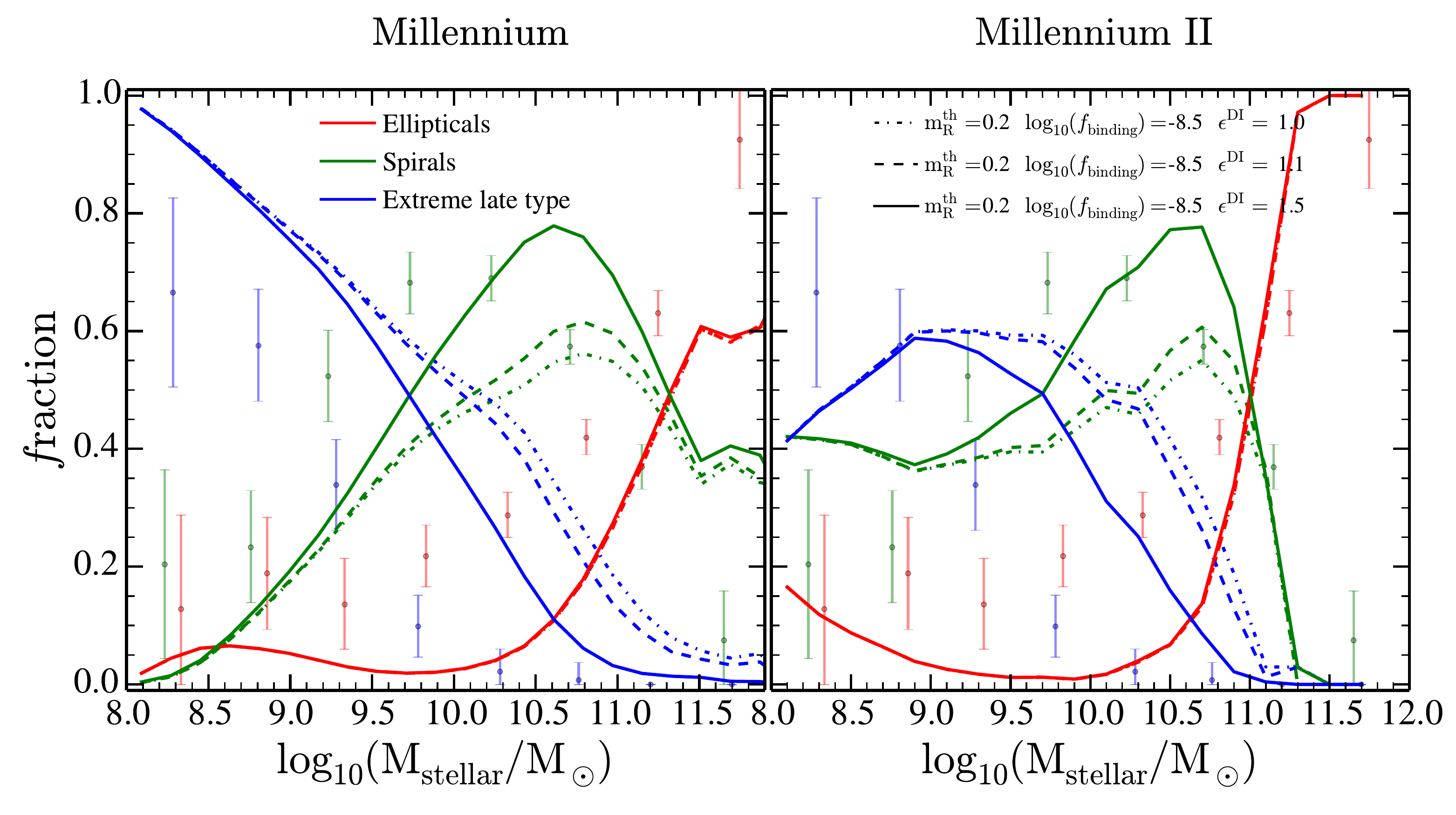}
    \caption{Same as Fig~\ref{fig:Morphology_merger_effect} but assuming a fixed values of $\rm m_R^{th}\,{=}\,0.2$ and $\rm log_{10}(\mathit{f}_{binding}^{th})\,{=}\,-8.5$ and varying $\rm \epsilon^{DI}$.}
\label{fig:Morphology_DI_effect}
\end{figure*}

We make use of the ratio $f_{\rm binding}$ between the binding energies of the merging structures to disentangle between \textit{normal} minor mergers and \textit{smooth accretion} episodes. We assume that the interacting (sub-)systems are i) the whole stellar satellite galaxy and ii) the central galaxy stellar disk (gas+stars), we compute the satellite $f_{\rm binding}$ by considering the entire satellite stellar mass, and only the disk mass (gas+stars) for the central galaxy, respectively:
\begin{equation}\label{eq:binding}
f_{\rm binding} = \frac{\rm E_{binding}^{Satellite}}{\rm E_{binding}^{Central}} = \frac{\rm M^2_{Sat,Stellar}}{\rm M^2_{\rm Cent, disk}}\frac{\rm R_{disk}^{Central}}{\rm R_{Stellar}^{Sat}},
\end{equation}
where $\rm R_{Stellar}^{Sat}$ is the mass-weighted average half-mass radii of the satellite bulge and disk, while $\rm R_{disk}^{Central}$ is the same quantity for the disk of the central galaxy (as it is composed by both gas and stars). The larger the value of $f_{\rm binding}$, the closer are the binding energies of the merging galaxies, so the remnant of the satellite galaxy might survive the interaction with the central disk and reach the centre of its massive companion (\textit{usual} minor merger). On the opposite case, we assume that the central galaxy can easily unbound the satellite stellar system, which will be incorporated in the central galaxy disk (\textit{smooth accretion}). Following this approach, the best agreement with observational data is obtained by imposing $ f_{\rm binding}^{\rm th} \,{=}\, 10^{-8.5}$ as a threshold value to discriminate between the two scenarios. This low value of $ f_{\rm binding}^{\rm th}$ corresponds more or less to a cut in satellite stellar mass $\rm {\sim}\, 10^7M_{\odot}$, as it is shown in Fig~\ref{fig:binding_energy} of  \hyperref[Appendix:Smooth_accretion]{Appendix~\ref{Appendix:Smooth_accretion}}. 
As can be seen in Fig~\ref{fig:all_n_density}, by imposing $ f_{\rm binding}^{\rm th} \,{=}\, 10^{-8.5}$  we obtain a remarkable agreement in the  minor merger predictions for both MS and MSII merger trees (see blue solid lines). Concerning the \textit{smooth accretion} events (blue dotted lines), MSII and MS merger trees display different predictions. In the former, the \textit{smooth accretion} has 1 dex larger number density than in the latter, being the dominant type of interaction at any redshift.\\ 
\noindent To summarize:
\[
  \begin{cases}
               \; \mathrm{minor \, merger} \;\;\;\;\;\;\;\; \mathit{f}_{\rm binding} > \mathit{f}_{\rm binding}^{\rm th} \,\, \& \,\, \mathrm{m_R < m_R^{th}} \\
               \; \mathit{smooth \,accretion} \;\;\; \mathit{f}_{\rm binding} < \mathit{f}_{\rm binding}^{\rm th} \,\, \& \,\, \mathrm{m_R < m_R^{th}}\; 
            \end{cases}
\]
with $\mathit{f}_{\rm binding}^{\rm th}\,{=}\,10^{-8.5}$ and  $\rm m_R^{th}\,{=}\,0.2$. The results obtained with this new recipe for both MS and MSII are shown with solid lines in Fig~\ref{fig:Morphology_merger_effect}. Our new prescription  leaves the morphology distributions almost unchanged in the case of the MS, while it improves them for the MSII, providing a better agreement between data and model predictions. A detailed analysis showing the morphology evolution with respect to $\rm m_R^{th}$ and $f_{\rm binding}^{\rm th}$ parameters can be found in Fig \ref{fig:all_distribution_morph} of \hyperref[Appendix:Smooth_accretion]{Appendix~\ref{Appendix:Smooth_accretion}}.\\

Finally, as we can see in Fig~\ref{fig:Morphology_merger_effect}, in spite of the morphological improvements achieved at low stellar masses by changing the merger recipe of \LGalaxies, we can not find a significant improvement in the intermediate population $\rm 10^{10}\,{-}\,10^{11} M_{\odot}$. From this, we can draw a simple conclusion: mergers do not have the dominant role in this range of masses. In the next section we will explore the effects of the other bulge formation channel (disk instabilities) in the galaxy morphology.

\subsection{Disk instabilities: the growth of pseudobulges and classical bulges} \label{sec:DI}
In addition to mergers, the \textit{disk instabilities} (DI) channel is an important pathway for bulge growth in \LGalaxies. Within the context of this work, DI refers to the process by which the stellar disk becomes massive enough to be prone to non-axisymmetric instabilities which ultimately lead to the formation of a central ellipsoidal component via the buckling of nuclear stellar orbits \citep[see references in][]{MoWhite2010}. During this process, a possible result is the formation of a \textit{bar structure} \citep{Kalnajs1972,Ostriker1973,Combes&Sanders1981,Efstathio1982,Pfenniger1990,MoMaoWhite1997,Athanassoula2005,Sellwood2016}. Galactic bars have a deep impact the morphology of the nuclear parts. On one hand, they can efficiently modify the gas disk structure via gravitational torques able to produce strong nuclear gas inflows which can be transformed into stars inducing the formation of disc-like pseudobulge structure. On the other, shortly after the bar formation the structure can experience a bending mode that thickens it and forms a boxy/peanut pseudobulge \citep[see e.g.][]{Pfenniger1990,KormendyKennicutt2004,Saha2015,spinoso2017}. The \LGalaxies model accounts for galactic DI with a simple analytic stability criterion, based on the \cite{Efstathio1982} and \cite{MoMaoWhite1997} 2d simulations:
\begin{equation}\label{equation:diskinesta}
\rm  \frac{V_{max}}{(G M_{\star,d}/R_{\star,d})^{1/2}} \leqslant \epsilon^{\rm DI},
\end{equation}
where $\rm V_{max}$ is the maximum circular velocity of the host dark matter\footnote{We found no significant differences in our results when using the \textit{disk} circular velocity $\rm V_c(r\,{=}\,2.2R^d)\,{=}\,\sqrt{ \rm GM_{DM}(r)/r + G M_{bulge}(r)/r + V^2_{disk}(r)}$. This definition is obtained by assuming an Hernquist \citep{Hernquist1990} and NFW \citep{NFW1996} profile for the bulge and DM halo, respectively.}, $\rm R_{\star,d}$ and  $\rm M_{\star,d}$ are the exponential scale-length and stellar mass of the stellar disc respectively and $\epsilon^{\rm DI}$ a parameter which determines the importance of the disk self-gravity (set to $1.0$ in the standard version of \LGalaxies). If this stability criterion is met, an amount
\begin{equation}\label{equation:diskinestaMass}
\rm \Delta M_{\star}^{DI} =  M_{\rm disk} -  M_{crit}^{DI} = M_{\rm disk} - \frac{V_{max}^2 R_{\star,d}}{G \epsilon_{\rm DI}^{2}},
\end{equation}
of the disk stellar mass is transferred to the bulge in order to restore the disk (marginal) stability. Despite the  limitations of Eq.\eqref{equation:diskinesta} \citep[see][]{Athanassoula2008}, this criterion to follow disc stability has the advantage of being simple and to depend only on global galaxy properties, accessible by the model.\\

According to \cite{Efstathio1982,MoMaoWhite1997} $\epsilon^{\rm DI} \,{\approx}\, 1.1$ for a family of exponential-profile stellar disk models. Nevertheless, in order to improve the morphology at intermediate stellar masses and following the approach of other SAMs \citep{Hirschmann2012,Menci2014,Lacey2016,Lagos2018} we have tested the model with different values of DI stability parameter, $\rm \epsilon^{DI}$. The results are presented in Fig~\ref{fig:Morphology_DI_effect}. As we can see, we found that a slightly higher value (namely $\epsilon^{\rm DI}\,{=}\,1.5$) provides a better agreement with observations in the mass range $\rm 10^9 \,{<}\, M_{stellar} M_{\odot} \,{<}\, 10^{11}$. The change of parameter value causes galaxies to be more easily prone to instabilities, thus a larger fraction of stars is transferred from the disk to the bulge component, increasing the fraction of spirals and reducing the one of extreme late types in this mass range. Notice that the change of $\epsilon^{\rm DI}$ does not have any impact in the elliptical population. In a recent paper, \cite{Irodotou2018} achieved a better improvement of the spiral and elliptical population in \LGalaxies by imposing angular momentum losses during the gas cooling and allowing DI in the galaxy gaseous disk. Nevertheless, the results were not checked in MSII. Here we decide no to use that approach and keep our independent merger/disk instability analysis which lets us reach the convergence between MS and MSII and update/ improve the \LGalaxies standard merger recipe. Besides, we have checked that the increase of the stability parameter has a similar effect in the spiral galaxy population that the one achieved by adding angular momentum losses during the gas cooling.\\

\subsubsection{A discretization effect: Linking different DI events as a single episode}\label{sec:Linking_DI}

\begin{figure} 
	\centering
    \includegraphics[width=1.\columnwidth]{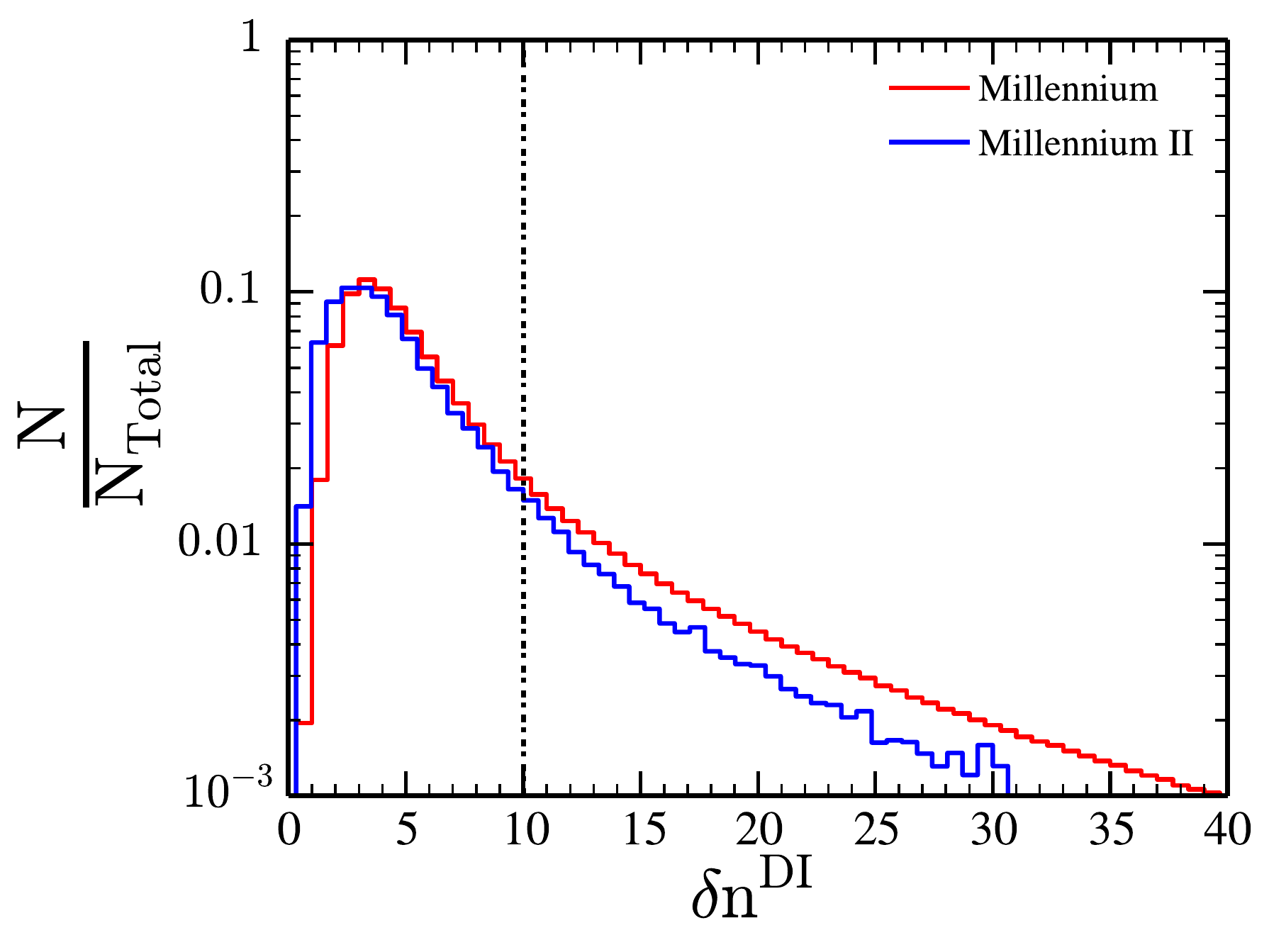}
    \caption{Number of dynamical times ($\rm \delta n^{DI}$) that a galaxy experiences between two no consecutive disk instabilities. Red histogram represents the results for MS while in blue the ones for MSII. Dotted vertical line represents our threshold to consider two no consecutive DIs as the same episode.}
	\label{fig:Disk_inestabilities_delay_and_minor_merger}
\end{figure}

The adopted approach to treat DIs in \LGalaxies (see Section~\ref{sec:DI}), is such that galaxy equilibrium is restored by  transferring the minimum amount of mass from the disk to the bulge. This means that the disk easily becomes unstable again in the one (or more) of the subsequent \textit{sub-step}. This generates a series of disk instabilities in a galaxy which are, effectively, all connected. This is especially true in systems in which the cooling rate is high enough to quickly replenish the stellar disk \citep{Porter2014}. While two consecutive disk instability events (in two subsequent \textit{sub-steps}) can be easily assumed to be part of the same event, connecting events which are more spread in time is less straightforward.
In order to join separate DI events, thus erasing the discretization effects of the time resolution of the simulation, we start by studying the typical time difference between two no-consecutive DIs in the same galaxy. 
For this, we define the quantity $\rm \delta n^{DI}$, defined as the time-difference between two events and normalized by the dynamical time of the galaxy: 
\begin{equation}
\rm \delta n^{DI} = \frac{t^{ Last \, DI} - t^{Current \, DI}}{t^{\star, disk}_{dyn}|_{Last\, DI}},
\end{equation}
where $ \rm t^{Current \, DI}$ is the lookback time of the current DI, $\rm t^{ Last \, DI}$ is the lookback time in which the galaxy experienced the last disk instability and $\rm t^{\star, disk}_{dyn}|_{Last\, DI}$ is the dynamical time of the stellar disk at the epoch of the last DI. 
The distributions of $\rm  \delta n^{DI}$ for MS (red) and MSII (blue) are presented in Fig~\ref{fig:Disk_inestabilities_delay_and_minor_merger}. As we can see, both distributions present a clear peak at values of $\rm \delta n^{DI}\,{\sim}\,5$, indicating that a large fraction of DI events are separated by few dynamical times, and are likely causally connected. We then assume that DI events which are separated by less then  $\rm \delta n^{DI}_{th} \times t^{\star, disk}_{dyn}|_{Last\, DI}$ are causally connected. In what follows, we will assume a  threshold value of  $\rm \delta n^{DI}_{th}\,{=}\,10$. We have checked that the results presented in this paper have a very weak dependence on the exact value of the threshold for $\rm \delta n^{DI}$, as long as the peak of the distribution is included in the sample.\\

  \begin{figure*} 
  \centering
  \includegraphics[width=2.\columnwidth]{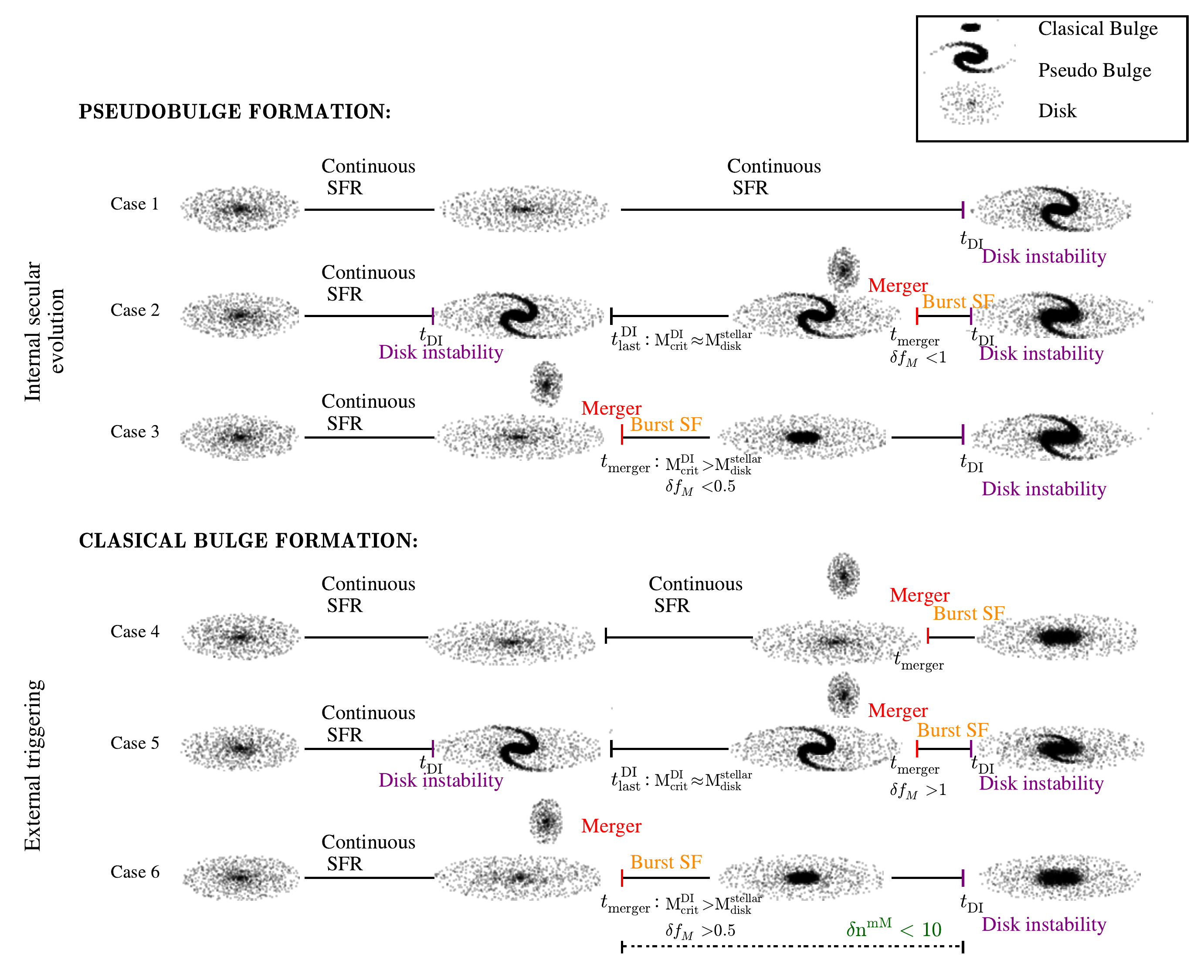}
  \caption{Possible paths of pseudobulge and classical bulge build-up. Following \protect\citet{KormendyandKennicutt2004} we have assumed that the build-up of pseudobulges is trigger by internal evolution (Case 1, 2 and 3). In the case of classical bulges, we have assumed an external mechanism of formation (Case 4, 5 and 6).}
  \label{fig:Pseudo_bulges_Pahways}
\end{figure*}

\subsubsection{From disk instabilities to bulges: \textit{merger-induced} vs. secular processes} \label{sec:DI_merger_induced}

Disk instabilities presented in \LGalaxies have been already used to study spheroidal components \citep[see][]{Shankar2012,Shankar2013}. In the model presented here, we re-visit the way DI are treated, by linking DI to the formation of both classical bulges and pseudobulges. Following the history and the physical conditions of the galaxy in which a DI takes place, we are able to distinguish between instabilities that are \textit{merger-induced} and the ones that are a consequence of the slow, \textit{secular} evolution of galaxies. Here we describe the details of how to discriminate between different instability events, and how these events lead to the build up of classical bulges and pseudobulges. 

On one hand,  \textit{merger-induced} DIs are produced as a consequence of the fast increase of stellar disk mass after the collisional starburst or \textit{smooth} satellite galaxy accretion. On the other hand, \textit{secular} DIs result from the slow, but continuous, mass growth of the disk, playing an important role in galaxies evolving in isolation. 
Under the assumption that bars are a consequence of the secular evolution of galaxies (\cite{Debattista2004,Debattista2006,MendezAbreu2010,Kormendy2013,Kim2016,Moetazedian2017,Zana2018A,Zana2018B}) and that bars lead to the formation of pseudobulges\footnote{Numerical simulations have shown that shortly after the bar formation the structure suffers a bending mode that thickens it and forms the boxy/peanut pseudobulge shape \citep[see][]{Combes1999,MendezAbreu2019}. Also galactic bars can produce, via gravitational torques, strong nuclear gas inflows which can be transformed into stars inducing the formation of disc-like pseudobulge structure.}, we link \textit{secular} DIs to the formation of galactic bars and pseudobulges. Therefore, we assume that the mass removed to the disk during the DI phase (according to Equation~\ref{equation:diskinestaMass}), is transferred to the pseudobulge, which we treat as a new component of the galaxy. On the other side, \textit{merger-induced} DI, closely associated with injection of external stars or/and SF burst triggered during the interaction, are assumed to be the ones that lead to the formation of a classical bulge structure. In Fig~\ref{fig:Pseudo_bulges_Pahways} we present an illustrative scheme of the scenarios that lead to the growth of both the pseudobulge and classical bulge component of the galaxy. \underline{Case 1} and \underline{Case 6} are the two simplest scenarios, as described above. In the first case, the galaxy experiences continuous star formation until the disk becomes unstable, forms a bar and the stellar component removed from the disk is transferred to the pseudobulge. In \underline{Case 6}, the galaxy, starting from a stable configuration, experiences a  merger (either a minor merger or a \textit{smooth accretion} as described in Section~\ref{sec:mergers_and_smooth_accretion}), which triggers a burst of SF that causes the disk to become unstable, and the stellar component removed from the disk to restore stability is effectively transferred to the classical bulge component.\\

\begin{figure} 
	\centering
    \includegraphics[width=1.0\columnwidth]{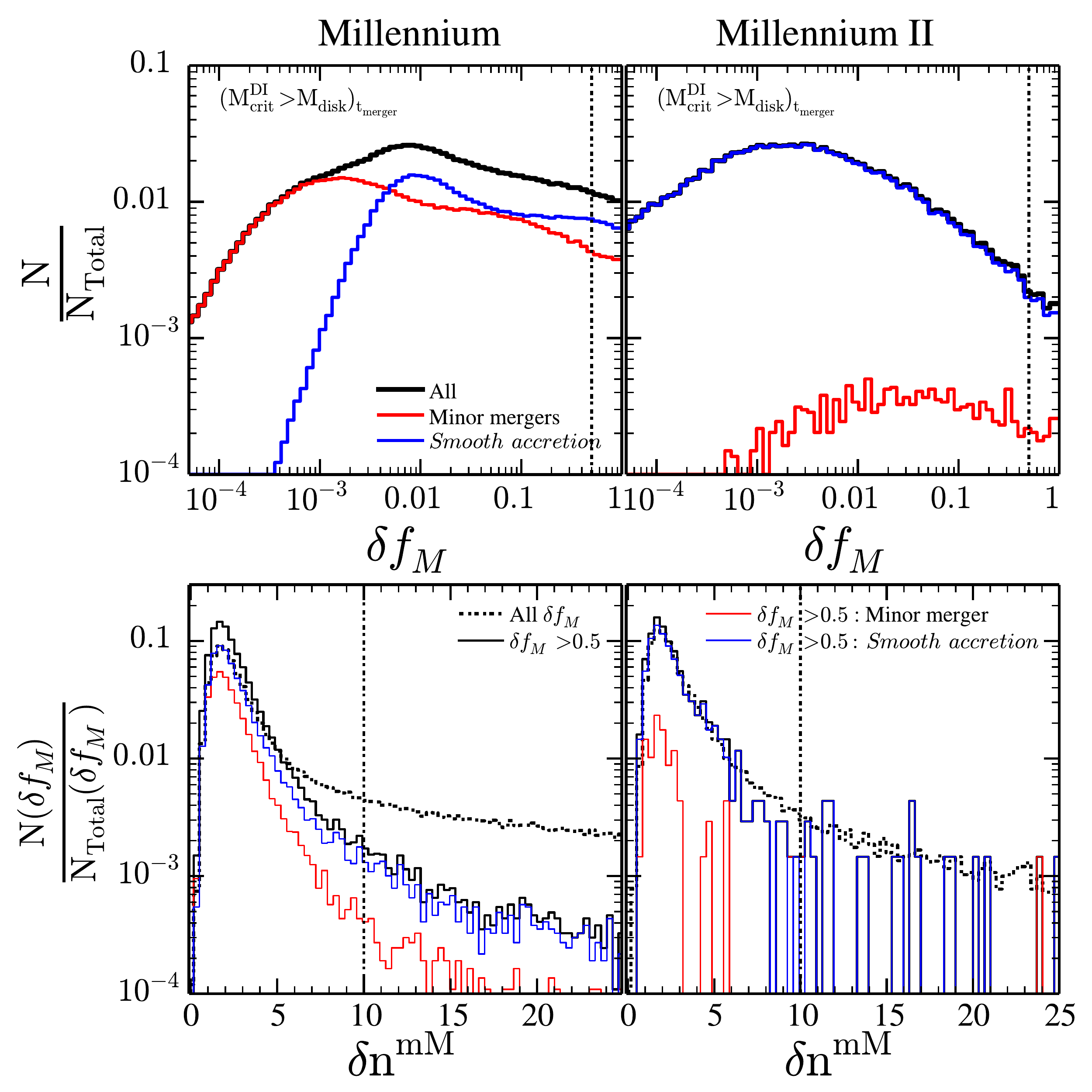}
 \caption{\textbf{Upper panels}: Distribution of $\delta f_{\rm M}$ for all galaxies that suffered a DI after a merger in which the disk was stable, i.e  $\rm M_{crit}^{DI} > M_{disk}^{stellar}$ at $\rm t = t_{merger}$. Black line represents all the events while red and blue only the ones after a minor merger and \textit{smooth accretion}, respectively. \textbf{Bottom panels}: In dotted lines the $\rm \delta n^{mM}$ distribution of all the events which satisfy Eq.\eqref{eq:delta_f}. Solid lines represent the same but imposing the extra condition of $\delta f_{M}\,{>}\,0.5$: in red minor mergers and in blue \textit{smooth accretion}. In all the panels, left and right columns display, respectively, the results for MS and MSII.} 
\label{fig:Minor_merger_event}
\end{figure}

However, the life of a galaxy can be rather complicated, with continuous mergers and episodes of star formation, that make it more difficult to discriminate between the two scenarios. 
Naively we could think that a DI which takes place right after a merger is a consequence of it. Nevertheless, this is not necessarily true. In order to quantify the importance of a minor merger or a \textit{smooth accretion} in triggering a DI event, we are going to study how efficient is the interaction in injecting new stars in a  stable (or marginally stable) disk. To check that, we introduce the quantity $\delta f_M$, defined as follows:
   \begin{equation}
     \label{eq:delta_f}
     \delta f_M = \left\{
	       \begin{array}{cc}
	       
		 \rm \frac{\Delta M_{stars}^{Burst} + \left[ M_{stellar}^{Satellite}\,\mathit{H}(\mathit{f}_{\rm binding}) \right]}{ \left(M_{crit}^{DI} - M_{disk} \right)}  \; \rm \;\; for \;  M_{crit}^{DI} > M_{disk} \; at \;\ t = t_{merger} \\ \\
		 \rm \frac{SFR_{isnt}^{merger}}{SFR_{isnt}^{sec}} \;\;\;\;\;\;\;\;\;\;\;\;\;\;\;\;\;\;\;\;\;\; 
              for \;    M_{crit}^{DI} \approx M_{disk} \; at  \;  t = t_{merger}, \\ 
	       \end{array}
	     \right.
   \end{equation}


where $\mathit{H}(\mathit{f}_{\rm binding})$ is a unit step function, whose value depend on the type of interaction: $ H\,{=}\,0$ for minor mergers ($\mathit{f}_{\rm binding}\,{>}\,\mathit{f}_{\rm binding}^{\rm th}$) and $ H\,{=}\,1$ for smooth accretion events ($\mathit{f}_{\rm binding}\,{<}\,\mathit{f}_{\rm binding}^{\rm th}$), as explained in Section~\ref{sec:mergers_and_smooth_accretion}. The first condition refers to events in which the disk is stable ($\rm M_{crit}^{DI}\,{>}\,M_{disk}^{stellar}$) at the time the merger takes place ($\rm t = t_{merger}$). In this case, $\delta f_{M}$ indicates how much the stellar disk grows with respect to how stable the disk is ($\rm M_{crit}^{DI} - M_{disk}$).  If the interaction is a minor merger the entire stellar component of the satellite is transferred to the bulge of the central, and the only new contribution to the disk is given by the burst of SF, $\rm \Delta M_{stars}^{Burst}$. In the case of \textit{smooth accretion}, the stellar disk of the central galaxy increases its mass not only through the SF burst, but also by incorporating the stellar component of the satellite, as described in Section~\ref{sec:mergers_and_smooth_accretion}.
The larger is the $\delta f_{M}$, the stronger the impact of the merger on the next DI event. In the  upper panels of Fig~\ref{fig:Minor_merger_event} we present the distribution of $\delta f_M$ for the MS (left panels) and MSII (right panels) for all the events which satisfy the first case of Eq.\eqref{eq:delta_f}. The values of $\delta f_{M}$ are shown separately for  \textit{smooth accretion} (blue) and minor merger events (red). The differences in the relative abundance between minor mergers and \textit{smooth accretion} in MS and MSII is just a consequence of resolution, as already discussed in Section~\ref{sec:mergers_and_smooth_accretion}. Except for the differences due to resolution effects, distributions of $\delta f_M$  for both minor merger and \textit{smooth accretion} peak at low values (${\sim}\, 0.001 - 0.01$) for both simulations. This points to the conclusion that most of the  interactions have a minimum contribution in the DIs happening after mergers.  The  small fraction of events characterized by high $\delta f_M$ values, however, can have a considerable impact on a subsequent disk instability.
To differentiate between interactions that are responsible to a DI and the ones that do not, we set the limit to $\delta f_M \,{=}\, 0.5$: only the minor mergers/\textit{smooth accretions} that reduce the galaxy disk stability ${\sim}\, 50\%$ are assumed to be responsible for the following DIs.\\


In addition to the $\delta f_{M}$ condition, we have also imposed that the mergers/\textit{smooth accretions} have to be causally connected to the subsequent DI, by imposing a maximum time scale that can pass between the two events, defined to be a multiple of the dynamical time of the galaxy stellar disk at the moment of the interaction. To set this value, we analyze the quantity $\rm \delta n^{mM}$,  defined as:
\begin{equation}
\rm \delta n^{mM} = \frac{t^{ Last}_{minor \, Merger} - t^{First}_{DI}}{t^{\star, disk}_{dyn}|_{minor\, Merger\,time}},
\end{equation}
where  $\rm t^{\rm Last} _{\rm merger}$ is the lookback time of the last minor merger or \textit{smooth} accretion, $\rm t^{\rm First}_{\rm DI}$ is the lookback time of the first DI after the galaxy interaction and $\rm t_{\rm dyn}^{\rm \star \, disk}|_{\rm minor \, Merger \, time}$ is the dynamical time of the galaxy stellar disk at the moment of the interaction. The distribution of $\rm \delta n^{mM}$ for both the MS and the MSII are shown in the lower panels of Fig~\ref{fig:Minor_merger_event}. 
As we can see, the distribution of $\rm \delta n^{mM}$ is more concentrated towards lower values when we consider only events with $\delta f_M \,{>}\, 0.5$ (solid black lines). Interestingly, the distribution of $\rm \delta n^{mM}$ peaks close to few number of dynamical times ($\rm \delta n^{mM}\,{\sim}\,2 -5$) with a very sharp decrease at large $\rm \delta n^{mM}$. This is a clear signal that the \textit{smooth accretion} and minor merger are responsible for the triggering of a subsequent instability. Moreover, when we distinguish between minor merger and \textit{smooth accretion}  (red and blue lines respectively), we find that the distribution of  $\rm \delta n^{mM}$ for minor mergers is more concentrated towards lower values than the one for  \textit{smooth accretion} events. This suggests that minor mergers are typically  able to destabilize the galaxy disk in shorter times scales than \textit{smooth accretion}. 

Based on these results, we set that all instabilities happening within $10$ dynamical times from the interaction ($\rm \delta n^{mM}= 10$), and for which $\delta f_M\,{>}\, 0.5$, are \textit{merger-induced}. These events thus lead to the growth of the classical bulge component (\underline{Case 6} in Fig~\ref{fig:Pseudo_bulges_Pahways}). On contrary, all DI events for which $\delta f_M < 0.5$ or $\delta f_M > 0.5$ and $\rm \delta n^{mM} \,{>}\, 10$ are assumed to be secular processes which contribute to the formation of a bar and a pseudobulge (\underline{Case 3} in Fig~\ref{fig:Pseudo_bulges_Pahways}).\\

Finally, the second case in Eq.\eqref{eq:delta_f} addresses the peculiar case in which a merger event happens in a galaxy characterized by a marginally stable disk after a DI, i.e $\rm M_{crit}^{DI}\,{\approx}\,M_{disk}^{stellar}$ (with $\rm M_{crit}^{DI}\,{\gtrsim}\,M_{disk}^{stellar}$). In this cases the DI is induced immediately ($\rm \delta n^{mM}\,{=}\,0$). However, it is difficult to say if the merger was a necessary phenomena to trigger a DI in the galaxy given that any event producing stars (either internal SF or merger burst) would rise the disk stellar mass over the critical threshold. Despite this cases are not as common as the ones described by the first case in Eq.\eqref{eq:delta_f} (less than the $10\%$ of the whole DI \textit{merger-induced}) we still take into account them by studying the relative importance of the interaction with respect to the continuous star formation happening in the disk in the triggering of the subsequent DI. When $\rm SFR_{isnt}^{merger}\,{>}\,SFR_{isnt}^{sec}$, i.e $\delta f_M\,{>}\,1$, we assume that the minor merger/\textit{smooth accretion} dominates the disk growth and the subsequent DI is \textit{merger-induced} (\underline{Case 5} in Fig~\ref{fig:Pseudo_bulges_Pahways}). Otherwise we assume the DI to  be  of \textit{secular} origin (\underline{Case 2} in Fig~\ref{fig:Pseudo_bulges_Pahways}).\\ 

\begin{figure}
\centering
\includegraphics[width=1.\columnwidth]{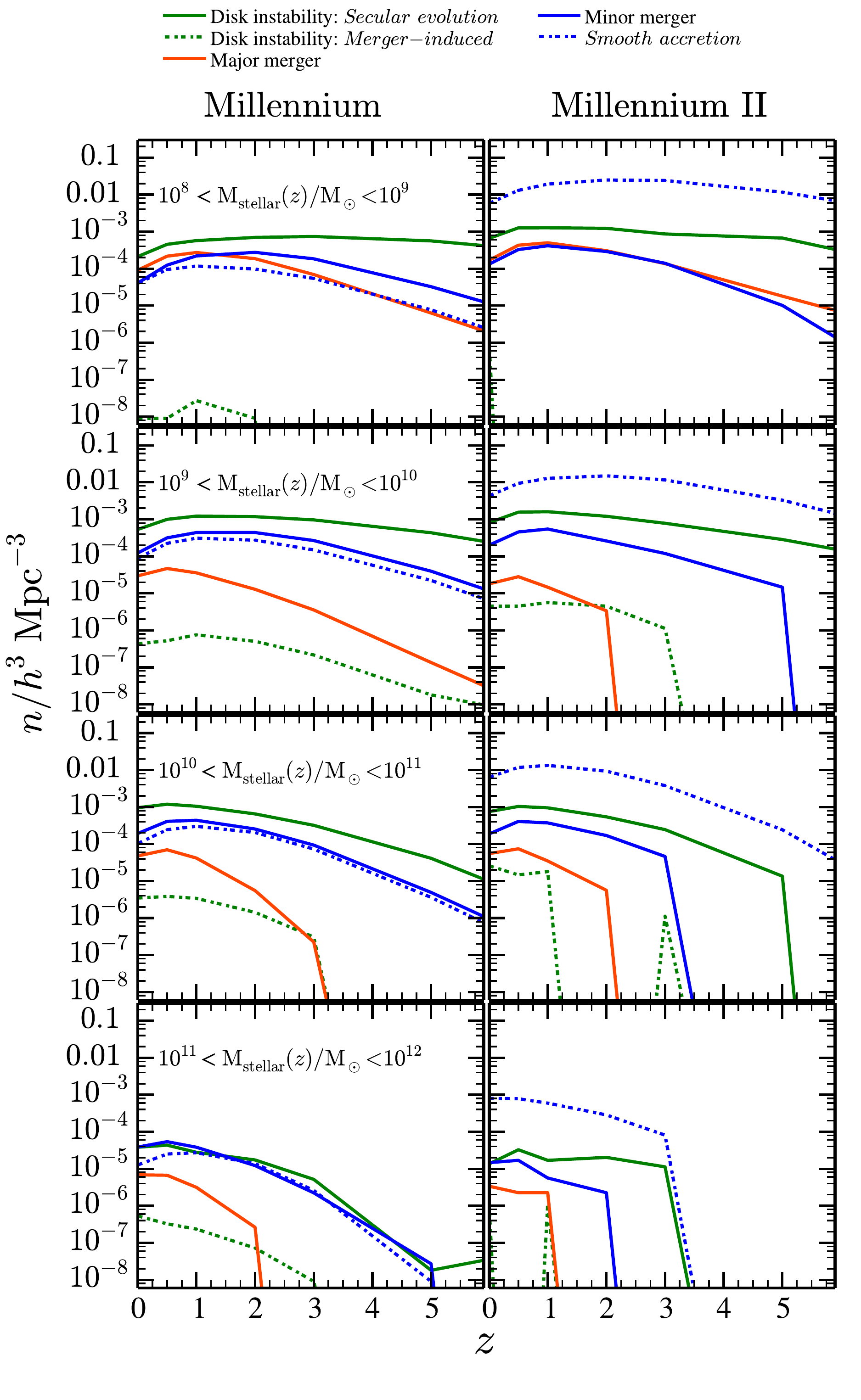}
\caption{Number density of 5 different type of events: \textit{secular} disk instability(solid green line), \textit{merger-induced} disk instability (dashed green line), major mergers (red solid line), minor mergers (solid blue line) and \textit{smooth accretion} (dashed blue line). The left and right columns display the results for the MS and MSII simulations, respectively. The different rows represent the number density for different stellar mass bins at a given redshift.} 
\label{fig:all_n_density_DI}
\end{figure}

In Fig~\ref{fig:all_n_density_DI} we present the predicted number density of \textit{secular} and \textit{merger-induced} DIs\footnote{In both types of DIs, we  correct for time-discretization effects using  the procedure presented in Section~\ref{sec:Linking_DI}}
(solid and dashed green lines, respectively) for different stellar mass ranges. Left and right panels are, respectively, the \LGalaxies predictions run on top of MS and MSII merger trees. For completeness, we have added the predictions of major/minor mergers and \textit{smooth accretion}. As we can see, \textit{secular} DIs evolution dominates the DI number density, being \textit{merger-induced} DIs  $\rm \,{\sim}\, 3 \, dex$ less abundant. Even though Fig~\ref{fig:all_n_density_DI} shows that DI \textit{secular} events dominate over mergers in all mass bins, this does not mean that the importance of such events is the same. For instance, galaxies with stellar mass in the range $\rm 10^8-10^9\,M_{\odot}$ \textit{secular} DI contributes with $\rm 10^5-10^6\,M_{\odot}$ to the bulge per event while galaxies with $\rm 10^{10} - 10^{11}\,M_{\odot}$ the DI \textit{secular} events are characterized by $\rm 10^8 - 10^{10}\,M_{\odot}$ of mass transferred (these numbers corresponds to DI events defined as in Section~\ref{sec:Linking_DI}).\\

About the redshift distribution, \textit{secular} evolution DIs take place at any redshift and they are the main mechanisms of bulge formation/growth at high-$z$. On contrary, \textit{merger-induced} DIs occur at $z\,{\sim}\,1$ with a sharp cut-off towards higher redshifts. Even more, they are rare events at $z\,{>}\,3$. Besides, we can see that \textit{merger-induced} DIs can not compete at any redshift with mergers (major/minor) in the classical bulge formation/growth given that their number density is a factor $100$ smaller and the amount of mass transferred per event is less than the $\sim 0.1$ \% of the whole galaxy stellar mass. Nevertheless, these events can  complement  classical bulge build-up at low redshifts. As can be seen, both MS and MSII predicts similar redshift distributions for the  \textit{merger-induced} DIs and \textit{secular} DIs, and similar values of number densities, even though the MSII  predicts slightly larger  number densities of \textit{merger-induced} DIs, as the number of \textit{smooth accretion} is much larger than in the MS.\\ 

We want to highlight that, in a hierarchical universe, each case presented in Fig~\ref{fig:Pseudo_bulges_Pahways} does not live in isolation. Due to the complex merger history that a galaxy can experience, the final bulge can be the result of a multiple physical processes, being a composite structure formed by both classical and pseudobulge component \citep[see,][]{Erwin2015,DiMatteoP2015,Fragkoudi2017,Balana2018}. 
For simplicity, we are going to use the following criteria to define galaxy bulge morphology: 
\begin{itemize}
\item {\bf Pseudobulge:} We assume that a galaxy hosts a pseudobulge when the fraction of bulge formed via  \textit{secular induced} DI is at least 2/3. This cut allows us to be confident about the fact that the pseudobulge is the dominant structure in the bulge. Galaxies hosting this type of structures are going to be tagged as \textit{pseudobulge galaxies}.\\

\item{\bf Classical bulge:} The fraction of bulge formed via  \textit{secular induced} disk instabilities is smaller than 2/3 of the total bulge mass and its \textit{bulge-to-total} is $\rm 0.01\,{<}\,B/T\,{<}\,0.7$. Galaxies with $\rm B/T\,{<}\,0.01$ are considered bulgeless galaxies. Galaxies hosting this type of bulges are going to be called \textit{classical bulge galaxies}.\\

\item{\bf Ellipticals:} The fraction of bulge formed via \textit{secular induced} disk instabilities is smaller than 2/3 of the total bulge mass the \textit{bulge-to-total} ratio is $\rm B/T\,{>}\,0.7$. Galaxies hosting this type of bulges are \textit{elliptical galaxies}.
\end{itemize}

\section{Results} \label{sec:results}

In this section we present the main findings of this work. We first focus on characterizing the properties of pseudobulges and host galaxies at different cosmic times. We then explore the structural properties of pseudobulges predicted for the local universe and compare with available data.

\subsection{Pseudobulges across cosmic time} \label{sec:occupancy}

In  Fig~\ref{fig:all_n_density_DI}  we have shown that \textit{secular} DIs are quite frequent at all cosmic times and for a broad range of stellar masses, although we discussed that the amount of mass transferred to the pseudobulge component is very modest for galaxies with  $\rm M_{stellar}\,{<}\,10^{9} M_{\odot}$. We thus do not expect a significant pseudobulge component in small galaxies.  Using the criteria described at the end of the last section to select pseudobulges, we study the properties of their hosts across cosmic time.

  \begin{figure} 
  \centering
  \includegraphics[width=1.0\columnwidth]{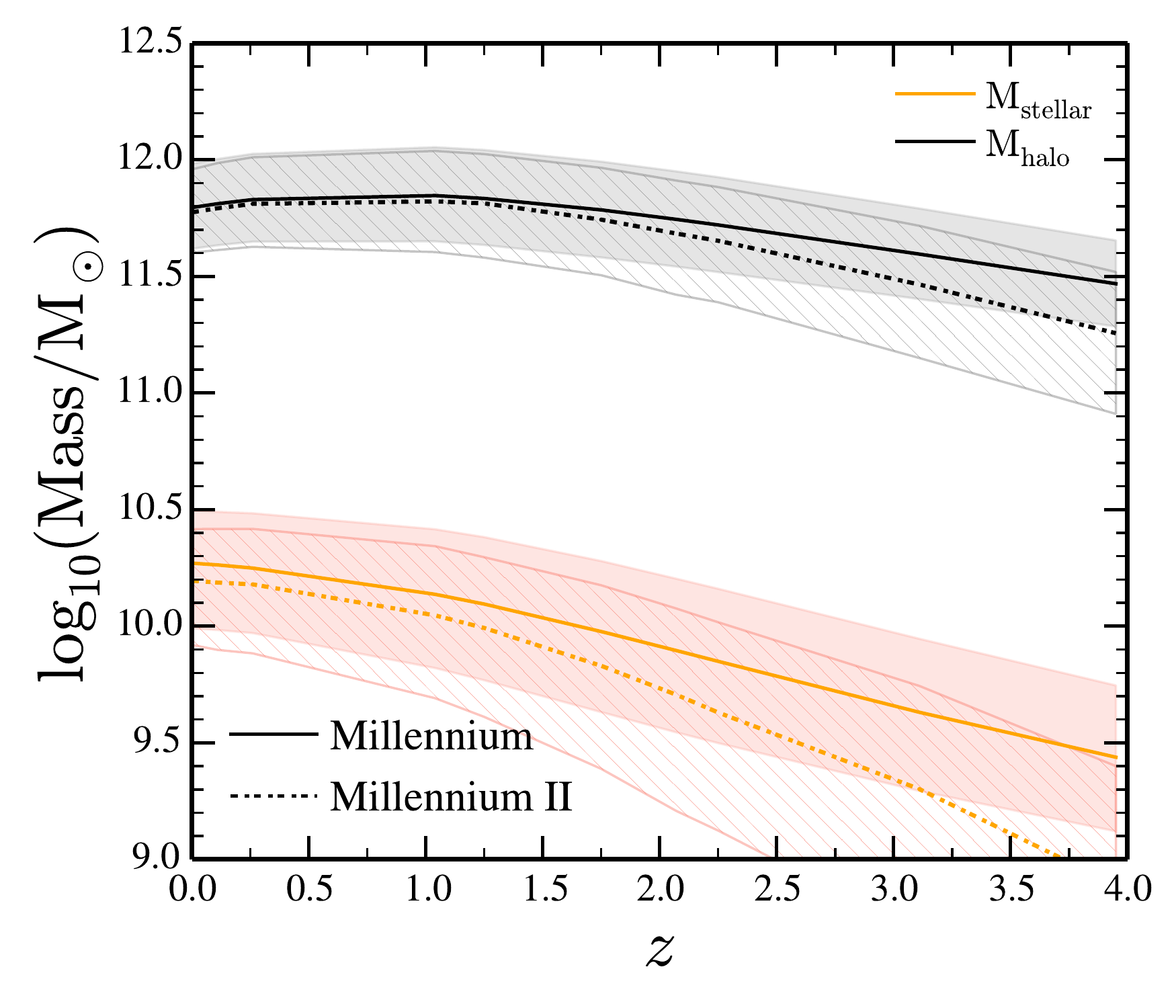}
  \caption{Median halo mass ($\rm M_{halo}$, black) and stellar mass ($\rm M_{stellar}$, orange) of galaxies classified as \textit{pseudobulge galaxies} at different redshift. The selection of this population at any redshift were done following the definition presented in Section~\ref{sec:DI_merger_induced}. The solid line and shadow region represents the median and $\rm 1\sigma$ values for MS. The dashed line and the lined area with lines symbolizes the same but for MSII.}
  \label{fig:typical_stellar_halo_pseudo}
  \end{figure}
  
In Fig~\ref{fig:typical_stellar_halo_pseudo} we present the typical halo and stellar masses of galaxies hosting pseudobulges at different redshifts, for both MS (solid line) and MSII (dashed line).  As expected, galaxies hosting pseudobulges are tipically more massive than $\rm M_{stellar}\, {\sim}\, 10^{9} M_{\odot}$. In particular, we find that pseudobulges tend to be hosted by galaxies in a relatively small range of stellar masses, with values mildly evolving with time. In the MS the typical host stellar mass grows from $\rm M_{stellar} \,{\sim}\, 10^{9.5} M_{\odot}$ at $z\,{>}\,2.5$ to $\rm M_{stellar}\,{\sim}\, 10^{10.3} M_{\odot}$ at $z\,{=}\,0$. A similar trend is shown also by MSII, even though smaller masses are reached at higher redshifts. Despite this little difference, the MS and MSII simulations agree within $1\sigma$ confidence level at any redshift. On the halo mass side, pseudobulges are hosted in Milky Way-like halos (i.e $\rm \,{\sim}\, 10^{11.8} M_{\odot}$) at $z\,{=}\, 0$ in both MS and MSII. Moreover, the typical halo mass evolution seems to be truncated at $z\,{\sim}\,1.25$, where the increasing trend exhibited from $\rm M_{halo}\,{\sim}\,10^{11.4} M_{\odot}$ at $z\,{=}\,4$ up to $\rm \,{\sim}\, 10^{11.9} M_{\odot}$ at $z\,{=}\,1.5$, changes into a decreasing tendency. We interpret this as a consequence of the hierarchical growth of structures: pseudobulge galaxies are less likely to be hosted by very massive halos at low-$z$, as these halos are closely related to major mergers events who deeply impact the host galaxy structure erasing any secular evolution characteristic.\\

   \begin{figure} 
  \centering
  \includegraphics[width=1.\columnwidth]{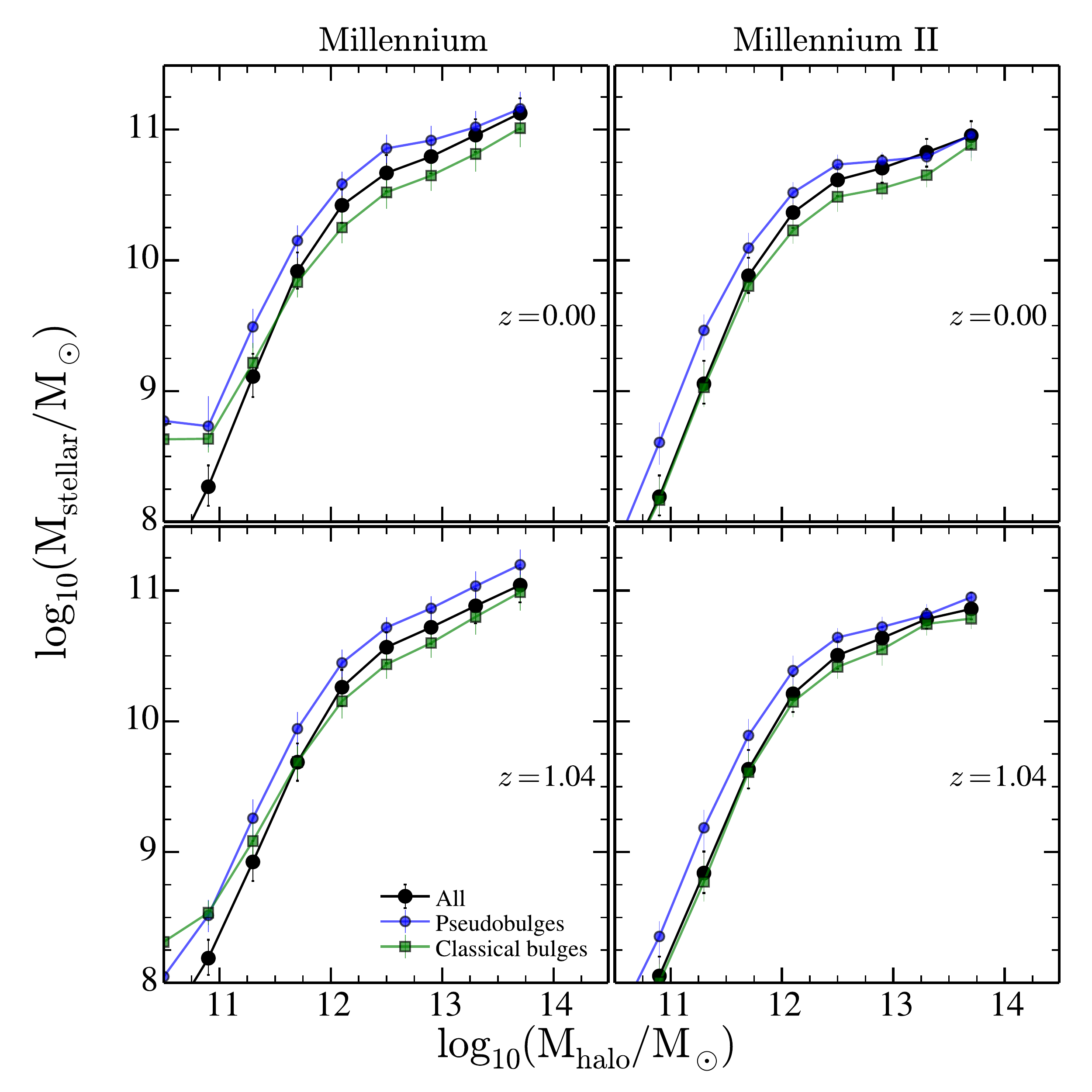}
  \caption{Halo - Stellar mass relation for pseudobulges (blue) classical bulge (green) and all (black) galaxies in MS (left) and MSII (right) at different redshifts ($z \,{=}\, 0$ (top), $z\,{=}\,1$ (bottom)). The error bar represents the $1\sigma$ dispersion. Here we present the relation for central and satellite galaxies. By dividing the galaxies in central and satellites we find the same trend.}
  \label{fig:halo_stellar_relation}
  \end{figure}
  

To understand if pseudobulge are hosted in peculiar type of galaxies with respect to the standard population, in Fig~\ref{fig:halo_stellar_relation} we show the $\rm M_{halo}\,{-}\,M_{stellar}$ plane at different redshifts for MS and MSII (right and left respectively)\footnote{We have done the same plot dividing between central and satellite galaxies. No difference with the Fig~\ref{fig:halo_stellar_relation} has been found.}. As can be seen in the relation, galaxies which display a dominant pseudobulge structure (blue dots) are systematically above the galaxy median relation (black dots) at any redshift, i.e., at fixed halo mass, pseudobulge structures are hosted by galaxies more massive than the median population. On contrary, when we place the classical bulge galaxies (green dots) on the plane, it is evident that they populate a different region. While for  $\rm M_{halo}\,{<}\,10^{12} M_{\odot}$ classical bulges lie on the median relation, in the most massive halos ($\rm M_{halo}\,{>}\,10^{12} M_{\odot}$) their host galaxies are characterized by systematically smaller stellar masses (notice that the results for the range $\rm M_{halo}\,{<}\,10^{11} M_{\odot}$  suffer of low-resolution statistics in the case of the MS). As we will see later in Section~\ref{sec:Pseudo_char}, pseudobulges are typically hosted by star forming galaxies, while classical bulges tend to live in more quenched systems, explaining why pseudobulges tend to have higher stellar content than classical bulges, at a fixed halo mass.

\begin{figure} 
  \centering
  \includegraphics[width=1.0\columnwidth]{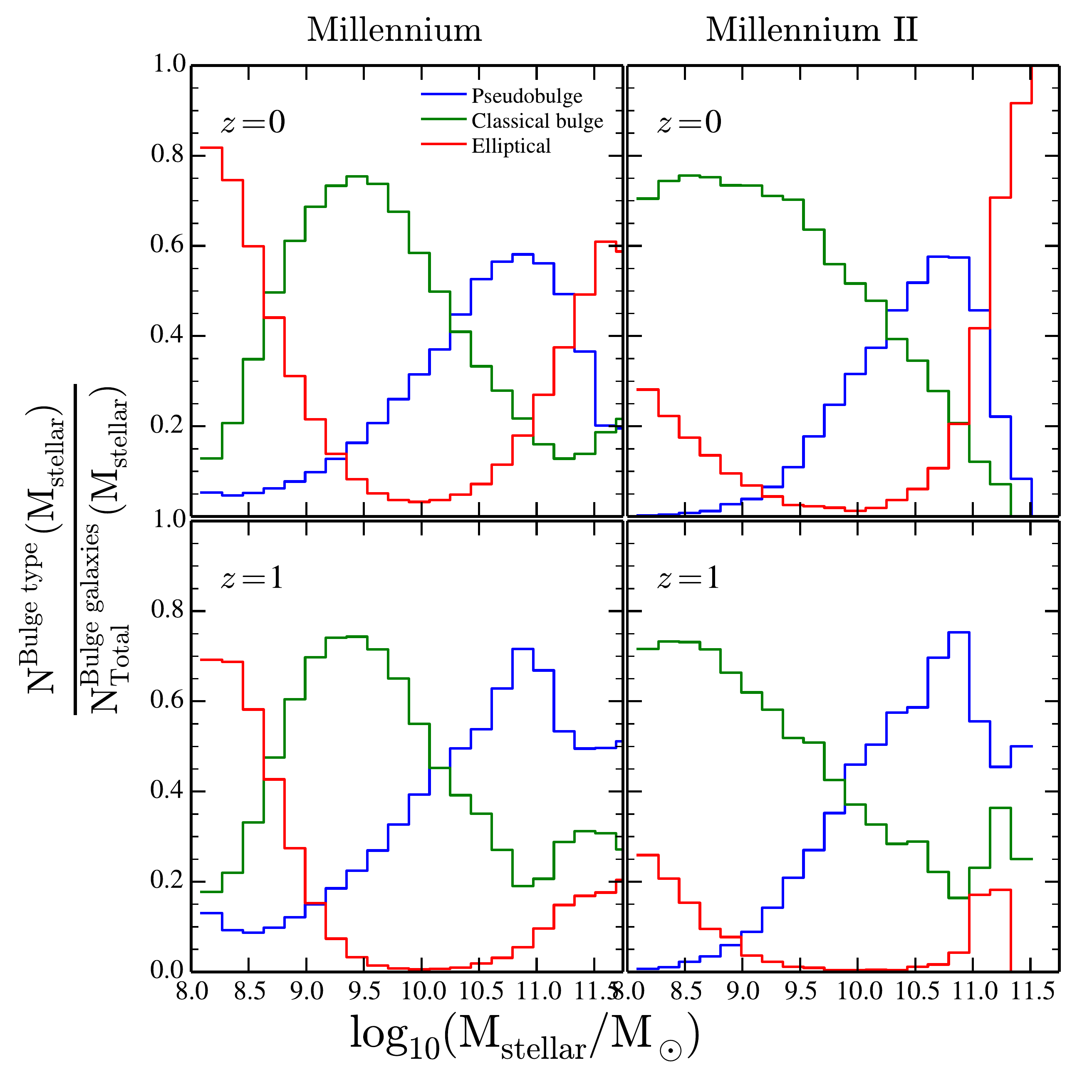} 
  \caption{Relative contribution of pseudobulge (blue), classical bulges (green) and elliptical (red) galaxies in the bulge galaxy population. Upper and lower rows represents the results at $z\,{=}\,$ 0 and 1, respectively. In the left the results for MS and in the right for MSII.}
  \label{fig:Bulges_ocupancy}
  \end{figure}

\begin{figure} 
\centering
\includegraphics[width=1.0\columnwidth]{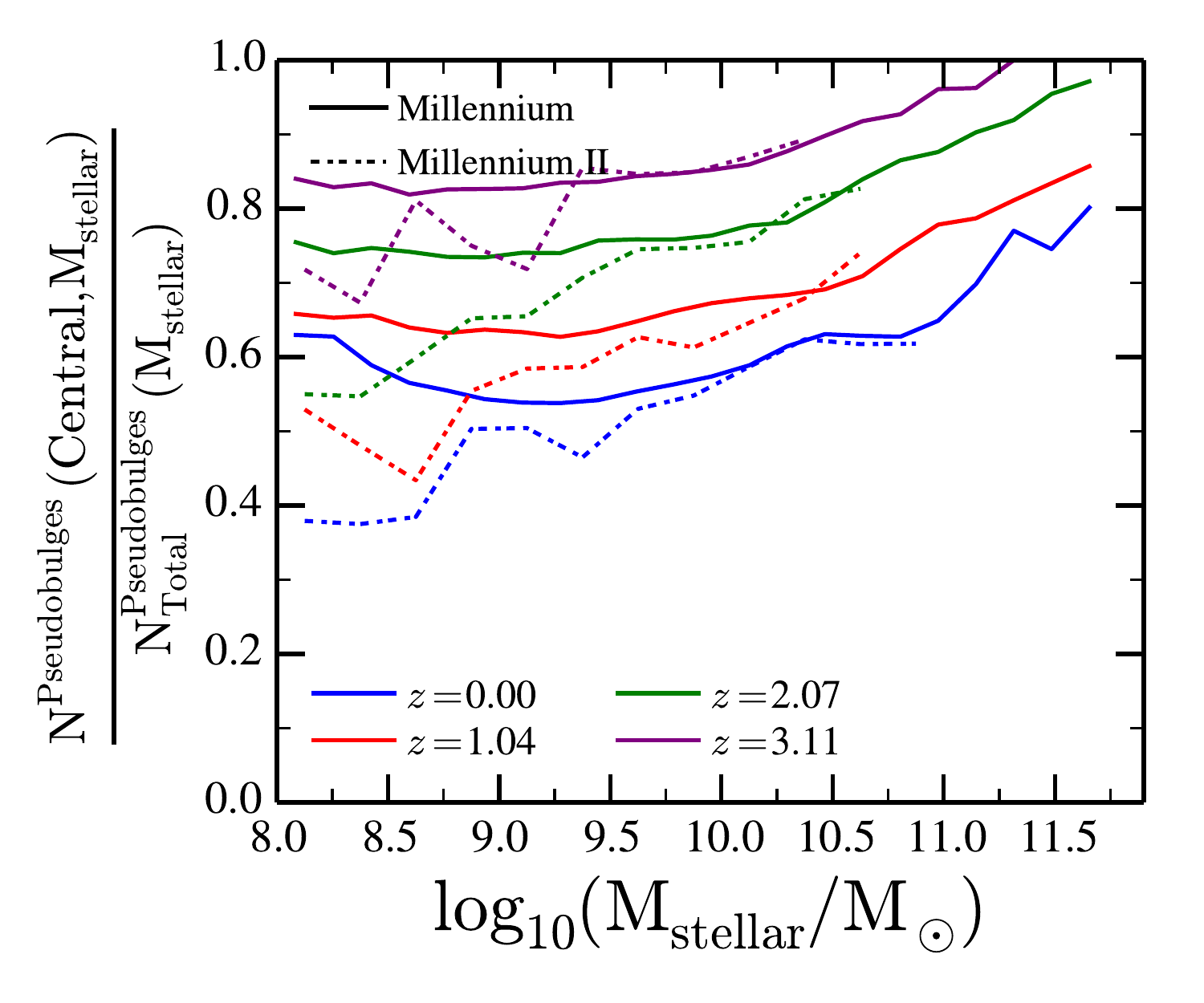} 
\caption{Fraction of pseudobulge galaxies that are centrals of their \textit{friend-of-friend} halo. Solid and dashed lines represent the results for MS and MSII, respectively. Pseudobulges galaxies at different redshift were selected following the definition presented in Section~\ref{sec:DI_merger_induced}. For MSII it is plotted until the stellar masses in which the simulation predicts, at that redshift, a total number of objects larger than 10.}
\label{fig:Ocupacy_pseudo}
\end{figure}

To see the relative importance of different classes of galaxies, in Fig~\ref{fig:Bulges_ocupancy} we show the relative contribution of pseudobulges (blue), classical bulges (green) and elliptical (red) galaxies to the total population of galaxies with a bulge (i.e. $\rm B/T\,{>}\,0.01$), at different stellar masses. Results at $z \,{=}\, 0$ and $z\,{=}\, 1$ are shown respectively in first and second row. 
At $z\,{=}\,1$ both MS and MSII show that pseudobulges are the main type of galaxies at large stellar masses (i.e, $\rm M_{stellar}\,{>}\,10^{10.5} M_{\odot}$) while classical bulges and ellipticals are the main ones for $\rm M_{stellar}\,{<}\,10^{10.5} M_{\odot}$. While at low masses there is little evolution between $z\,{=}\,1$ and $z\,{=}\,0$, at high masses we find that, by $z\,{=}\,0$, ellipticals dominate the galaxy population. As previously discussed, this is a result of the hierarchical growth of structures: pseudobulges hosted in the most massive galaxies at high-z are subsequently destroyed by major mergers which turn galaxies into pure bulges (see example $f$ of Fig~\ref{fig:Merger_Trees}). Additional support to this picture can be gained by studying the fraction of pseudobulge galaxies who are  \textit{centrals} of  their FoF group, as a function of redshift. Fig~\ref{fig:Ocupacy_pseudo} shows that $\,{\sim}\, 80\%$ of pseudobulges at $z\,{=}\,3$ were centrals (with larger fractions reached for more massive galaxies stellar mass), and the percentage  drops to $\,{\sim}\, 60\%$ at $z \,{=}\, 0$. This points out that pseudobulge galaxies are less likely to be hosted in the central subhalo of their FOF at low $z$, independently of their stellar mass. This trend is well followed by both MS and MSII at high stellar masses, while at small ones MSII predicts slightly larger fractions of satellite pseudobulges. Note that this difference is consequence of the MS halo mass resolution issues at $\rm M_{stellar} < 10^9 M_{\odot}$. Therefore, we rely in the MSII predictions whose limitation is at $\rm M_{stellar}\,{\sim}\,10^{8} M_{\odot}$.\\

\begin{figure} 
 \centering
 \includegraphics[width=0.75\columnwidth]{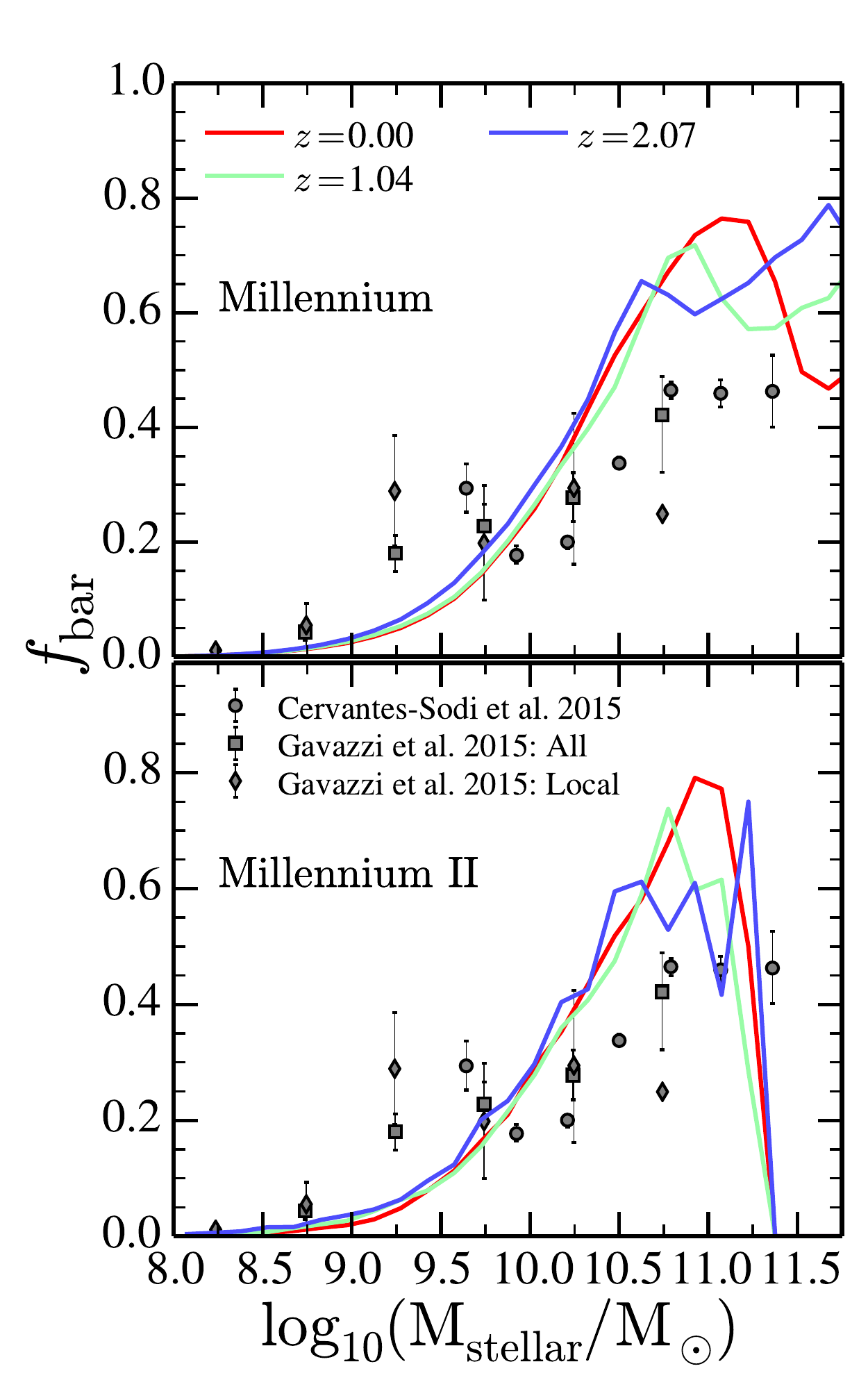} 
 \caption{Bar fraction $f_{\rm bar}$ in the MS (top) and MSII (bottom). We define $f_{\rm bar}$ as the number of galaxies hosting a pseudobulge (bar, boxy/peanut or disc-like structure) over the total spiral galaxies population ($\rm B/T\,{<}\,0.7$) in a given mass bin and redshift. Colors encodes different redshifts. We compare this with recent observation by \protect\cite{Cervantes2015} and \protect\citet{Gavazzi2015}. }
\label{fig:bar_fraction}
 \end{figure}

Finally, in Fig~\ref{fig:bar_fraction} we present the bar fraction $f_{\rm bar}$ as a function of redshift and stellar mass for the two simulations. We have defined $f_{\rm bar}$ as the number of galaxies hosting a pseudobulge 
over the total number of spiral galaxies ($\rm B/T\,{<}\,0.7$) in a given bin of mass and redshift. The fraction  of pseudobulge in spiral galaxies has a peak at $\rm M_{stellar} \,{\sim}\,10^{10.5} M_{\odot}$ with a sharp cut-off towards low stellar masses. This trend is broadly in agreement with the  observational results of \citet{Cervantes2015} and \cite{Gavazzi2015}. The fact that our predictions lie above is reassuring, as we regard our fraction as upper limits, given that a fraction of galaxies that we tag as pseudobulges might not have a clear detectable bar.\\

\subsection{Pseudobulges and their hosts in the local universe}
In this last part we analyze the properties of $z\,{=}\,0$ \textit{pseudobulge galaxies} such as star formation and stellar age (Section~\ref{sec:Pseudo_char}), structural properties (Section~\ref{sec:Gadotti}), redshift of the last major/minor interaction and pseudobulge structure formation (Section~\ref{sec:merger_interaction_P_E_C}).


\subsubsection{Star formation in pseudobulge galaxies} \label{sec:Pseudo_char}
\begin{figure} 
 \centering
 \includegraphics[width=1.0\columnwidth]{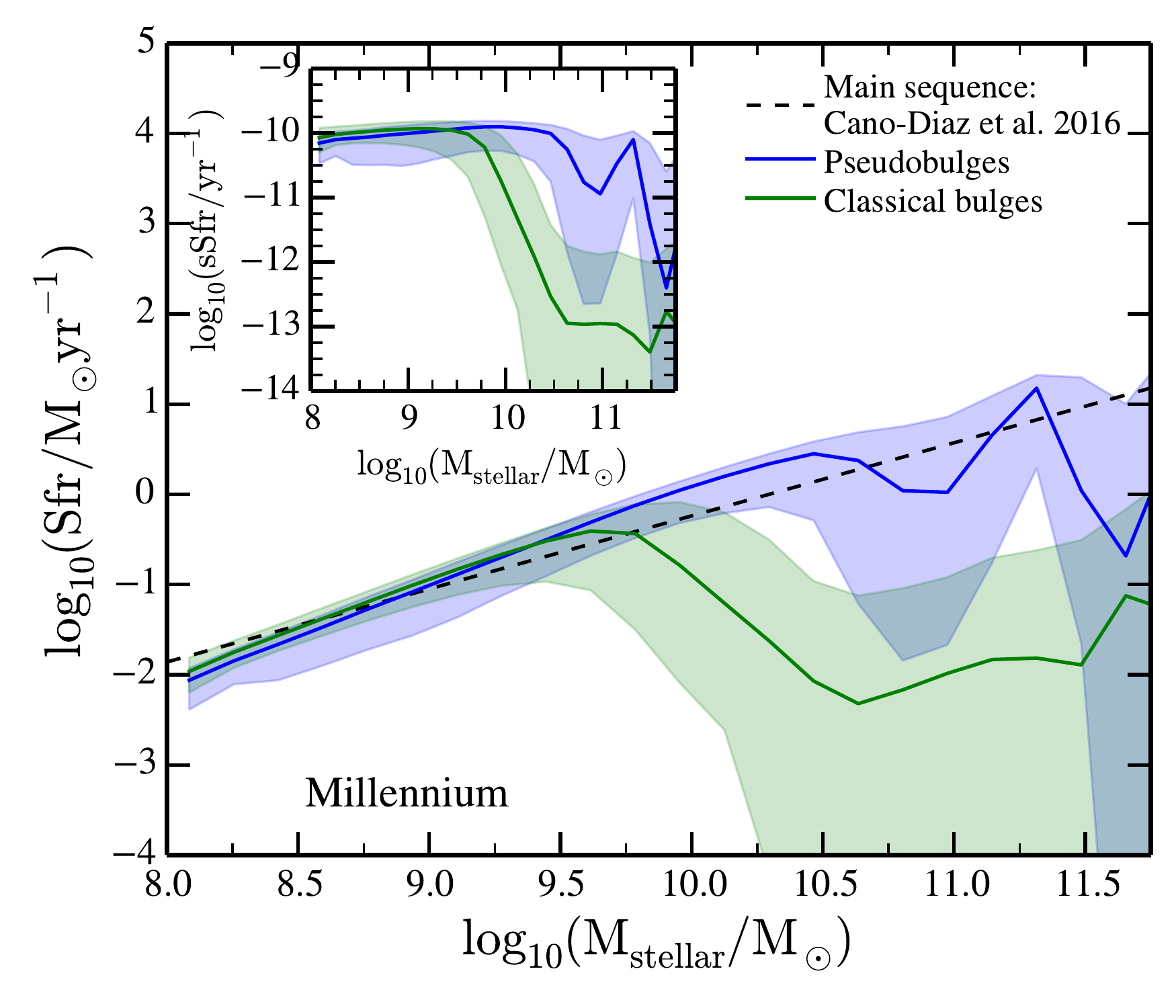} 
 \caption{Star formation rate (Sfr) - $\rm M_{stellar}$ plane for $z\,{=}\,0$ pseudobulges (blue) and classical bulges (green) galaxies predicted by MS. The solid lines represent the median of the distribution while the shaded area represents the $\rm  1 \sigma$ dispersion. Black dashed line is the main sequence fit of \protect\cite{CanoDiaz2016}. The inset plot represents the same but for the plane specific star formation rate (Ssfr) - $\rm M_{stellar}$.}
 \label{fig:Main_sequence}
 \end{figure}

In the previous section we have seen that, at fixed host halo mass, pseudobulges tend to live in galaxies more massive than what predicted by the median $\rm M_{halo}$-$\rm M_{stellar}$ relation. Pseudobulge galaxies thus seem to not suffer from the same quenching mechanisms that other galaxies experience (e.g., AGN feedback) and that cause massive galaxies to be inefficient star forming engines. When looking at the star formation properties of local pseudobulges predicted by the model, we find indeed that galaxies hosting a pseudobulge are efficient in producing stars. 
We show this in Fig~\ref{fig:Main_sequence}, where pseudobulges and classical bulges predicted by the MS are shown in the Sfr-$\rm M_{stellar}$ plane. To guide the reader we have added in dashed black line the \textit{main sequence}\footnote{The main sequence is defined as the relation of actively star-forming galaxies which relates their star formation rate and their stellar mass \citep{Brinchmann2004,Noeske2007,CanoDiaz2016}} of star formation form \cite{CanoDiaz2016}. At $\rm M_{stellar} \,{<}\, 10^{9.5} M_{\odot}$ the hosts of both classical and pseudobulges follow the main sequence. 
At higher stellar masses, however ($\rm M_{stellar} > 10^{9.5} M_{\odot}$), pseudobulges and classical bulges follow two different trends. While the former population remains on the main sequence and only starts deviating for very massive systems, classical bulges present a clear shift in their relation,  falling in the \textit{red sequence} region with $\rm \,{\sim}\, 2 \, dex$ of lower star formation than pseudobulges. In the inset of Fig~\ref{fig:Main_sequence} we show the plane specific star formation rate (sSFR) - $\rm M_{stellar}$. As we can see, the trend is similar to the $\rm Sfr\,{-}\,M_{stellar}$ one.\\

\begin{figure}
\centering
\includegraphics[width=1.0\columnwidth]{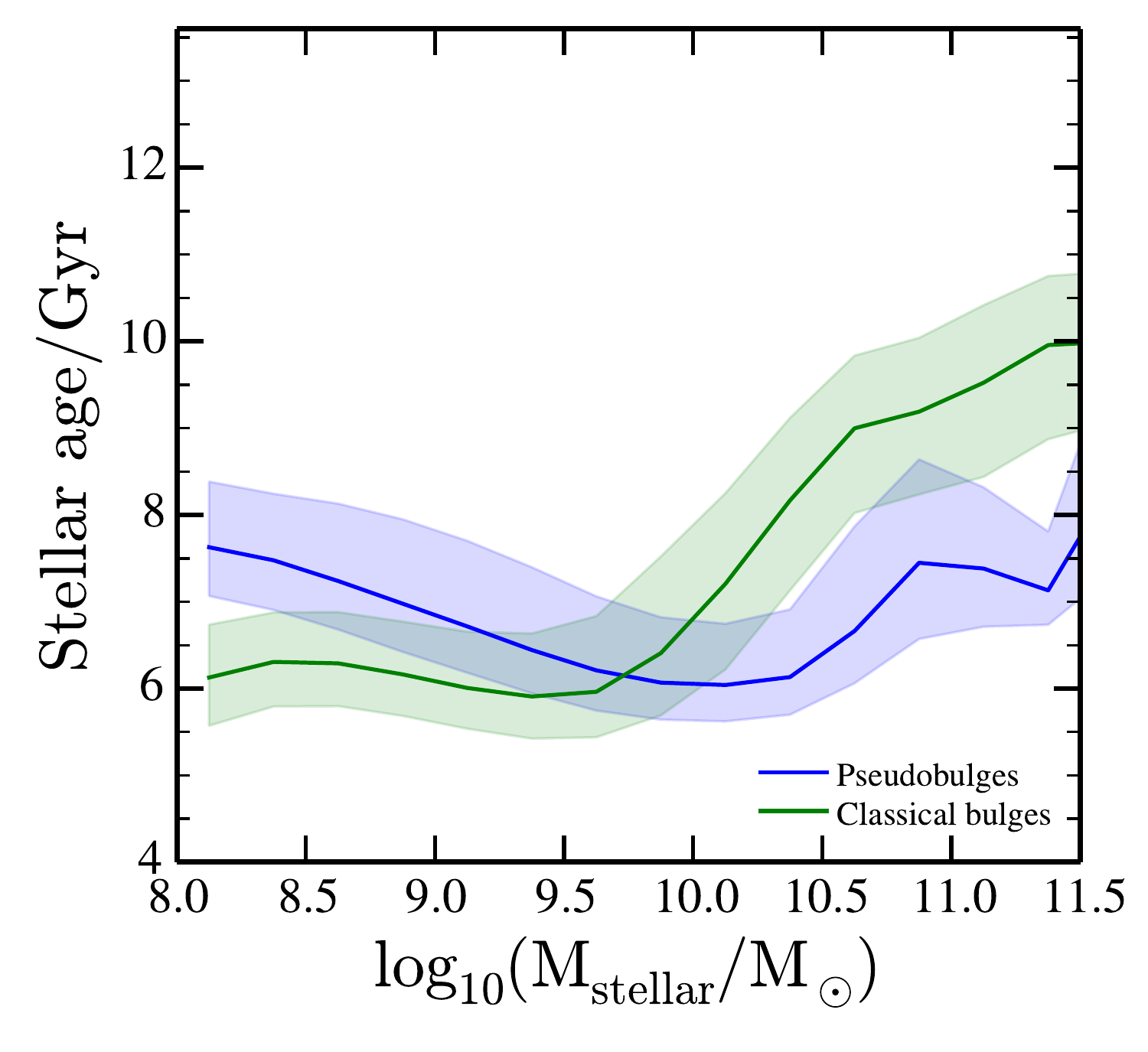}
\caption{Stellar population age (mass weighted age) of $z = 0$ pseudobulge (blue) and classical bulge (green) galaxies predicted by MS simulation (MSII display similar behavior). The solid lines represent the median of the distribution while the shaded area represents the $1 \sigma$ dispersion.}
\label{fig:age}
\end{figure}

Fig~\ref{fig:age} shows instead the mass-weighted age of the stellar population in pseudobulge and classical bulge galaxies. While the typical age of stars in pseudobulge galaxies seems shows a very weak dependence with stellar mass,  classical bulges hosted in massive galaxies $\rm M_{stellar} \,{>}\,10^{10.5} M_{\odot}$ are significantly older. 
At low masses (i.e, $\rm M_{stellar} \,{<}\, 10^{9.5} M_{\odot}$), instead, classical bulges are hosted by galaxies with slightly younger average stellar populations. This is due to the different merger history of classical bulges and pseudobulges as we will show in Fig~\ref{fig:Time_mergers}. In this mass range, in fact, almost all the classical bulges experienced at least one minor (major) mergers at $z\,{\sim}\,0.5$ which rejuvenated the galaxy population via SF burst. On contrary, pseudobulges did not suffered any (major) minor merger and their stellar population only grew via internal star formation. 
We remark here that also classical bulges in massive galaxies experience mergers at recent times, but, as discussed above, star formation in these galaxies is quenched by AGN feedback and the cold gas fraction in these massive systems is lower. 
The MSII gives very similar results, do not shown here to avoid redundancy.

\subsubsection{Structural properties of pseudobulge galaxies}\label{sec:Gadotti}

 \begin{figure*} 
  \centering
  \includegraphics[width=0.49\textwidth]{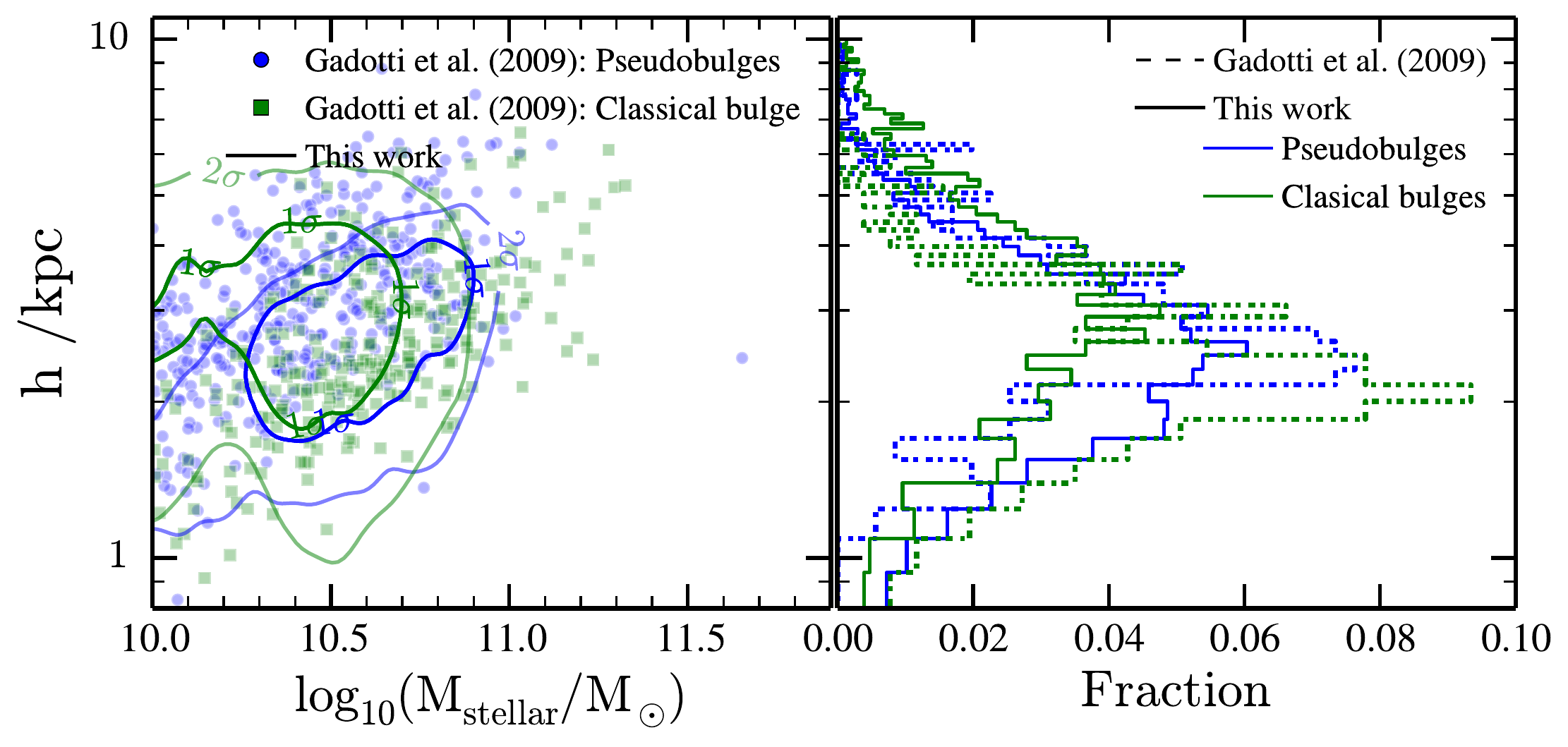}
  \includegraphics[width=0.49\textwidth]{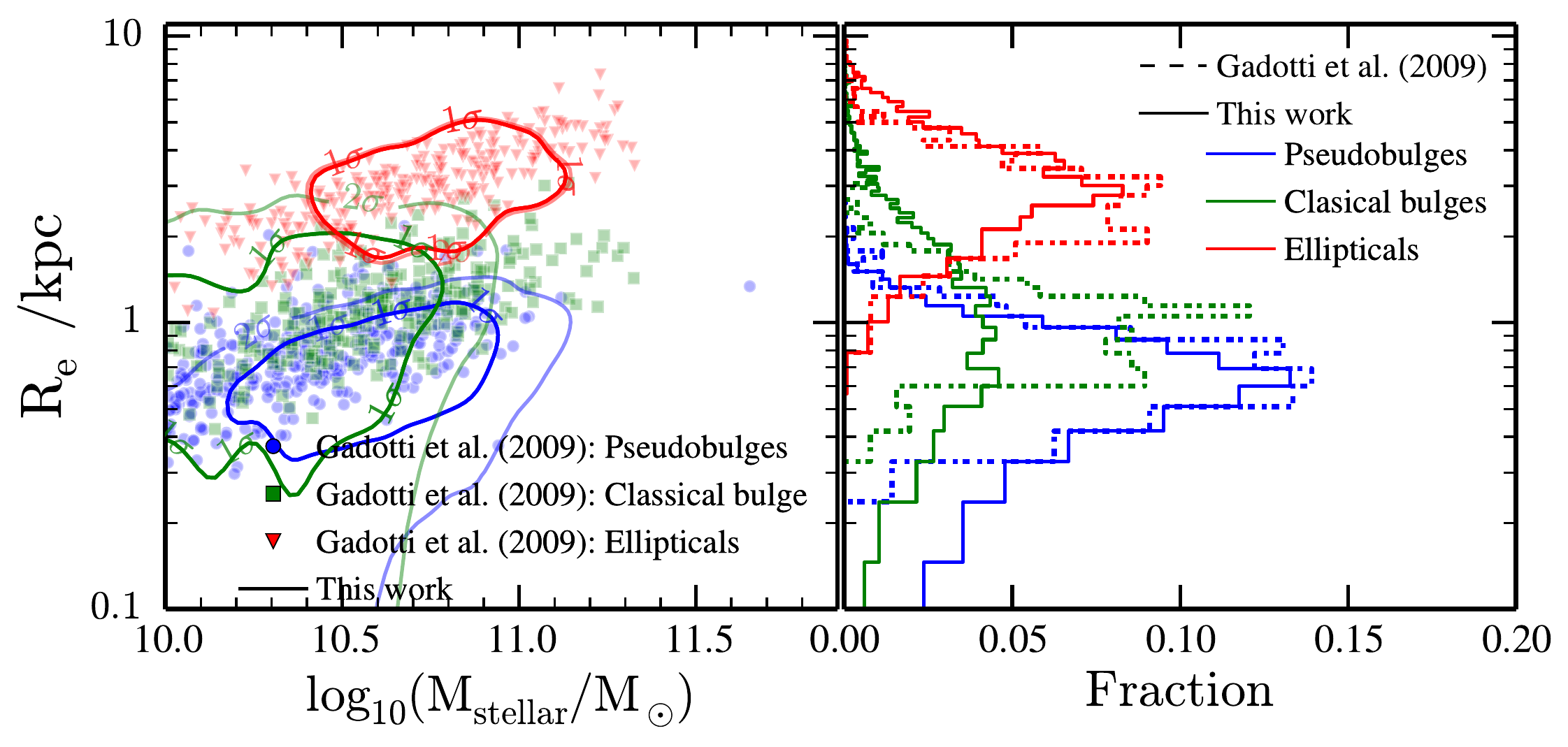}
  \includegraphics[width=0.49\textwidth]{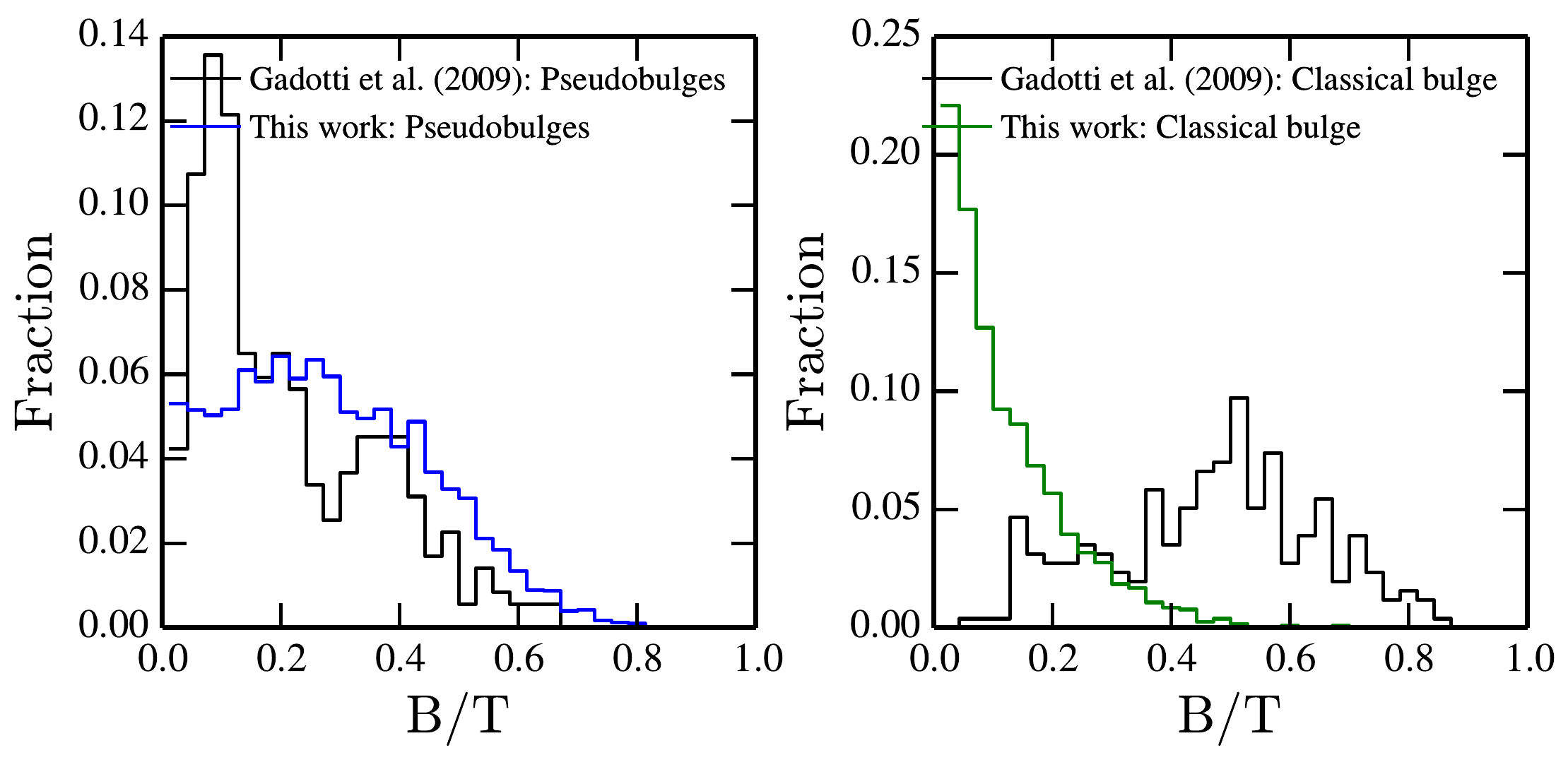}
  \includegraphics[width=0.49\textwidth]{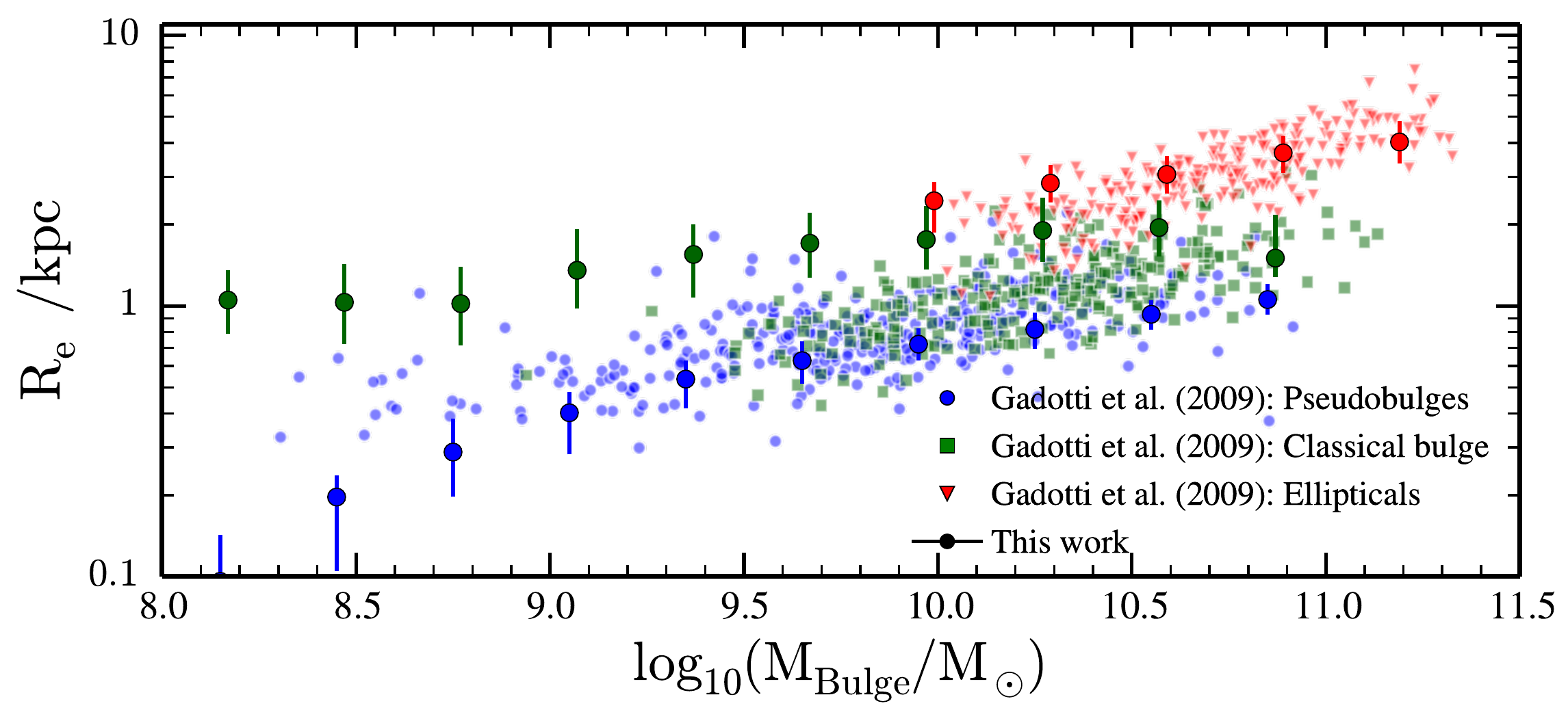}
  \caption{\textbf{Upper left}: In the left panel the distribution of disk scale length ($\rm h$) as a function of the stellar mass ($\rm M_{stellar}$) for pseudobulges (blue points) and classical bulges (green squares) galaxies. The contours represent the 1 and 2 $\sigma$ dispersion of the relation $\rm h-M_{stellar}$ predicted in this work in the MS simulation. In the right panel we represent the $\rm h$ distribution for pseudobulges (blue) and classical bulges (green) galaxies. Dashed and solid histogram display respectively the observed and predicted distribution. \textbf{Upper right}: In the left, the distribution of bulge effective radius ($\rm R_e$) as a function of the stellar mass for pseudobulges (blue points), classical bulges (green squares) and ellipticals (red triangles) galaxies. The contours represent the 1 and 2 $\sigma$ dispersion of the relation $\rm R_e-M_{stellar}$ predicted in this work. In the right, the $\rm h$ distribution for pseudobulges (blue), classical bulges (green) and ellipticals (red) galaxies. Dashed histogram display the observed distribution while the solid one the predicted. \textbf{Lower left}: In the left and right, the distribution of \textit{bulge-to-total} ratio B/T for pseudobulges and classical bulges galaxies. In black the observed distribution and in blue and green the predicted one for pseudobulges (blue) and classical bulges (green) galaxies. \textbf{Lower right}: Relation of bulge mass ($\rm M_{bulge}$) vs effective radius of the bulge ($\rm R_e$) for pseudobulges (blue), classical bulges (green) and ellipticals (red) galaxies. The points represent the median of each sample while the error bars the $\rm 1\sigma$ dispersion.}
  \label{fig:Gadotti_comparation}
\end{figure*}

We now move to the analysis of the structural properties of the pseudobulges that our model predicts and compare our predictions with the observational  results of \cite{Gadotti2009}. In that work Gadotti studied bulge properties, such as B/T ratios, bulge masses and scale length parameters, for a sample of 963 galaxies with masses $\rm 10^{10} \,{\lesssim}\, M_{stellar} \,{\lesssim}\, 10^{11.5} M_{\odot}$ and redshift range $0.02\,{\leq}\,z\,{\leq}\, 0.07$. The sample was divided in disk galaxies and ellipticals. The former ones were further sub-divided in galaxies hosting a pseudobulge, classical bulge or bulgeless. To compare with the observations, we  generated a galaxy sample using the MS which reproduces the exact stellar mass selection of  \cite{Gadotti2009}, but with a much larger number of galaxies (about a factor of ten). We could not do the same exercise with the MSII, as the smaller box does not allow to properly sample the most massive galaxies. The definition of pseudobulges, classical bulges and ellipticals is the one used in this work, as presented at the end of subsection~\ref{sec:DI_merger_induced}. \\

Results for pseudobulge, classical bulge and elliptical galaxies in MS are presented in Fig~\ref{fig:Gadotti_comparation}. We show the distribution of disk scale length ($\rm h$) as a function of the stellar mass (top left), the distribution of the bulge effective radius ($\rm R_e$) versus stellar mass (top right), the $\rm B/T$ distribution and the effective bulge radius as a function of the bulge mass (bottom left and right, respectively). Overall, the structural parameters $\rm h$ and $\rm R_e$ for pseudobulges, classical bulges and elliptical galaxies are reasonably well reproduced by the model. Nevertheless, classical bulges show $h$ values which are slightly offset, i.e. disks are larger than the observed one, and the number of massive disks is also larger than observed.\\ 
Regarding bulge to total ratios, pseudobulges broadly follow the distribution found by \cite{Gadotti2009}, even though we seem to lack pseudobulges with very small B/T ratios. Other studies, however, found that most pseudobulges are hosted by galaxies with $\rm B/T>0.2$ \citep[see][]{Fisher2008,Fisher2010}, which is consistent with our results. On the other hand, the  $\rm B/T $ distribution for hosts of classical bulges peaks at $\rm {\sim}\,0.1$ with a fast decrease towards large B/T values. Once again, this points towards typically too-massive stellar disks being hosted by  galaxies with classical bulges in our model. A population of overly-large disks was already present in the  \cite{Henriques2015} version of the model. This could be due to the delayed growth of black holes hosted in classical bulges: as these objects accreted most of their mass at low redshift, their associated AGN feedback has been very modest at high-$z$, allowing for a significant and prolonged growth of the stellar disc. An improved version of the black hole growth model and its impact on galaxy morphology is going to be presented in a following paper (Izquierdo-Villalba et al., in prep). 
Finally, the model reproduces well the typical values of $\rm R_e$ found by \cite{Gadotti2009} as a function of bulge mass for pseudobulges, classical bulges and ellipticals, even though classical bulges lie slightly above the observations. Despite this, we can confirm the \cite{Gadotti2009} findings: classical bulges appear to be offset in the mass-to-size relation with respect to ellipticals, pointing out to fundamental structural differences, and that they are not simply ellipticals surrounded by disks. Indeed, bulge formation is an extremely complex phenomenon, which is shaped by both mergers and secular processes. As we explore in the next section, while in pseudobulge galaxies the process that dominates  galaxy evolution is \textit{secular} DI and in classical bulges is minor mergers, both population experience the two bulge formation mechanisms during their complex cosmological growth.

\subsubsection{Merger history of galaxies} \label{sec:merger_interaction_P_E_C}

The main assumptions of our model for bar and pseudobulge formation, is that those are linked to the secular growth of disks, while violent events, such are mergers and their consequent starbursts are responsible for the growth of classical bulges. We expect the merger history of galaxies hosting pseudobulges to be different from the one of elliptical and classical bulge hosts. In the reaming of this section we explore the (major/minor) merger history of $z\,{=}\,0$ pseudobulge, classical bulges and ellipticals galaxies and we present some archetypal examples of their merger trees. Finally, we explore the imprints that the galaxy interaction history leaves in the pseudobulge structure.\\

\begin{figure*} 
  \centering
  \includegraphics[width=.48\textwidth]{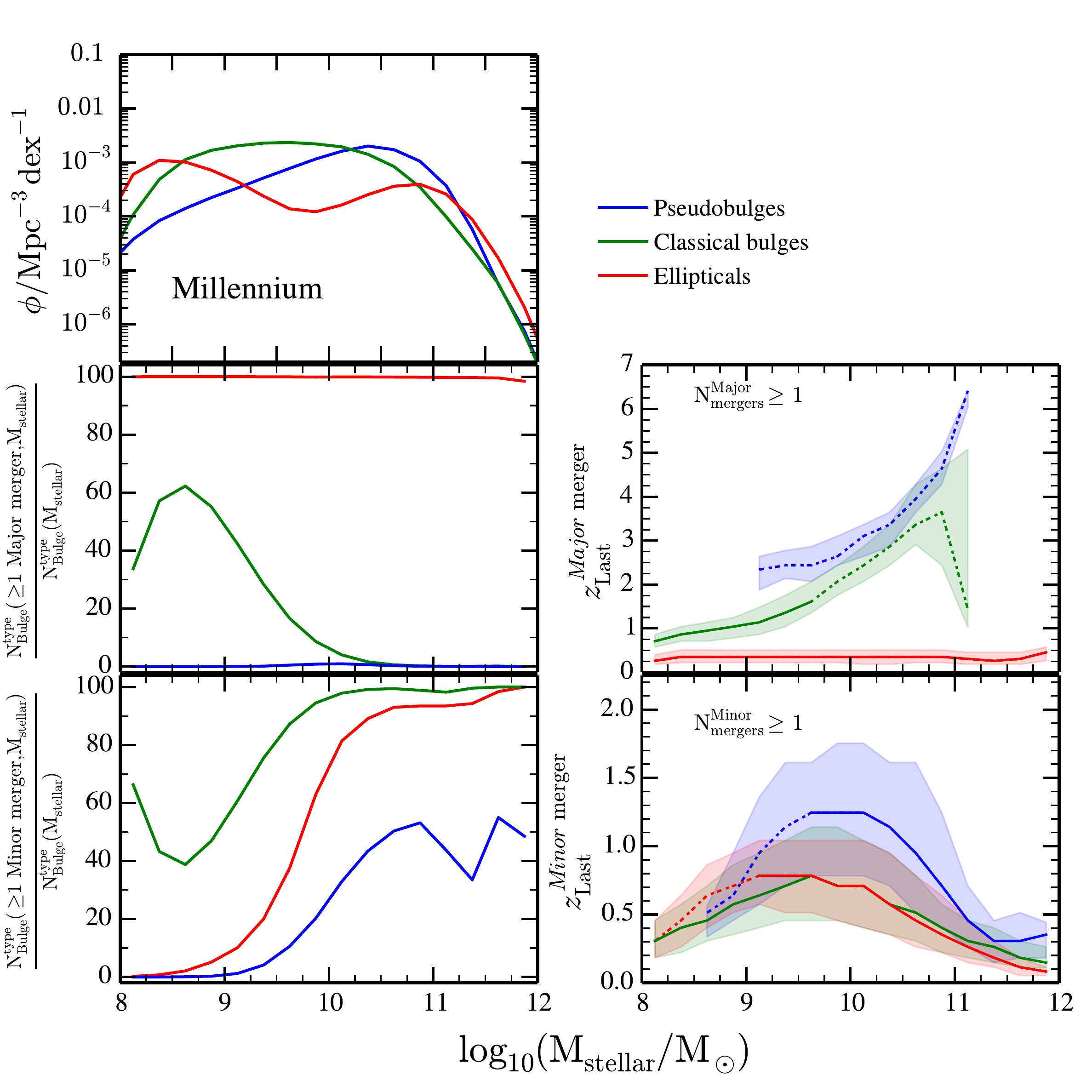}
   \includegraphics[width=.48\textwidth]{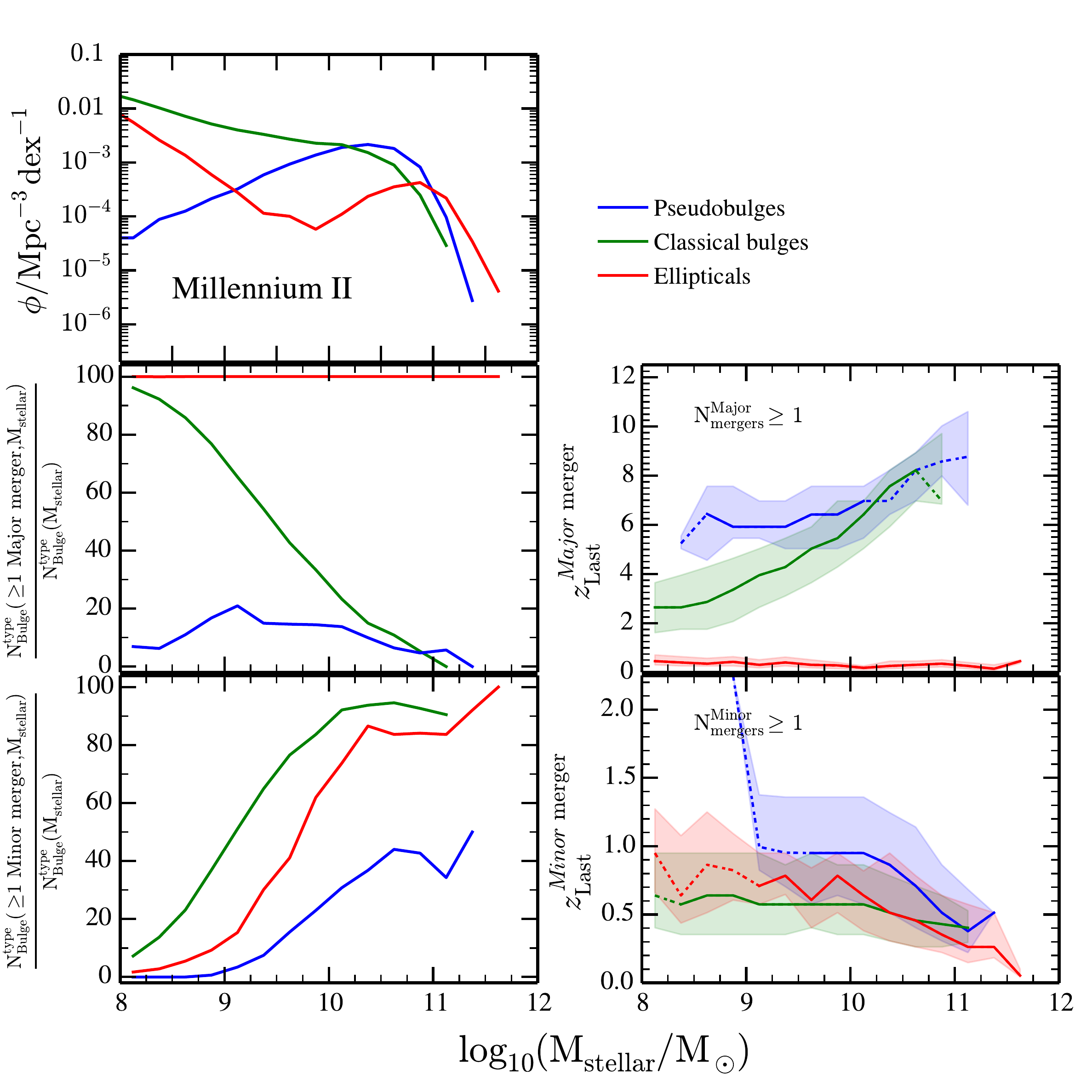}
  \caption{\textbf{Left panels}: \textit{First row}: Stellar mass function of pseudobulges (blue), classical bulges (green) and ellipticals (red) galaxies at $z\,{=}\,0$ in the MS simulation. \textit{Second row}: Percentage of $z\,{=}\,0$ pseudobulges, classical bulges and ellipticals galaxies that experience at least one major merger (left). From those ones we have presented the redshift in which the last one took place (right). \textit{Third row}: Percentage of $z\,{=}\,0$ pseudobulges, classical bulges and ellipticals galaxies that experience at least one minor merger (left). From those ones we have presented the redshift in which the last one took place. \textbf{Right panels}: The same as the left ones but for the MSII simulation. In all the plots, the shaded area represents the $\rm 1 \sigma$ dispersion of the distribution and dotted lines represents the regions in which the percentage of having suffered a
  minor/major merger is less than 10\%.}
\label{fig:Time_mergers}
\end{figure*}

In the second row of Fig~\ref{fig:Time_mergers} it is presented, per stellar mass bin, the percentage of $z\,{=}\,0$ pseudobulges (blue), classical bulges (green) and ellipticals (red) galaxies that experience at least one major merger (left plot) and from those ones the redshift in which the last major merger took place (right plot). The third row shows the same but for minor mergers. The left panels show the results for the MS and the right ones  for the  MSII. To guide the reader we have added the $z\,{=}\,0$ stellar mass function of pseudobulges, classical bulges and ellipticals. As we can see, all elliptical galaxies experienced at least one major merger, being the last one at $z\,{<}\,0.5$. On contrary, pseudobulges display a much more quiet major merger history. Only 0.5\% of them experienced one and only at very high redshift, being this higher for more massive galaxies: from $z\,{\sim}\,2$ at $\rm M_{stellar} \,{=}\, 10^9 M_{\odot}$ up to $z\,{\sim}\,6$ at $\rm 10^{11} M_{\odot}$. The MSII pseudobulges present a similar behaviour: for this simulation, the percentage of pseudobulges which underwent a merger is larger (${\sim}\,10\%$), because of the larger number of small galaxies resolved, and most of these mergers took place at very high redshifts.   
Regarding classical bulges, we can see that the ones hosted by galaxies with $\rm M_{stellar}\,{>}\,10^{10} M_{\odot}$ follow a similar trend that pseudobulges, i.e a quiet major merger history. On the other hand, at $\rm M_{stellar}\,{<}\,10^{10} M_{\odot}$ the majority of classical bulge galaxies suffered a recent major merger at $z \,{\lesssim}\, 2$. Notice that the drop in percentage presented in MS at $\rm M_{stellar} \,{<}\, 10^{8.5} M_{\odot}$ is mainly due to resolution. Actually, MSII predicts that the ${\sim}\,$100\% of those galaxies experienced at least one major merger.\\

Regarding minor mergers, all classes of bulges display a  similar trend: the fraction of galaxies that experienced at least one minor merger  increases with stellar  mass, although this fraction is never above $\rm 50\%$ for pseudobulges. Also, for both the MS and the MSII the typical redshift of the last minor interaction decreases with increasing stellar mass. For classical bulges and ellipticals, however, the typical redshift of the last minor merger is lower than for the pseudobulge population. 
However, the agreement of classical bulges and ellipticals between MS and MSII decreases when we study galaxies at $\rm M_{stellar} \,{<}\, 10^9 M_{\odot}$. While MS predicts a decreasing trend in the redshift distribution, MSII finds a flattening at $z\,{\sim}\,0.5$. Again, this is due to resolution effects which affect the MS in that stellar mass range.  

\begin{figure*}
\centering
a)\includegraphics[width=0.66\columnwidth]{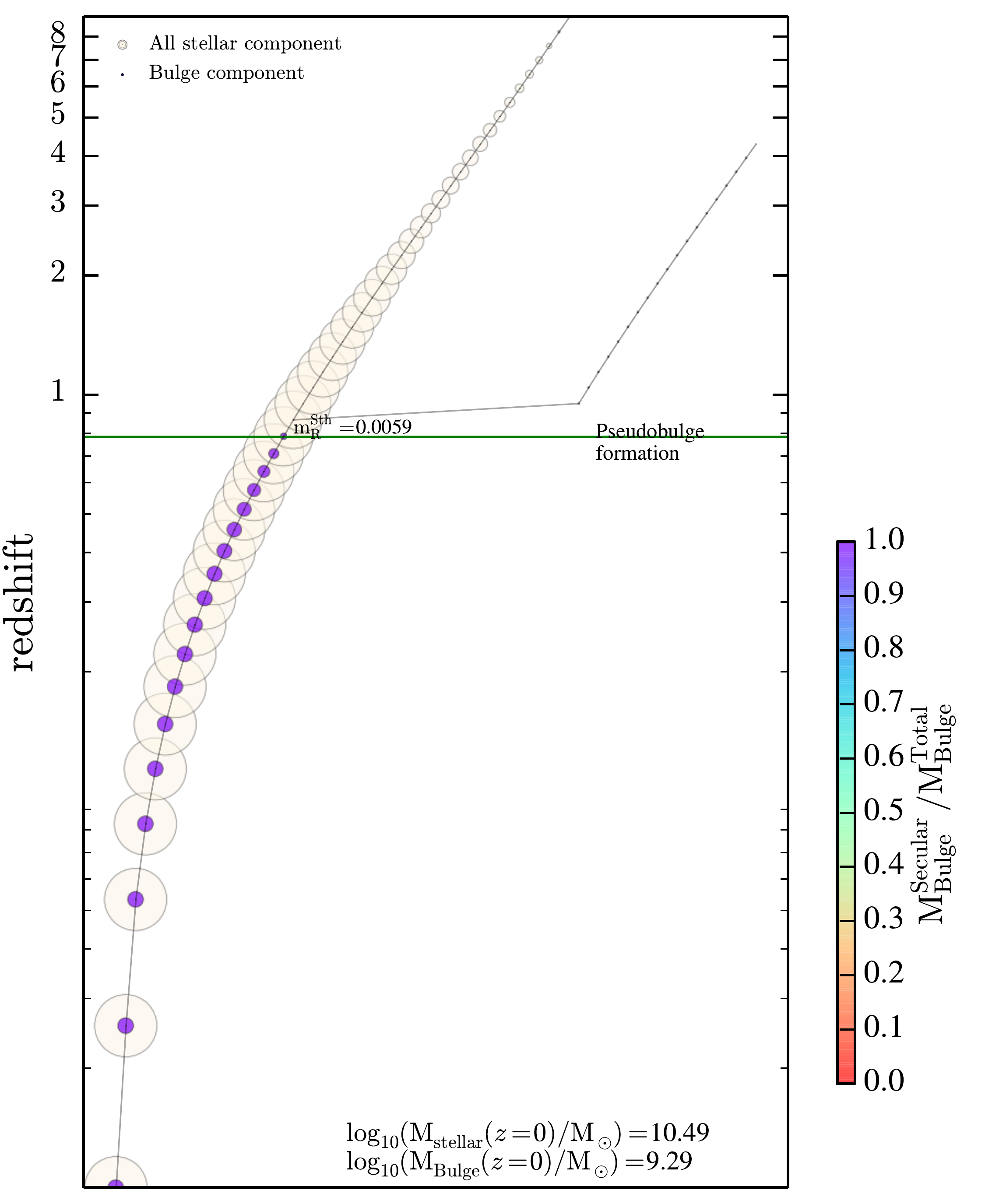}
b)\includegraphics[width=0.66\columnwidth]{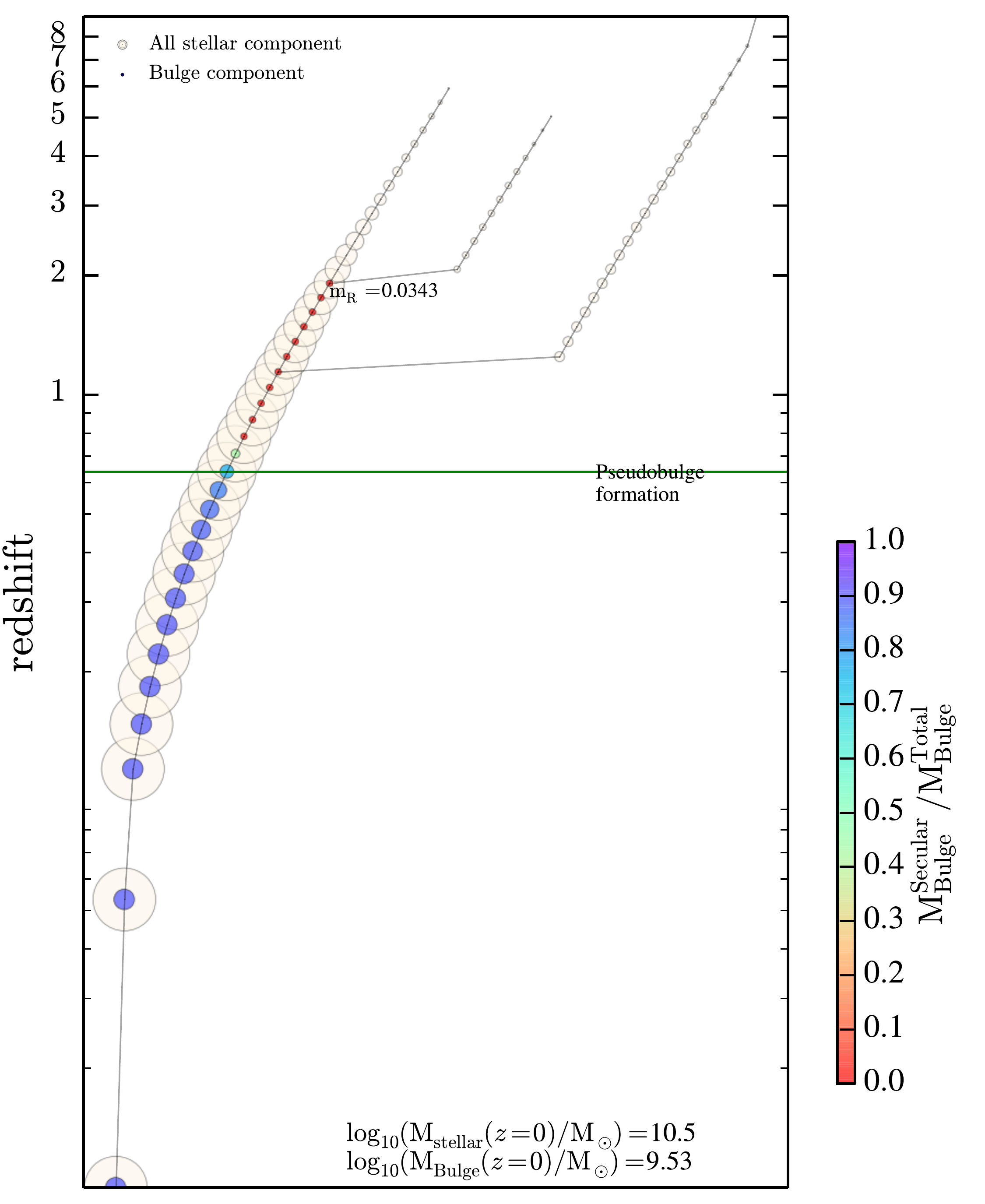}
c)\includegraphics[width=0.66\columnwidth]{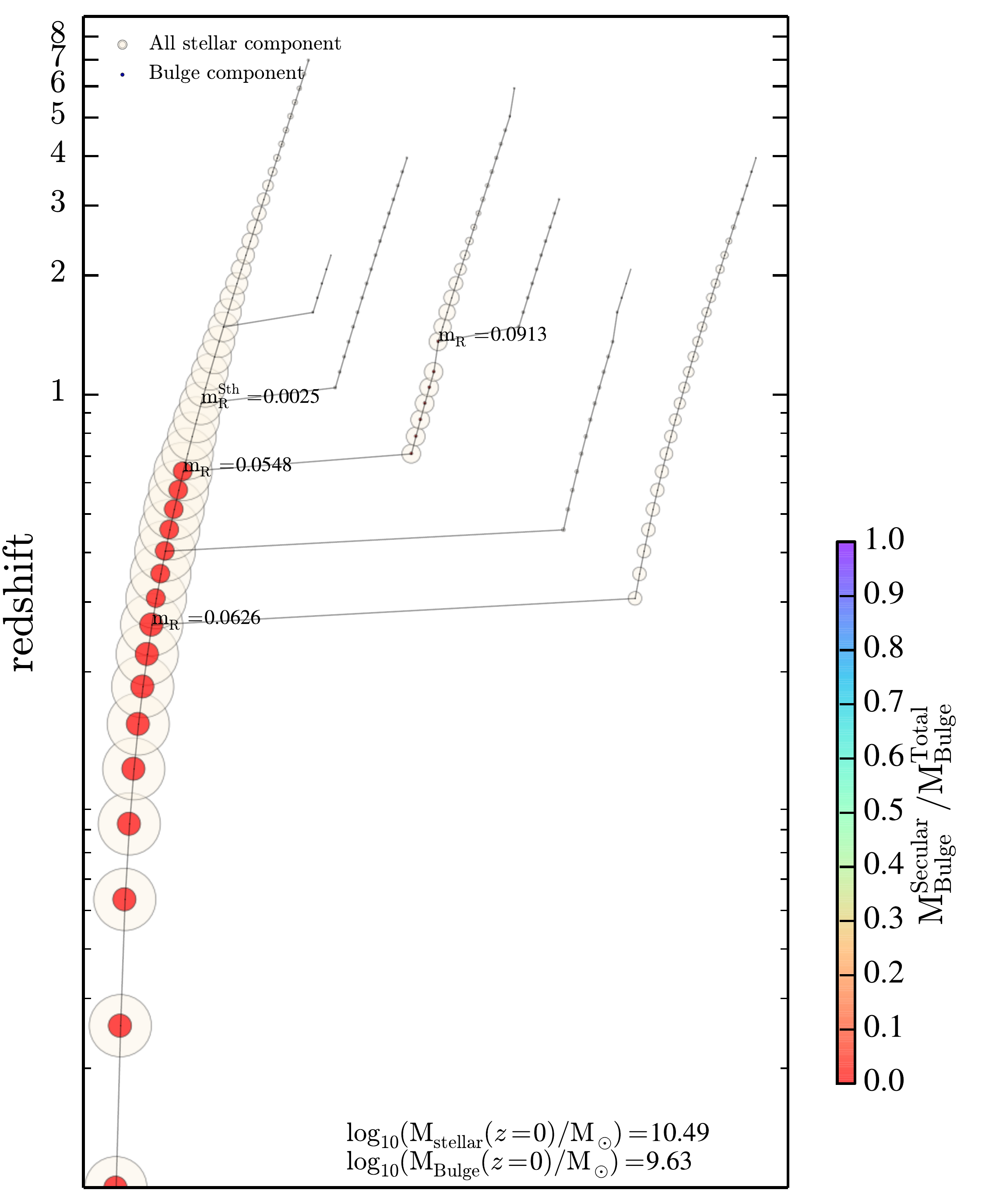}
d)\includegraphics[width=0.66\columnwidth]{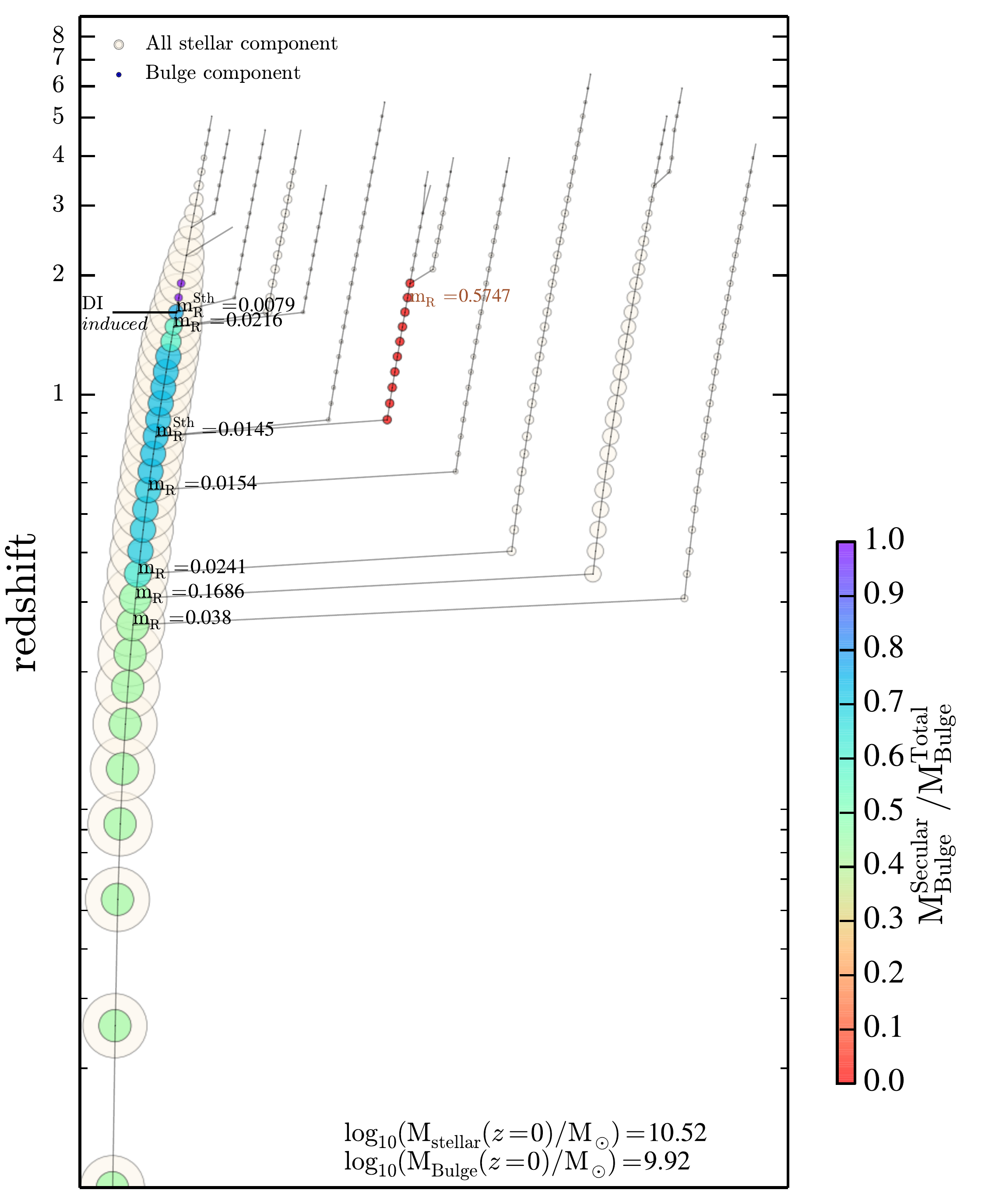}
e)\includegraphics[width=0.66\columnwidth]{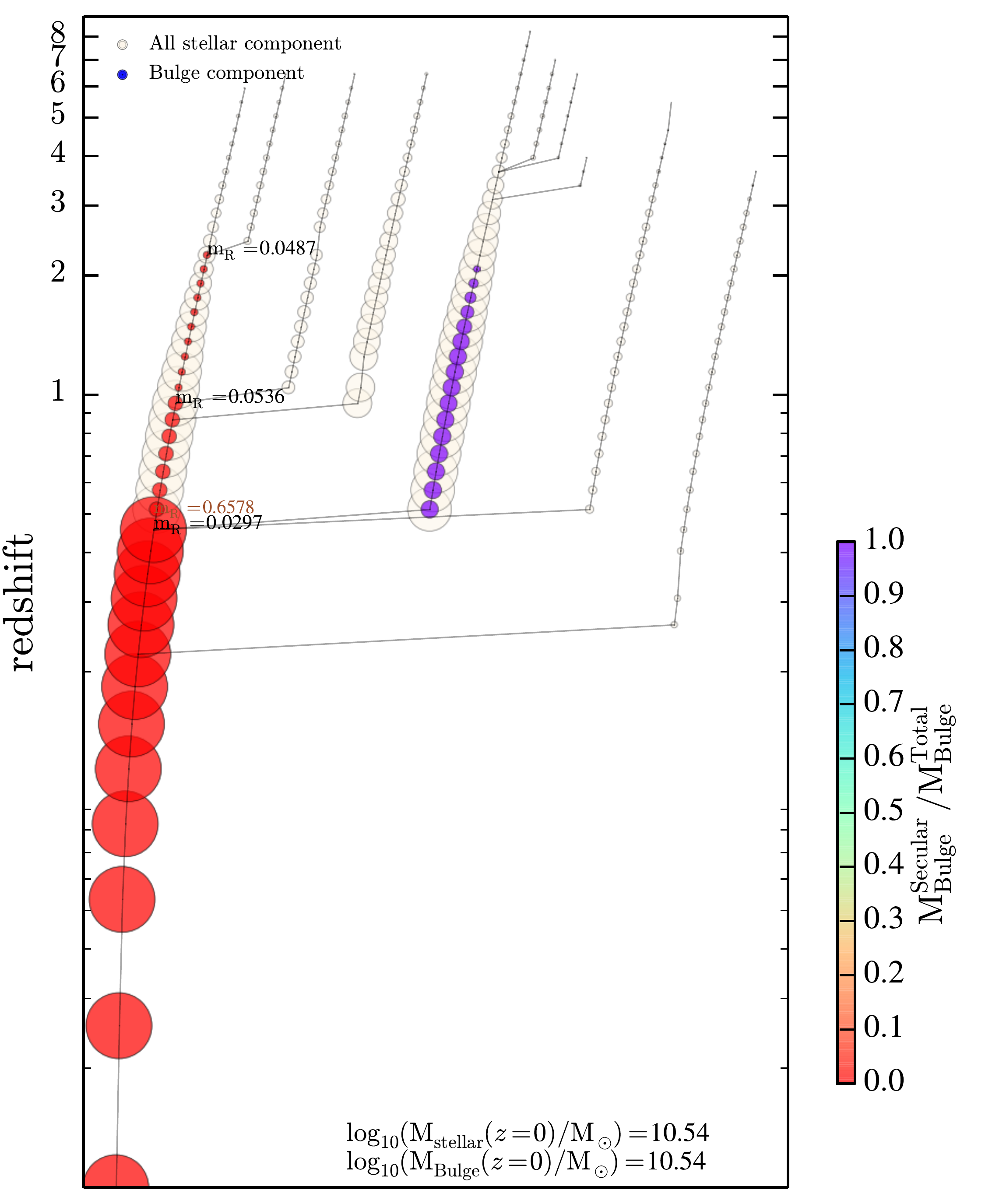}
f)\includegraphics[width=0.66\columnwidth]{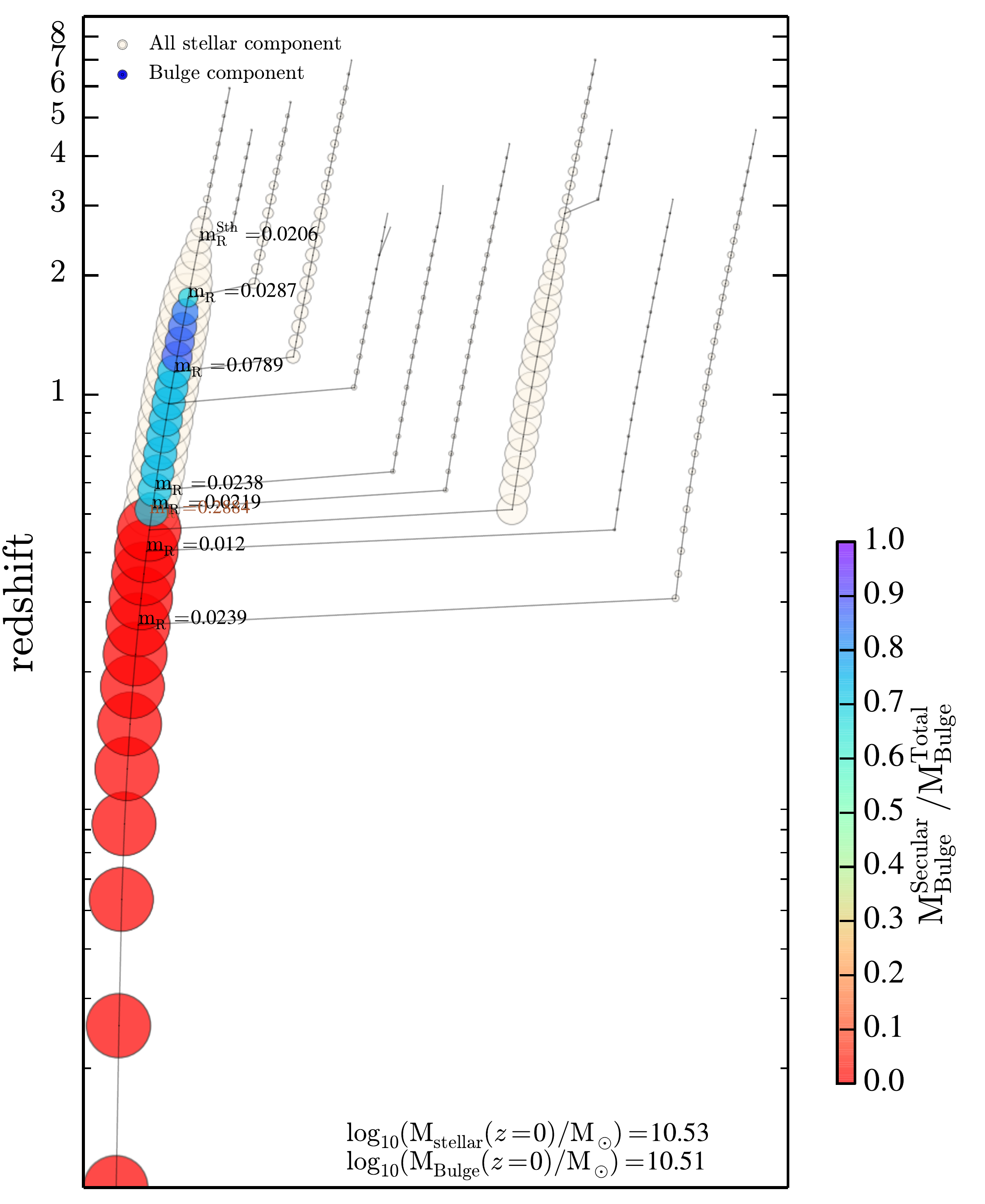}
\caption{Examples of the MS mergers trees for pseudobulges (examples \textit{a} and \textit{b}), classical bulges (examples \textit{c} and \textit{d}) and ellipticals (examples \textit{e} and \textit{f}). We have selected the galaxies with $\rm M_{stellar} \,{\sim}\, 10^{10.5} M_{\odot}$. In the plot we represent the stellar and bulge components (empty and filled circles respectively). The size of the circles is proportional to its respective mass. The bulge color represents the fraction of bulge mass coming from DI \textit{secular evolution}. In each merging branch we have added the merger ratio. Ticks with $\rm m_R$ corresponds to major/minor merger while ticks with $\rm m_R^{Sth}$ refers to \textit{smooth} accretion. Merging branch without any $\rm m_R$ value means that the galaxy was disrupted by environmental processes before the merger. Most of those galaxies are the ones close to $\rm M_{stellar} \,{\sim}\, 10^7 M_{\odot}$. In this cases, the satellite stellar mass was added in the \textit{Intra-Cluster Medium} (ICM).}
\label{fig:Merger_Trees}
\end{figure*}

In order to illustrate the different build-up of pseudobulges, in Fig~\ref{fig:Merger_Trees} \textit{a} and \textit{b} we present two typical examples of $z \,{=}\, 0$ pseudobulges mergers trees in in the \texttt{Millennium} simulation. Examples of classical bulges and ellipticals can bee seen in \textit{c}-\textit{d} and \textit{e}-\textit{f}, respectively. We have selected galaxies with $z\,{=}\,0$ stellar mass $\rm {\sim}\, 10^{10.5} M_{\odot}$, i.e the peak of $z\,{=}\,0$ pseudobulges stellar mass function (see Fig~\ref{fig:Time_mergers}). In the plot we represent the stellar and bulge components (empty and filled circles respectively). The size of the circles is proportional to mass. The color of the symbols represents the fraction of bulge mass coming from DI \textit{secular evolution}. In each merging branch we have added the mass ratio of the merger. Ticks with $\rm m_R$ corresponds to major/minor merger while ticks with $\rm m_R^{Sth}$ refers to \textit{smooth} accretion. Merging branch without any $\rm m_R$ value means that the galaxy was disrupted by environmental processes before the merger, and its mass added in the \textit{Intra-Cluster Medium}   (most of these galaxies are close to $\rm M_{stellar} \,{\sim}\, 10^7 M_{\odot}$).\\

As it was discussed above, pseudobulges have a very \textit{quiet} merger history. For instance, the first pseudobulge merger tree (\textit{example a}) just displays one \textit{smooth accretion} at $z\,{\sim}\,0.9$ with $\rm m_R = 6\,{\times}\,10^{-3}$. No other interaction takes place in its cosmological evolution. The pseudobulge structure appears after the \textit{smooth accretion} as a consequence of a DI \textit{secular evolution} causally disconnected from the satellite interaction. Therefore, the pseudobulge evolved thought \textit{internal secular evolution} represented by \underline{Case 1} and \underline{Case 2} in Fig~\ref{fig:Pseudo_bulges_Pahways}. In the case of the second pseudobulge galaxy (\textit{example b}), its bulge formation history is slightly more complicated, resulting from a combination of \underline{Case 4} and \underline{Case 1} from in Fig~\ref{fig:Pseudo_bulges_Pahways}. The galaxy developed a small bulge component as a consequence of a minor merger with $\rm m_R \,{=}\, 3.4\times10^{-2}$. After $\rm \sim 1 \, Gyr$ from the merger  (at $z\,{\sim}\,0.7$), an important disk instability took place, blurring the classical bulge structure and resulting in the birth of a prominent pseudobulge.\\

Regarding classical bulges and ellipticals galaxies, we can see that all the merger trees (\textit{example c,d,e,f}) are much more complicated than in the previous two cases. For instance, in the \textit{example c} it is presented  an archetype of classical bulge build-up (\underline{Case 4} Fig~\ref{fig:Pseudo_bulges_Pahways}): the bulge structure was generated by a minor merger at $z\,{\sim}\,0.7$ and strengthened by a more recent minor merger ($z\,{\sim}\,0.3$). No signatures or secular evolution in the bulge can be seen. \textit{Example d} represents another type of classical bulge galaxy evolution. In this case, the bulge structure was not completely build-up by mergers but by a combination of DI and minor mergers. The bulge was born via DI \textit{secular} evolution at $z\,{\sim}\,2 $ but at $z\,{\sim}\,1.5$ a DI \textit{merger-induced} (\underline{Case 5} in Fig~\ref{fig:Pseudo_bulges_Pahways}), consequence of a \textit{smooth accretion}, triggered the birth of a classical bulge component. The galaxy started to evolve and via secular DIs made the pseudobulge structure grow again. Nevertheless, the constant minor mergers interactions that the galaxy experienced at $z\,{=}\,0.4,0.3$ and $0.2$ led to the growth of a prominent classical bulge, where the pseudobulge component is negligible ($\,{<}\,4\, \%$ of $\mathrm{M_{Bulge}}$ at $z\,{=}\,0$). Finally, \textit{examples e} and \textit{f} display some pathways of elliptical galaxy formation. In \textit{example e} the galaxy started as a classical bulge galaxy, with the bulge component being due to several minor mergers, while in \textit{example e} the galaxy hosted a prominent pseudobulge at high z.  In both cases, however, a major merger took place at $z \sim  0.5$, which erased the previous galaxy morphology and formed a pure elliptical galaxy. 

\begin{figure}
\centering
\includegraphics[width=1.0\columnwidth]{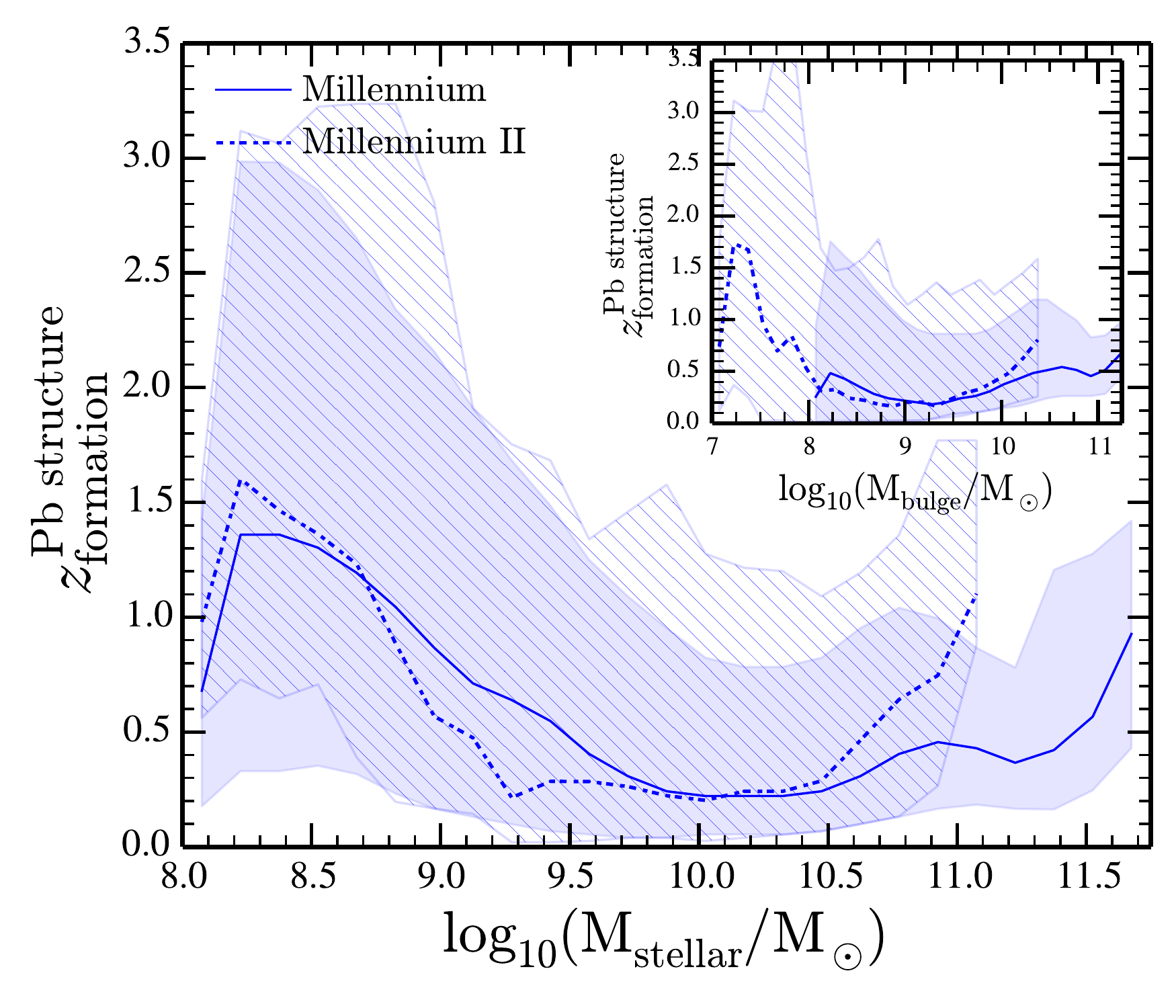}
\includegraphics[width=1.0\columnwidth]{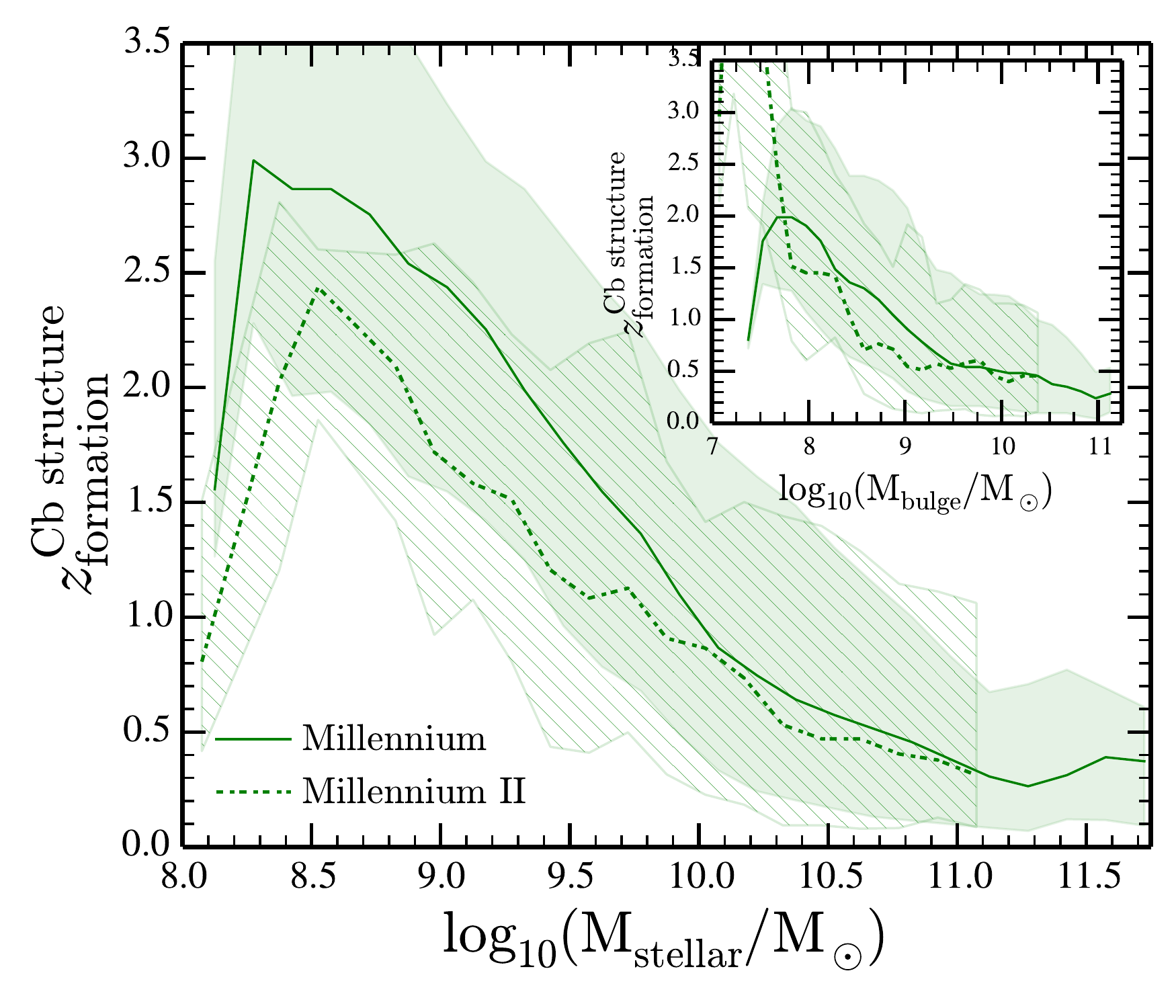}
\caption{\textbf{Upper panel}: Formation time of pseudobulge structure as a function of stellar mass ($\rm M_{stellar}$) in pseudobulge galaxies. Solid and dashed line represent respectively the median relation for MS and MSII. The shaded and lined areas represent the $\rm 1\sigma$ dispersion value. In the insert plot we present the same but as a function of the bulge mass ($\rm M_{bulge}$) \textbf{Lower panel}: Formation time of the classical component as a function of stellar mass ($\rm M_{stellar}$) for pseudobulge galaxies. In the insert plot we present the same but as a function of the bulge mass ($\rm M_{bulge}$). Line styles represent the same as the upper plot.}
\label{fig:formation_time}
\end{figure}

Finally, despite having quiet merger history, pseudobulge galaxies still have some minor/major merger interactions which can provide a classical component to the final bulge. In particular, the 31\% and 32.4\% of MS and MSII pseudobulge galaxies host a classical component, contributing typically with the ${\sim}\,$7\% of the whole bulge mass. To support the idea of last minor mergers being the main responsible for the strengthening of the classical bulge structure in pseudobulge galaxies at $z\,{=}\,0$, we define the formation redshift of the classical component (i.e. $z^{\rm Cb \, structure}_{\rm formation}$) as the moment in which it reached the 70\%\footnote{We have checked that the results do not suffer significant changes when we assume a value between 50\% - 90\%} of its final mass at $z\,{=}\, 0$. As shown in the lower panel of Fig~\ref{fig:formation_time}, the formation time of this component follows the same trend of $z^{\rm \mathit{Minor} \, merger}_{\rm Last}$ presented in Fig~\ref{fig:Time_mergers}. The last merger is thus responsible for the building (or strengthening) of  the classical bulge structure in $z \,{=}\,0$ pseudobulge galaxies.
We highlight that, in the $\rm 10^9\,{-}\,10^{10} M_{\odot}$ mass range, MSII predicts slightly lower time-formation values than MS. This is because MSII galaxies within this specific mass range are affected 5 times more frequently by \textit{merger-induced} DIs than MS (taking place at $z\,{\sim}\,1$, see Fig~\ref{fig:all_n_density_DI}).\\


Regarding the pseudobulge structure, in Fig~\ref{fig:formation_time} we present the distribution of formation times, $z^{\rm Pb \, structure}_{\rm formation}$,  as a function of stellar mass. This value has been defined as the moment in which the pseudobulge component reached the 2/3\footnote{i.e, the moment in which, independently of redshift, the galaxy would be always selected as a pseudobulge galaxy} of its $z\,{ =}\,0$ total bulge mass. As we can see, while pseudobulge galaxies with $\rm M_{stellar}\,{<}\,10^9 M_{\odot}$ formed their pseudobulge structure mainly at $z\,{\sim}\,1.5$, more massive galaxies (i.e, $\rm M_{stellar}\,{>}\,10^9 M_{\odot}$) formed it in the low-$z$ universe, although the  scatter is larger. Interestingly, this trend is broken for most massive pseudobulge galaxies ($\rm M_{stellar}\,{>}\,10^{10.5} M_{\odot}$) where the $z^{\rm Pb \, structure}_{\rm formation}$ rises again up to $z \,{\sim}\, 0.5{-}1.0$. We interpret this turn-over as an effect of the star-formation rate damping induced by AGN feedback in these galaxies. As a consequence of feedback, the DIs frequency is severely decreased and pseudobulge formation is suppressed in many massive galaxies at low-z. The sSfr distribution (Fig~\ref{fig:Main_sequence}) supported this scenario, by showing a clear drop at $\rm M_{stellar}\,{>}\,10^{10.5} M_{\odot}$. The insert plot in Fig~\ref{fig:formation_time} shows the same distribution but for fixed bulge mass. As we can see, the fixed-bulge trend is similar to the previous one: small-bulge galaxies formed their pseudobulge component at higher redshifts with respect to more massive-bulge ones. Our findings of pseudobulge formation at intermediate redshifts are supported by a recent work of \cite{Gadotti2015} who found that the bar in the Virgo galaxy NGC 4371 formed a pseudobulge at $z{\sim}1.8$ (with an uncertainty of $\rm \,{\sim}\, 1 Gyr$).


\section{Summary and conclusions} \label{sec:Conclusions}

In this paper we have studied the cosmological build-up of pseudobulges using the last public version of the \LGalaxies semi-analityical model \citep[SAM,][]{Henriques2015}. Taking advantage that \LGalaxies can be run on top the merger trees of both the \texttt{Millennium}  and the \texttt{Millennium II}  N-body simulations, we have been able to study the formation of pseudobulge structures and properties of their hosts across a wide range of stellar masses ($\rm 10^{8}\,{-}\,10^{11.5} M_{\odot}$). 
In order to reach a reasonable convergence between the MS and MSII in terms of distribution of galaxy morphology and number density of merger events across cosmic time, we first had to introduce some small modifications to the parameters that distinguish between major and minor mergers and that set the conditions for disk instability events. 
Moreover, to improve the predictions of the MSII for the morphological distribution of low stellar mass systems ($\rm M_{stellar}\,{<}\,10^{9.5} M_{\odot}$), we introduced a new prescription for the interactions  in which the binding energy of the satellite galaxy is very small compared to the one of the central galaxy. For these events, which we refer to with the term \textit{smooth accretion}, we assume that the stellar core of the satellite  gets incorporated by the disk of the central galaxy, being unable to reach its nucleus before being disrupted.  \\
 
Assuming that pseudobulges can only form and grow via secular evolution  \citep[e.g.,][]{Kormendy2013}, we have modified the treatment of galaxy disk instabilities (DI), distinguishing between two kinds of events: DI \textit{secular-induced}, which is a consequence of the slow  and continuous mass-growth of galaxies, and DI \textit{merger-induced}, linked to the fast growth of disks during galaxy interactions. 
The former are the events which we assume to lead to long-lasting bar structures  and the  formation/growth of pseudobulges, while \textit{merger-induced} instabilities contribute, together with mergers,  to the growth of classical bulges. 
Our SAM predicts that merger-induced instabilities have a number density $\rm \,{\sim}\, 2 -3 \, dex$ smaller than the DI \textit{secular evolution} at all cosmic epochs, and that it is a secondary channel in the growth of classical bulges, being classical-bulge growth during the merger event the primary channel at all redshifts.   
On the other hand, \textit{secular-induced} DIs are the most abundant events at any redshift and stellar masses. However, while in galaxies with $\rm M_{stellar} \,{=} \,10^{9}\,{-}\,10^{10}\,M_{\odot}$ these events are able to substantially contribute to the growth of the bulge, by transferring up  ${\sim}$10\% of the total stellar content to the pseudobulge, in galaxies with $\rm M_{stellar} \,{=}\,10^8-10^9\,M_{\odot}$ DIs can only lead to a small (sub-percent) transfer of mass from the disk to the pseudobulge. As a consequence, predominant pseudobulge structures are typically present in galaxies in the range  $\rm 10^{9.5}< M_{stellar}\,{<}\,10^{10} M_{\odot}$,  at high-$z$, moving to slightly higher values at more recent cosmic times. At $z=0$, in particular,  pseudobulges are hosted by galaxies in a very narrow stellar and halo mass window, $\rm 10^{10}\,{<}\, M_{stellar}\,{<}\,10^{10.5} M_{\odot}$ and $\rm 10^{11.5}\,{<}\, M_{halo}\,{<}\,10^{12} M_{\odot}$, i.e Milky-Way type galaxies. Moreover, while at high-$z$ pseudobulges are the dominant bulge structures in massive galaxies ($\rm M_{stellar}\,{>}\,10^{11} M_{\odot}$), they get systematically \textit{depleted} in such massive systems with decreasing redshift.  We interpret this as a consequence of the hierarchical growth of galaxies: pseudobulges are less likely to be hosted by very massive systems at late cosmic times, as  the  assembly of these galaxies is closely linked to numerous merger events which dramatically modify the dynamics of the galaxy, leading to deep morphological transformations.\\

When looking at the properties of the hosts, we find that pseudobulges  are hosted by actively star forming galaxies (in the main sequence of star formation), and with a relatively young stellar population (mass weighted age $\rm {\sim}\,6-8\, Gyr$, independently on the host stellar mass). Classical bulges, instead, reside in star forming galaxies only if the host mass is below $\rm {\sim}\, 10^{9.5}  M_{\odot}$, while more massive systems are quenched, or in the process of quenching, and are characterized by an older stellar population.

Tracing the history of galaxies hosting pseudobulges at $z\,{=}\,0$,  we found that they are characterized by an extremely-quiet merger history. The \texttt{Millennium} and \texttt{Millennium II} simulations predict, respectively, that only $\rm 0.5\%$ and $\rm 11\%$ of galaxies with a pseudobulge at $z \,{=}\, 0$ experienced a major merger, and this took place at very high redshifts $2\,{<}\,z\,{<}\,7$. Also minor mergers are rare in the history of today's pseudobulge galaxies, with less than 30\% of pseudobulges hosts having experienced a minor merger. Because of the minor mergers, however, these galaxies also contain a small classical bulge component (${\sim}\, 7\%$ of the total bulge mass). The two structures are characterized by  different formation times: while the pseudobulge was formed at $z\,{\lesssim}\,0.75$, the classical one did it at $z\,{\lesssim}\,1.5$, the time in which the last minor merger took place.\\

Finally, we have created mock samples of local pseudobulges and classical bulges, to compare with the observational results of \citet{Gadotti2009}, who analyzed the properties of pseudobulges in galaxies above $\rm {\sim}\, 10^{10} M_{\odot}$. We found that the pseudobulge structural properties  predicted by the model are broadly consistent with observations. In particular, we find a good agreement  in the effective radii distribution of different classes of bulges. The distribution of  \textit{bulge-to-total ratios} for pseudobulges is also consistent with the results of \citet{Gadotti2009}, while classical bulges are predicted to be in galaxies with disks larger than observed. These results are quite encouraging and give support to our main underlying assumption that pseudobulge structure can form mainly via secular evolution.\\

Despite the promising results, more investigation is needed to understand bar and pseudobulge formation in a broad cosmological context. Our simple approach is highly complementary to more sophisticated simulations which try to study the complex dynamical evolution of disk galaxies. More synergy among different theoretical approaches and  observations are certainly needed to reach a more clear picture on the different mechanisms that lead to formation  of different bulge classes.\\

\section*{Acknowledgements}
The authors thank Simon White, Dimitri Gadotti and Sergio Contreas for useful discussions and comments. We acknowledge the support from project \textit{AYA2015-66211-C2-2 MINECO/FEDER, UE} of the Spanish Ministerio de Economia, Industria y Competitividad. DIV particularly thanks the grant \textit{Programa Operativo Fondo Social Europeo de Arag\'{o}n 2014-2020. Construyendo Europa desde Arag\'{o}n}.
BMBH (ORCID 0000-0002-1392-489X) acknowledges support from a Zwicky Prize fellowship. YRG acknowledges support of the European Research Council through grant number ERC-StG/716151. This project has received funding from the European Unions Horizon 2020 Research and Innovation Programme under the Marie Sklodowska-Curie grant agreement No 734374.

\bibliography{references}

\begin{thebibliography}{}

\bibitem[{Abadi} et~al., 2003]{Abadi2003}
{Abadi}, M.~G., {Navarro}, J.~F., {Steinmetz}, M., and {Eke}, V.~R. (2003).
\newblock {Simulations of Galaxy Formation in a {$\Lambda$} Cold Dark Matter
  Universe. II. The Fine Structure of Simulated Galactic Disks}.
\newblock {\em \apj}, 597:21--34.

\bibitem[{Aguerri} et~al., 2001]{Aguerri2001}
{Aguerri}, J.~A.~L., {Balcells}, M., and {Peletier}, R.~F. (2001).
\newblock {Growth of galactic bulges by mergers. I. Dense satellites}.
\newblock {\em \aap}, 367:428--442.

\bibitem[{Angulo} and {White}, 2010]{AnguloandWhite2010}
{Angulo}, R.~E. and {White}, S.~D.~M. (2010).
\newblock {One simulation to fit them all - changing the background parameters
  of a cosmological N-body simulation}.
\newblock {\em \mnras}, 405:143--154.

\bibitem[{Athanassoula}, 2005]{Athanassoula2005}
{Athanassoula}, E. (2005).
\newblock {On the nature of bulges in general and of box/peanut bulges in
  particular: input from N-body simulations}.
\newblock {\em \mnras}, 358:1477--1488.

\bibitem[{Athanassoula}, 2008]{Athanassoula2008}
{Athanassoula}, E. (2008).
\newblock {Disc instabilities and semi-analytic modelling of galaxy formation}.
\newblock {\em \mnras}, 390:L69--L72.

\bibitem[{Athanassoula}, 2012]{Athanassoula2012A}
{Athanassoula}, E. (2012).
\newblock {Towards understanding the dynamics of the bar/bulge region in our
  Galaxy}.
\newblock In {\em European Physical Journal Web of Conferences}, volume~19 of
  {\em European Physical Journal Web of Conferences}, page 06004.

\bibitem[{Barnes}, 1999]{Barnes1999}
{Barnes}, J.~E. (1999).
\newblock {Galaxy Transformation by Merging}.
\newblock In {Beckman}, J.~E. and {Mahoney}, T.~J., editors, {\em The Evolution
  of Galaxies on Cosmological Timescales}, volume 187 of {\em Astronomical
  Society of the Pacific Conference Series}, pages 293--306.

\bibitem[{Baugh} et~al., 1996]{BaughFrenk1996}
{Baugh}, C.~M., {Cole}, S., and {Frenk}, C.~S. (1996).
\newblock {Faint galaxy counts as a function of morphological type in a
  hierarchical merger model}.
\newblock {\em \mnras}, 282:L27--L32.

\bibitem[{Benson} et~al., 2002]{Benson2002}
{Benson}, A.~J., {Ellis}, R.~S., and {Menanteau}, F. (2002).
\newblock {On the continuous formation of field spheroidal galaxies in
  hierarchical models of structure formation}.
\newblock {\em \mnras}, 336:564--576.

\bibitem[{Birnboim} and {Dekel}, 2003]{BirnboimandDekel2003}
{Birnboim}, Y. and {Dekel}, A. (2003).
\newblock {Virial shocks in galactic haloes?}
\newblock {\em \mnras}, 345:349--364.

\bibitem[{Bla{\~n}a D{\'{\i}}az} et~al., 2018]{Balana2018}
{Bla{\~n}a D{\'{\i}}az}, M., {Gerhard}, O., {Wegg}, C., {Portail}, M.,
  {Opitsch}, M., {Saglia}, R., {Fabricius}, M., {Erwin}, P., and {Bender}, R.
  (2018).
\newblock {Sculpting Andromeda - made-to-measure models for M31's bar and
  composite bulge: dynamics, stellar and dark matter mass}.
\newblock {\em \mnras}, 481:3210--3243.

\bibitem[{Bonoli} et~al., 2016]{Bonoli2016}
{Bonoli}, S., {Mayer}, L., {Kazantzidis}, S., {Madau}, P., {Bellovary}, J., and
  {Governato}, F. (2016).
\newblock {Black hole starvation and bulge evolution in a Milky Way-like
  galaxy}.
\newblock {\em \mnras}, 459:2603--2617.

\bibitem[{Bournaud} and {Combes}, 2002]{Bournaud2002}
{Bournaud}, F. and {Combes}, F. (2002).
\newblock {Gas accretion on spiral galaxies: Bar formation and renewal}.
\newblock {\em \aap}, 392:83--102.

\bibitem[{Bournaud} et~al., 2005]{Bournaud2005}
{Bournaud}, F., {Jog}, C.~J., and {Combes}, F. (2005).
\newblock {Galaxy mergers with various mass ratios: Properties of remnants}.
\newblock {\em \aap}, 437:69--85.

\bibitem[{Boylan-Kolchin} et~al., 2009]{Boylan-Kolchin2009}
{Boylan-Kolchin}, M., {Springel}, V., {White}, S.~D.~M., {Jenkins}, A., and
  {Lemson}, G. (2009).
\newblock {Resolving cosmic structure formation with the Millennium-II
  Simulation}.
\newblock {\em \mnras}, 398:1150--1164.

\bibitem[{Brinchmann} et~al., 2004]{Brinchmann2004}
{Brinchmann}, J., {Charlot}, S., {White}, S.~D.~M., {Tremonti}, C.,
  {Kauffmann}, G., {Heckman}, T., and {Brinkmann}, J. (2004).
\newblock {The physical properties of star-forming galaxies in the low-redshift
  Universe}.
\newblock {\em \mnras}, 351:1151--1179.

\bibitem[{Bureau} and {Freeman}, 1999]{Bureau1999}
{Bureau}, M. and {Freeman}, K.~C. (1999).
\newblock {The Nature of Boxy/Peanut-Shaped Bulges in Spiral Galaxies}.
\newblock {\em \aj}, 118:126--138.

\bibitem[{Cano-D{\'{\i}}az} et~al., 2016]{CanoDiaz2016}
{Cano-D{\'{\i}}az}, M., {S{\'a}nchez}, S.~F., {Zibetti}, S., {Ascasibar}, Y.,
  {Bland-Hawthorn}, J., {Ziegler}, B., {Gonz{\'a}lez Delgado}, R.~M.,
  {Walcher}, C.~J., {Garc{\'{\i}}a-Benito}, R., {Mast}, D.,
  {Mendoza-P{\'e}rez}, M.~A., {Falc{\'o}n-Barroso}, J., {Galbany}, L.,
  {Husemann}, B., {Kehrig}, C., {Marino}, R.~A., {S{\'a}nchez-Bl{\'a}zquez},
  P., {L{\'o}pez-Cob{\'a}}, C., {L{\'o}pez-S{\'a}nchez}, {\'A}.~R., and
  {Vilchez}, J.~M. (2016).
\newblock {Spatially Resolved Star Formation Main Sequence of Galaxies in the
  CALIFA Survey}.
\newblock {\em \apjl}, 821:L26.

\bibitem[{Carpineti} et~al., 2012]{Carpineti2012}
{Carpineti}, A., {Kaviraj}, S., {Darg}, D., {Lintott}, C., {Schawinski}, K.,
  and {Shabala}, S. (2012).
\newblock {Spheroidal post-mergers in the local Universe}.
\newblock {\em \mnras}, 420:2139--2146.

\bibitem[{Cervantes Sodi} et~al., 2015]{Cervantes2015}
{Cervantes Sodi}, B., {Li}, C., and {Park}, C. (2015).
\newblock {Dark Matter Halos of Barred Disk Galaxies}.
\newblock {\em \apj}, 807:111.

\bibitem[{Colless} et~al., 2001]{Colless2001}
{Colless}, M., {Dalton}, G., {Maddox}, S., {Sutherland}, W., {Norberg}, P.,
  {Cole}, S., {Bland-Hawthorn}, J., {Bridges}, T., {Cannon}, R., {Collins}, C.,
  {Couch}, W., {Cross}, N., {Deeley}, K., {De Propris}, R., {Driver}, S.~P.,
  {Efstathiou}, G., {Ellis}, R.~S., {Frenk}, C.~S., {Glazebrook}, K.,
  {Jackson}, C., {Lahav}, O., {Lewis}, I., {Lumsden}, S., {Madgwick}, D.,
  {Peacock}, J.~A., {Peterson}, B.~A., {Price}, I., {Seaborne}, M., and
  {Taylor}, K. (2001).
\newblock {The 2dF Galaxy Redshift Survey: spectra and redshifts}.
\newblock {\em \mnras}, 328:1039--1063.

\bibitem[{Combes}, 2009]{Combes2009}
{Combes}, F. (2009).
\newblock {Secular Evolution and the Assembly of Bulges}.
\newblock In {Jogee}, S., {Marinova}, I., {Hao}, L., and {Blanc}, G.~A.,
  editors, {\em Galaxy Evolution: Emerging Insights and Future Challenges},
  volume 419 of {\em Astronomical Society of the Pacific Conference Series},
  page~31.

\bibitem[{Combes} et~al., 1990]{Combes1999}
{Combes}, F., {Debbasch}, F., {Friedli}, D., and {Pfenniger}, D. (1990).
\newblock {Box and peanut shapes generated by stellar bars}.
\newblock {\em \aap}, 233:82--95.

\bibitem[{Combes} and {Sanders}, 1981]{Combes&Sanders1981}
{Combes}, F. and {Sanders}, R.~H. (1981).
\newblock {Formation and properties of persisting stellar bars}.
\newblock {\em \aap}, 96:164--173.

\bibitem[{Conselice}, 2006]{Conselice2006}
{Conselice}, C.~J. (2006).
\newblock {The fundamental properties of galaxies and a new galaxy
  classification system}.
\newblock {\em \mnras}, 373:1389--1408.

\bibitem[{C{\^o}t{\'e}} et~al., 1998]{Cote1998}
{C{\^o}t{\'e}}, P., {Marzke}, R.~O., and {West}, M.~J. (1998).
\newblock {The Formation of Giant Elliptical Galaxies and Their Globular
  Cluster Systems}.
\newblock {\em \apj}, 501:554--570.

\bibitem[{Covington} et~al., 2011]{Covington2011}
{Covington}, M.~D., {Primack}, J.~R., {Porter}, L.~A., {Croton}, D.~J.,
  {Somerville}, R.~S., and {Dekel}, A. (2011).
\newblock {The role of dissipation in the scaling relations of cosmological
  merger remnants}.
\newblock {\em \mnras}, 415:3135--3152.

\bibitem[{Croton}, 2006]{Croton2006}
{Croton}, D.~J. (2006).
\newblock {Evolution in the black hole mass-bulge mass relation: a theoretical
  perspective}.
\newblock {\em \mnras}, 369:1808--1812.

\bibitem[{Dasyra} et~al., 2006]{Dasyra2006A}
{Dasyra}, K.~M., {Tacconi}, L.~J., {Davies}, R.~I., {Genzel}, R., {Lutz}, D.,
  {Naab}, T., {Burkert}, A., {Veilleux}, S., and {Sanders}, D.~B. (2006).
\newblock {Dynamical Properties of Ultraluminous Infrared Galaxies. I. Mass
  Ratio Conditions for ULIRG Activity in Interacting Pairs}.
\newblock {\em \apj}, 638:745--758.

\bibitem[{Dasyra} et~al., 2007]{Dasyra2007}
{Dasyra}, K.~M., {Tacconi}, L.~J., {Davies}, R.~I., {Genzel}, R., {Lutz}, D.,
  {Peterson}, B.~M., {Veilleux}, S., {Baker}, A.~J., {Schweitzer}, M., and
  {Sturm}, E. (2007).
\newblock {Host Dynamics and Origin of Palomar-Green QSOs}.
\newblock {\em \apj}, 657:102--115.

\bibitem[{De Lucia} et~al., 2004]{DeLucia2004}
{De Lucia}, G., {Kauffmann}, G., and {White}, S.~D.~M. (2004).
\newblock {Chemical enrichment of the intracluster and intergalactic medium in
  a hierarchical galaxy formation model}.
\newblock {\em \mnras}, 349:1101--1116.

\bibitem[{Debattista} et~al., 2004]{Debattista2004}
{Debattista}, V.~P., {Carollo}, C.~M., {Mayer}, L., and {Moore}, B. (2004).
\newblock {Bulges or Bars from Secular Evolution?}
\newblock {\em \apjl}, 604:L93--L96.

\bibitem[{Debattista} et~al., 2006]{Debattista2006}
{Debattista}, V.~P., {Mayer}, L., {Carollo}, C.~M., {Moore}, B., {Wadsley}, J.,
  and {Quinn}, T. (2006).
\newblock {The Secular Evolution of Disk Structural Parameters}.
\newblock {\em \apj}, 645:209--227.

\bibitem[{Dekel} et~al., 2009]{Dekel2009}
{Dekel}, A., {Sari}, R., and {Ceverino}, D. (2009).
\newblock {Formation of Massive Galaxies at High Redshift: Cold Streams, Clumpy
  Disks, and Compact Spheroids}.
\newblock {\em \apj}, 703:785--801.

\bibitem[{Di Matteo} et~al., 2015]{DiMatteoP2015}
{Di Matteo}, P., {G{\'o}mez}, A., {Haywood}, M., {Combes}, F., {Lehnert},
  M.~D., {Ness}, M., {Snaith}, O.~N., {Katz}, D., and {Semelin}, B. (2015).
\newblock {Why the Milky Way's bulge is not only a bar formed from a cold thin
  disk}.
\newblock {\em \aap}, 577:A1.

\bibitem[{Doyon} et~al., 1994]{Doyon1994}
{Doyon}, R., {Wells}, M., {Wright}, G.~S., {Joseph}, R.~D., {Nadeau}, D., and
  {James}, P.~A. (1994).
\newblock {Stellar velocity dispersion in ARP 220 and NGC 6240: Elliptical
  galaxies in formation}.
\newblock {\em \apjl}, 437:L23--L26.

\bibitem[{Drory} and {Fisher}, 2007]{DroryandFisher2007}
{Drory}, N. and {Fisher}, D.~B. (2007).
\newblock {A Connection between Bulge Properties and the Bimodality of
  Galaxies}.
\newblock {\em \apj}, 664:640--649.

\bibitem[{Dutton} and {Macci{\`o}}, 2014]{Dutton2014}
{Dutton}, A.~A. and {Macci{\`o}}, A.~V. (2014).
\newblock {Cold dark matter haloes in the Planck era: evolution of structural
  parameters for Einasto and NFW profiles}.
\newblock {\em \mnras}, 441:3359--3374.

\bibitem[{Efstathiou} et~al., 1982]{Efstathio1982}
{Efstathiou}, G., {Lake}, G., and {Negroponte}, J. (1982).
\newblock {The stability and masses of disc galaxies}.
\newblock {\em \mnras}, 199:1069--1088.

\bibitem[{Eliche-Moral} et~al., 2006]{Eliche2006}
{Eliche-Moral}, M.~C., {Balcells}, M., {Aguerri}, J.~A.~L., and
  {Gonz{\'a}lez-Garc{\'{\i}}a}, A.~C. (2006).
\newblock {Growth of galactic bulges by mergers. II. Low-density satellites}.
\newblock {\em \aap}, 457:91--108.

\bibitem[{Erwin} et~al., 2015]{Erwin2015}
{Erwin}, P., {Saglia}, R.~P., {Fabricius}, M., {Thomas}, J., {Nowak}, N.,
  {Rusli}, S., {Bender}, R., {Vega Beltr{\'a}n}, J.~C., and {Beckman}, J.~E.
  (2015).
\newblock {Composite bulges: the coexistence of classical bulges and discy
  pseudo-bulges in S0 and spiral galaxies}.
\newblock {\em \mnras}, 446:4039--4077.

\bibitem[{Fisher} and {Drory}, 2008a]{FisherandDrory2008}
{Fisher}, D.~B. and {Drory}, N. (2008a).
\newblock {The Structure of Classical Bulges and Pseudobulges: the Link Between
  Pseudobulges and S{\'E}RSIC Index}.
\newblock {\em \aj}, 136:773--839.

\bibitem[{Fisher} and {Drory}, 2008b]{Fisher2008}
{Fisher}, D.~B. and {Drory}, N. (2008b).
\newblock {The Structure of Classical Bulges and Pseudobulges: the Link Between
  Pseudobulges and S{\'E}RSIC Index}.
\newblock {\em \aj}, 136:773--839.

\bibitem[{Fisher} and {Drory}, 2010]{Fisher2010}
{Fisher}, D.~B. and {Drory}, N. (2010).
\newblock {Bulges of Nearby Galaxies with Spitzer: Scaling Relations in
  Pseudobulges and Classical Bulges}.
\newblock {\em \apj}, 716:942--969.

\bibitem[{Fisher} et~al., 2009]{Fisher2009}
{Fisher}, D.~B., {Drory}, N., and {Fabricius}, M.~H. (2009).
\newblock {Bulges of Nearby Galaxies with Spitzer: The Growth of Pseudobulges
  in Disk Galaxies and its Connection to Outer Disks}.
\newblock {\em \apj}, 697:630--650.

\bibitem[{Forcada-Miro} and {White}, 1997]{ForcadaandWhite1997}
{Forcada-Miro}, M.~I. and {White}, S.~D.~M. (1997).
\newblock {Radiative shocks in galaxy formation. I: Cooling of a primordial
  plasma with no sources of heating}.
\newblock {\em ArXiv Astrophysics e-prints}.

\bibitem[{Fragkoudi} et~al., 2017]{Fragkoudi2017}
{Fragkoudi}, F., {Di Matteo}, P., {Haywood}, M., {G{\'o}mez}, A., {Combes}, F.,
  {Katz}, D., and {Semelin}, B. (2017).
\newblock {Bars and boxy/peanut bulges in thin and thick discs. I. Morphology
  and line-of-sight velocities of a fiducial model}.
\newblock {\em \aap}, 606:A47.

\bibitem[{Gadotti}, 2009]{Gadotti2009}
{Gadotti}, D.~A. (2009).
\newblock {Structural properties of pseudo-bulges, classical bulges and
  elliptical galaxies: a Sloan Digital Sky Survey perspective}.
\newblock {\em \mnras}, 393:1531--1552.

\bibitem[{Gadotti} et~al., 2015]{Gadotti2015}
{Gadotti}, D.~A., {Seidel}, M.~K., {S{\'a}nchez-Bl{\'a}zquez}, P.,
  {Falc{\'o}n-Barroso}, J., {Husemann}, B., {Coelho}, P., and {P{\'e}rez}, I.
  (2015).
\newblock {MUSE tells the story of NGC 4371: The dawning of secular evolution}.
\newblock {\em \aap}, 584:A90.

\bibitem[{Gargiulo} et~al., 2015]{Gargiulo2015}
{Gargiulo}, I.~D., {Cora}, S.~A., {Padilla}, N.~D., {Mu{\~n}oz Arancibia},
  A.~M., {Ruiz}, A.~N., {Orsi}, A.~A., {Tecce}, T.~E., {Weidner}, C., and
  {Bruzual}, G. (2015).
\newblock {Chemoarchaeological downsizing in a hierarchical universe: impact of
  a top-heavy IGIMF}.
\newblock {\em \mnras}, 446:3820--3841.

\bibitem[{Gavazzi} et~al., 2015]{Gavazzi2015}
{Gavazzi}, G., {Consolandi}, G., {Viscardi}, E., {Fossati}, M., {Savorgnan},
  G., {Fumagalli}, M., {Gutierrez}, L., {Hernandez Toledo}, H., {Boselli}, A.,
  {Giovanelli}, R., and {Haynes}, M.~P. (2015).
\newblock {H{$\alpha$}3: an H{$\alpha$} imaging survey of HI selected galaxies
  from ALFALFA . V. The Coma supercluster survey completion}.
\newblock {\em \aap}, 576:A16.

\bibitem[{Guo} et~al., 2011]{Guo2011}
{Guo}, Q., {White}, S., {Boylan-Kolchin}, M., {De Lucia}, G., {Kauffmann}, G.,
  {Lemson}, G., {Li}, C., {Springel}, V., and {Weinmann}, S. (2011).
\newblock {From dwarf spheroidals to cD galaxies: simulating the galaxy
  population in a {$\Lambda$}CDM cosmology}.
\newblock {\em \mnras}, 413:101--131.

\bibitem[{Hammer} et~al., 2005]{Hammer2005}
{Hammer}, F., {Flores}, H., {Elbaz}, D., {Zheng}, X.~Z., {Liang}, Y.~C., and
  {Cesarsky}, C. (2005).
\newblock {Did most present-day spirals form during the last 8 Gyr?. A
  formation history with violent episodes revealed by panchromatic
  observations}.
\newblock {\em \aap}, 430:115--128.

\bibitem[{Hatton} et~al., 2003]{Hatton2003}
{Hatton}, S., {Devriendt}, J.~E.~G., {Ninin}, S., {Bouchet}, F.~R.,
  {Guiderdoni}, B., and {Vibert}, D. (2003).
\newblock {GALICS- I. A hybrid N-body/semi-analytic model of hierarchical
  galaxy formation}.
\newblock {\em \mnras}, 343:75--106.

\bibitem[{Henriques} et~al., 2015]{Henriques2015}
{Henriques}, B.~M.~B., {White}, S.~D.~M., {Thomas}, P.~A., {Angulo}, R., {Guo},
  Q., {Lemson}, G., {Springel}, V., and {Overzier}, R. (2015).
\newblock {Galaxy formation in the Planck cosmology - I. Matching the observed
  evolution of star formation rates, colours and stellar masses}.
\newblock {\em \mnras}, 451:2663--2680.

\bibitem[{Hernquist}, 1990]{Hernquist1990}
{Hernquist}, L. (1990).
\newblock {An analytical model for spherical galaxies and bulges}.
\newblock {\em \apj}, 356:359--364.

\bibitem[{Hirschmann} et~al., 2012]{Hirschmann2012}
{Hirschmann}, M., {Somerville}, R.~S., {Naab}, T., and {Burkert}, A. (2012).
\newblock {Origin of the antihierarchical growth of black holes}.
\newblock {\em \mnras}, 426:237--257.

\bibitem[{Hopkins} et~al., 2009a]{Hopkins2009}
{Hopkins}, P.~F., {Cox}, T.~J., {Younger}, J.~D., and {Hernquist}, L. (2009a).
\newblock {How do Disks Survive Mergers?}
\newblock {\em \apj}, 691:1168--1201.

\bibitem[{Hopkins} et~al., 2009b]{Hopkins_and_Somerville2009}
{Hopkins}, P.~F., {Somerville}, R.~S., {Cox}, T.~J., {Hernquist}, L., {Jogee},
  S., {Kere{\v s}}, D., {Ma}, C.-P., {Robertson}, B., and {Stewart}, K.
  (2009b).
\newblock {The effects of gas on morphological transformation in mergers:
  implications for bulge and disc demographics}.
\newblock {\em \mnras}, 397:802--814.

\bibitem[{Irodotou} et~al., 2018]{Irodotou2018}
{Irodotou}, D., {Thomas}, P.~A., {Henriques}, B.~M., and {Sargent}, M.~T.
  (2018).
\newblock {Morphological evolution and galactic sizes in the L-Galaxies SA
  model}.
\newblock {\em ArXiv e-prints}.

\bibitem[{Kalnajs}, 1972]{Kalnajs1972}
{Kalnajs}, A.~J. (1972).
\newblock {The Equilibria and Oscillations of a Family of Uniformly Rotating
  Stellar Disks}.
\newblock {\em \apj}, 175:63.

\bibitem[{Kauffmann} et~al., 1993]{KauffmannWhiteGuiderdoni1993}
{Kauffmann}, G., {White}, S.~D.~M., and {Guiderdoni}, B. (1993).
\newblock {The Formation and Evolution of Galaxies Within Merging Dark Matter
  Haloes}.
\newblock {\em \mnras}, 264:201.

\bibitem[{Kazantzidis} et~al., 2008]{Kazantzidis2008}
{Kazantzidis}, S., {Bullock}, J.~S., {Zentner}, A.~R., {Kravtsov}, A.~V., and
  {Moustakas}, L.~A. (2008).
\newblock {Cold Dark Matter Substructure and Galactic Disks. I. Morphological
  Signatures of Hierarchical Satellite Accretion}.
\newblock {\em \apj}, 688:254--276.

\bibitem[{Kim} et~al., 2016]{Kim2016}
{Kim}, T., {Gadotti}, D.~A., {Athanassoula}, E., {Bosma}, A., {Sheth}, K., and
  {Lee}, M.~G. (2016).
\newblock {Evidence of bar-induced secular evolution in the inner regions of
  stellar discs in galaxies: what shapes disc galaxies?}
\newblock {\em \mnras}, 462:3430--3440.

\bibitem[{Kormendy} and {Ho}, 2013]{Kormendy2013}
{Kormendy}, J. and {Ho}, L.~C. (2013).
\newblock {Coevolution (Or Not) of Supermassive Black Holes and Host Galaxies}.
\newblock {\em \araa}, 51:511--653.

\bibitem[{Kormendy} and {Kennicutt}, 2004a]{KormendyandKennicutt2004}
{Kormendy}, J. and {Kennicutt}, Jr., R.~C. (2004a).
\newblock {Secular Evolution and the Formation of Pseudobulges in Disk
  Galaxies}.
\newblock {\em \araa}, 42:603--683.

\bibitem[{Kormendy} and {Kennicutt}, 2004b]{KormendyKennicutt2004}
{Kormendy}, J. and {Kennicutt}, Jr., R.~C. (2004b).
\newblock {Secular Evolution and the Formation of Pseudobulges in Disk
  Galaxies}.
\newblock {\em \araa}, 42:603--683.

\bibitem[{Lacey} et~al., 2016]{Lacey2016}
{Lacey}, C.~G., {Baugh}, C.~M., {Frenk}, C.~S., {Benson}, A.~J., {Bower},
  R.~G., {Cole}, S., {Gonzalez-Perez}, V., {Helly}, J.~C., {Lagos}, C.~D.~P.,
  and {Mitchell}, P.~D. (2016).
\newblock {A unified multiwavelength model of galaxy formation}.
\newblock {\em \mnras}, 462:3854--3911.

\bibitem[{Lagos} et~al., 2011]{Lagos2011}
{Lagos}, C.~D.~P., {Baugh}, C.~M., {Lacey}, C.~G., {Benson}, A.~J., {Kim},
  H.-S., and {Power}, C. (2011).
\newblock {Cosmic evolution of the atomic and molecular gas contents of
  galaxies}.
\newblock {\em \mnras}, 418:1649--1667.

\bibitem[{Lagos} et~al., 2008]{Lagos2008}
{Lagos}, C.~D.~P., {Cora}, S.~A., and {Padilla}, N.~D. (2008).
\newblock {Effects of AGN feedback on {$\Lambda$}CDM galaxies}.
\newblock {\em \mnras}, 388:587--602.

\bibitem[{Lagos} et~al., 2018]{Lagos2018}
{Lagos}, C.~d.~P., {Tobar}, R.~J., {Robotham}, A.~S.~G., {Obreschkow}, D.,
  {Mitchell}, P.~D., {Power}, C., and {Elahi}, P.~J. (2018).
\newblock {Shark: introducing an open source, free and flexible semi-analytic
  model of galaxy formation}.
\newblock {\em ArXiv e-prints}.

\bibitem[{Lange} et~al., 2015]{Lange2015}
{Lange}, R., {Driver}, S.~P., {Robotham}, A.~S.~G., {Kelvin}, L.~S., {Graham},
  A.~W., {Alpaslan}, M., {Andrews}, S.~K., {Baldry}, I.~K., {Bamford}, S.,
  {Bland-Hawthorn}, J., {Brough}, S., {Cluver}, M.~E., {Conselice}, C.~J.,
  {Davies}, L.~J.~M., {Haeussler}, B., {Konstantopoulos}, I.~S., {Loveday}, J.,
  {Moffett}, A.~J., {Norberg}, P., {Phillipps}, S., {Taylor}, E.~N.,
  {L{\'o}pez-S{\'a}nchez}, {\'A}.~R., and {Wilkins}, S.~M. (2015).
\newblock {Galaxy And Mass Assembly (GAMA): mass-size relations of z $\sim$ 0.1
  galaxies subdivided by S{\'e}rsic index, colour and morphology}.
\newblock {\em \mnras}, 447:2603--2630.

\bibitem[{Laurikainen} and {Salo}, 2016]{Laurikainen2016A}
{Laurikainen}, E. and {Salo}, H. (2016).
\newblock {Observed Properties of Boxy/Peanut/Barlens Bulges}.
\newblock In {Laurikainen}, E., {Peletier}, R., and {Gadotti}, D., editors,
  {\em Galactic Bulges}, volume 418 of {\em Astrophysics and Space Science
  Library}, page~77.

\bibitem[{Menci} et~al., 2004]{Menci2004}
{Menci}, N., {Cavaliere}, A., {Fontana}, A., {Giallongo}, E., {Poli}, F., and
  {Vittorini}, V. (2004).
\newblock {Early Hierarchical Formation of Massive Galaxies Triggered by
  Interactions}.
\newblock {\em \apj}, 604:12--17.

\bibitem[{Menci} et~al., 2014]{Menci2014}
{Menci}, N., {Gatti}, M., {Fiore}, F., and {Lamastra}, A. (2014).
\newblock {Triggering active galactic nuclei in hierarchical galaxy formation:
  disk instability vs. interactions}.
\newblock {\em \aap}, 569:A37.

\bibitem[{M{\'e}ndez-Abreu} et~al., 2019]{MendezAbreu2019}
{M{\'e}ndez-Abreu}, J., {de Lorenzo-C{\'a}ceres}, A., {Gadotti}, D.~A.,
  {Fragkoudi}, F., {van de Ven}, G., {Falc{\'o}n-Barroso}, J., {Leaman}, R.,
  {P{\'e}rez}, I., {Querejeta}, M., {S{\'a}nchez-Blazquez}, P., and {Seidel},
  M. (2019).
\newblock {Inner bars also buckle. The MUSE TIMER view of the double-barred
  galaxy NGC 1291}.
\newblock {\em \mnras}, 482:L118--L122.

\bibitem[{M{\'e}ndez-Abreu} et~al., 2010]{MendezAbreu2010}
{M{\'e}ndez-Abreu}, J., {S{\'a}nchez-Janssen}, R., and {Aguerri}, J.~A.~L.
  (2010).
\newblock {Which Galaxies Host Bars and Disks? A Study of the Coma Cluster}.
\newblock {\em \apjl}, 711:L61--L65.

\bibitem[{Mo} et~al., 2010]{MoWhite2010}
{Mo}, H., {van den Bosch}, F.~C., and {White}, S. (2010).
\newblock {\em {Galaxy Formation and Evolution}}.

\bibitem[{Mo} et~al., 1998]{MoMaoWhite1997}
{Mo}, H.~J., {Mao}, S., and {White}, S.~D.~M. (1998).
\newblock {The formation of galactic discs}.
\newblock {\em \mnras}, 295:319--336.

\bibitem[{Moetazedian} et~al., 2017]{Moetazedian2017}
{Moetazedian}, R., {Polyachenko}, E.~V., {Berczik}, P., and {Just}, A. (2017).
\newblock {Effects of galaxy-satellite interactions on bar formation}.
\newblock {\em \aap}, 604:A75.

\bibitem[{Moorthy} and {Holtzman}, 2006]{Moorthy2006}
{Moorthy}, B.~K. and {Holtzman}, J.~A. (2006).
\newblock {Stellar populations in bulges of spiral galaxies}.
\newblock {\em \mnras}, 371:583--608.

\bibitem[{Naab} et~al., 2006]{Naab2006}
{Naab}, T., {Jesseit}, R., and {Burkert}, A. (2006).
\newblock {The influence of gas on the structure of merger remnants}.
\newblock {\em \mnras}, 372:839--852.

\bibitem[{Navarro} et~al., 1996]{NFW1996}
{Navarro}, J.~F., {Frenk}, C.~S., and {White}, S.~D.~M. (1996).
\newblock {The Structure of Cold Dark Matter Halos}.
\newblock {\em \apj}, 462:563.

\bibitem[{Noeske} et~al., 2007]{Noeske2007}
{Noeske}, K.~G., {Weiner}, B.~J., {Faber}, S.~M., {Papovich}, C., {Koo}, D.~C.,
  {Somerville}, R.~S., {Bundy}, K., {Conselice}, C.~J., {Newman}, J.~A.,
  {Schiminovich}, D., {Le Floc'h}, E., {Coil}, A.~L., {Rieke}, G.~H., {Lotz},
  J.~M., {Primack}, J.~R., {Barmby}, P., {Cooper}, M.~C., {Davis}, M., {Ellis},
  R.~S., {Fazio}, G.~G., {Guhathakurta}, P., {Huang}, J., {Kassin}, S.~A.,
  {Martin}, D.~C., {Phillips}, A.~C., {Rich}, R.~M., {Small}, T.~A., {Willmer},
  C.~N.~A., and {Wilson}, G. (2007).
\newblock {Star Formation in AEGIS Field Galaxies since z=1.1: The Dominance of
  Gradually Declining Star Formation, and the Main Sequence of Star-forming
  Galaxies}.
\newblock {\em \apjl}, 660:L43--L46.

\bibitem[{Noguchi}, 1998]{Noguchi1998}
{Noguchi}, M. (1998).
\newblock {Clumpy star-forming regions as the origin of the peculiar morphology
  of high-redshift galaxies}.
\newblock {\em \nat}, 392:253.

\bibitem[{Noguchi}, 1999]{Noguchi1999}
{Noguchi}, M. (1999).
\newblock {Early Evolution of Disk Galaxies: Formation of Bulges in Clumpy
  Young Galactic Disks}.
\newblock {\em \apj}, 514:77--95.

\bibitem[{Obreja} et~al., 2013]{Obreja2013}
{Obreja}, A., {Dom{\'{\i}}nguez-Tenreiro}, R., {Brook}, C.,
  {Mart{\'{\i}}nez-Serrano}, F.~J., {Dom{\'e}nech-Moral}, M., {Serna}, A.,
  {Moll{\'a}}, M., and {Stinson}, G. (2013).
\newblock {A Two-phase Scenario for Bulge Assembly in {$\Lambda$}CDM
  Cosmologies}.
\newblock {\em \apj}, 763:26.

\bibitem[{Ostriker} and {Peebles}, 1973]{Ostriker1973}
{Ostriker}, J.~P. and {Peebles}, P.~J.~E. (1973).
\newblock {A Numerical Study of the Stability of Flattened Galaxies: or, can
  Cold Galaxies Survive?}
\newblock {\em \apj}, 186:467--480.

\bibitem[{Papovich} et~al., 2005]{Papovich2005}
{Papovich}, C., {Dickinson}, M., {Giavalisco}, M., {Conselice}, C.~J., and
  {Ferguson}, H.~C. (2005).
\newblock {The Assembly of Diversity in the Morphologies and Stellar
  Populations of High-Redshift Galaxies}.
\newblock {\em \apj}, 631:101--120.

\bibitem[{Pe{\~n}arrubia} et~al., 2006]{Peniarubia2006}
{Pe{\~n}arrubia}, J., {McConnachie}, A., and {Babul}, A. (2006).
\newblock {On the Formation of Extended Galactic Disks by Tidally Disrupted
  Dwarf Galaxies}.
\newblock {\em \apjl}, 650:L33--L36.

\bibitem[{Pfenniger} and {Norman}, 1990]{Pfenniger1990}
{Pfenniger}, D. and {Norman}, C. (1990).
\newblock {Dissipation in barred galaxies - The growth of bulges and central
  mass concentrations}.
\newblock {\em \apj}, 363:391--410.

\bibitem[{Planck Collaboration} et~al., 2014]{PlanckCollaboration2014}
{Planck Collaboration}, {Ade}, P.~A.~R., {Aghanim}, N., {Armitage-Caplan}, C.,
  {Arnaud}, M., {Ashdown}, M., {Atrio-Barandela}, F., {Aumont}, J.,
  {Baccigalupi}, C., {Banday}, A.~J., and et~al. (2014).
\newblock {Planck 2013 results. XVI. Cosmological parameters}.
\newblock {\em \aap}, 571:A16.

\bibitem[{Porter} et~al., 2014]{Porter2014}
{Porter}, L.~A., {Somerville}, R.~S., {Primack}, J.~R., and {Johansson}, P.~H.
  (2014).
\newblock {Understanding the structural scaling relations of early-type
  galaxies}.
\newblock {\em \mnras}, 444:942--960.

\bibitem[{Rahimi} et~al., 2010]{Rahimi2010}
{Rahimi}, A., {Kawata}, D., {Brook}, C.~B., and {Gibson}, B.~K. (2010).
\newblock {Chemodynamical analysis of bulge stars for simulated disc galaxies}.
\newblock {\em \mnras}, 401:1826--1831.

\bibitem[{Ribeiro} et~al., 2016]{Ribeiro2016}
{Ribeiro}, B., {Lobo}, C., {Ant{\'o}n}, S., {Gomes}, J.~M., and {Papaderos}, P.
  (2016).
\newblock {Red galaxies with pseudo-bulges in the SDSS: closer to disc galaxies
  or to classical bulges?}
\newblock {\em \mnras}, 456:3899--3914.

\bibitem[{Ryan} et~al., 2008]{Ryan2008}
{Ryan}, Jr., R.~E., {Cohen}, S.~H., {Windhorst}, R.~A., and {Silk}, J. (2008).
\newblock {Galaxy Mergers at z gtrsim 1 in the HUDF: Evidence for a Peak in the
  Major Merger Rate of Massive Galaxies}.
\newblock {\em \apj}, 678:751--757.

\bibitem[{Saha}, 2015]{Saha2015}
{Saha}, K. (2015).
\newblock {Lost in Secular Evolution: The Case of a Low-mass Classical Bulge}.
\newblock {\em \apjl}, 806:L29.

\bibitem[{Sales} et~al., 2007]{Sales2007}
{Sales}, L.~V., {Navarro}, J.~F., {Abadi}, M.~G., and {Steinmetz}, M. (2007).
\newblock {Satellites of simulated galaxies: survival, merging and their
  relationto the dark and stellar haloes}.
\newblock {\em \mnras}, 379:1464--1474.

\bibitem[{Sellwood}, 2016]{Sellwood2016}
{Sellwood}, J.~A. (2016).
\newblock {Bar Instability in Disk-Halo Systems}.
\newblock {\em \apj}, 819:92.

\bibitem[{Shankar} et~al., 2013]{Shankar2013}
{Shankar}, F., {Marulli}, F., {Bernardi}, M., {Mei}, S., {Meert}, A., and
  {Vikram}, V. (2013).
\newblock {Size evolution of spheroids in a hierarchical Universe}.
\newblock {\em \mnras}, 428:109--128.

\bibitem[{Shankar} et~al., 2012]{Shankar2012}
{Shankar}, F., {Marulli}, F., {Mathur}, S., {Bernardi}, M., and {Bournaud}, F.
  (2012).
\newblock {Black holes in pseudobulges: demography and models}.
\newblock {\em \aap}, 540:A23.

\bibitem[{Shen} et~al., 2003]{Shen2003}
{Shen}, S., {Mo}, H.~J., {White}, S.~D.~M., {Blanton}, M.~R., {Kauffmann}, G.,
  {Voges}, W., {Brinkmann}, J., and {Csabai}, I. (2003).
\newblock {The size distribution of galaxies in the Sloan Digital Sky Survey}.
\newblock {\em \mnras}, 343:978--994.

\bibitem[{Simien} and {de Vaucouleurs}, 1986]{SimienandVaucouleurs1986S}
{Simien}, F. and {de Vaucouleurs}, G. (1986).
\newblock {Systematics of bulge-to-disk ratios}.
\newblock {\em \apj}, 302:564--578.

\bibitem[{Somerville} et~al., 2008]{Somerville2008}
{Somerville}, R.~S., {Hopkins}, P.~F., {Cox}, T.~J., {Robertson}, B.~E., and
  {Hernquist}, L. (2008).
\newblock {A semi-analytic model for the co-evolution of galaxies, black holes
  and active galactic nuclei}.
\newblock {\em \mnras}, 391:481--506.

\bibitem[{Somerville} et~al., 2001]{Somerville2001}
{Somerville}, R.~S., {Primack}, J.~R., and {Faber}, S.~M. (2001).
\newblock {The nature of high-redshift galaxies}.
\newblock {\em \mnras}, 320:504--528.

\bibitem[{Spinoso} et~al., 2017]{spinoso2017}
{Spinoso}, D., {Bonoli}, S., {Dotti}, M., {Mayer}, L., {Madau}, P., and
  {Bellovary}, J. (2017).
\newblock {Bar-driven evolution and quenching of spiral galaxies in
  cosmological simulations}.
\newblock {\em \mnras}, 465:3729--3740.

\bibitem[{Springel}, 2005]{Springel2005}
{Springel}, V. (2005).
\newblock {The cosmological simulation code GADGET-2}.
\newblock {\em \mnras}, 364:1105--1134.

\bibitem[{Springel} et~al., 2001]{Springel2001}
{Springel}, V., {White}, S.~D.~M., {Tormen}, G., and {Kauffmann}, G. (2001).
\newblock {Populating a cluster of galaxies - I. Results at [formmu2]z=0}.
\newblock {\em \mnras}, 328:726--750.

\bibitem[{Tacconi} et~al., 2002]{Tacconi2002}
{Tacconi}, L.~J., {Genzel}, R., {Lutz}, D., {Rigopoulou}, D., {Baker}, A.~J.,
  {Iserlohe}, C., and {Tecza}, M. (2002).
\newblock {Ultraluminous Infrared Galaxies: QSOs in Formation?}
\newblock {\em \apj}, 580:73--87.

\bibitem[{Tamburri} et~al., 2014]{Tamburri2014}
{Tamburri}, S., {Saracco}, P., {Longhetti}, M., {Gargiulo}, A., {Lonoce}, I.,
  and {Ciocca}, F. (2014).
\newblock {The population of early-type galaxies: how it evolves with time and
  how it differs from passive and late-type galaxies}.
\newblock {\em \aap}, 570:A102.

\bibitem[{Tonini} et~al., 2016]{Tonini2016}
{Tonini}, C., {Mutch}, S.~J., {Croton}, D.~J., and {Wyithe}, J.~S.~B. (2016).
\newblock {The growth of discs and bulges during hierarchical galaxy formation
  - I. Fast evolution versus secular processes}.
\newblock {\em \mnras}, 459:4109--4129.

\bibitem[{Toomre}, 1981]{Toomre1981}
{Toomre}, A. (1981).
\newblock {What amplifies the spirals}.
\newblock In {Fall}, S.~M. and {Lynden-Bell}, D., editors, {\em Structure and
  Evolution of Normal Galaxies}, pages 111--136.

\bibitem[{Ueda} et~al., 2014]{Ueda2014}
{Ueda}, J., {Iono}, D., {Yun}, M.~S., {Crocker}, A.~F., {Narayanan}, D.,
  {Komugi}, S., {Espada}, D., {Hatsukade}, B., {Kaneko}, H., {Matsuda}, Y.,
  {Tamura}, Y., {Wilner}, D.~J., {Kawabe}, R., and {Pan}, H.-A. (2014).
\newblock {Cold Molecular Gas in Merger Remnants. I. Formation of Molecular Gas
  Disks}.
\newblock {\em \apjs}, 214:1.

\bibitem[{van der Wel} et~al., 2014]{vanderWel2014}
{van der Wel}, A., {Franx}, M., {van Dokkum}, P.~G., {Skelton}, R.~E.,
  {Momcheva}, I.~G., {Whitaker}, K.~E., {Brammer}, G.~B., {Bell}, E.~F., {Rix},
  H.-W., {Wuyts}, S., {Ferguson}, H.~C., {Holden}, B.~P., {Barro}, G.,
  {Koekemoer}, A.~M., {Chang}, Y.-Y., {McGrath}, E.~J., {H{\"a}ussler}, B.,
  {Dekel}, A., {Behroozi}, P., {Fumagalli}, M., {Leja}, J., {Lundgren}, B.~F.,
  {Maseda}, M.~V., {Nelson}, E.~J., {Wake}, D.~A., {Patel}, S.~G., {Labb{\'e}},
  I., {Faber}, S.~M., {Grogin}, N.~A., and {Kocevski}, D.~D. (2014).
\newblock {3D-HST+CANDELS: The Evolution of the Galaxy Size-Mass Distribution
  since z = 3}.
\newblock {\em \apj}, 788:28.

\bibitem[{van Dokkum}, 2005]{vanDokkum2005}
{van Dokkum}, P.~G. (2005).
\newblock {The Recent and Continuing Assembly of Field Elliptical Galaxies by
  Red Mergers}.
\newblock {\em \aj}, 130:2647--2665.

\bibitem[{White} and {Frenk}, 1991]{WhiteFrenk1991}
{White}, S.~D.~M. and {Frenk}, C.~S. (1991).
\newblock {Galaxy formation through hierarchical clustering}.
\newblock {\em \apj}, 379:52--79.

\bibitem[{White} and {Rees}, 1978]{WhiteandRees1978}
{White}, S.~D.~M. and {Rees}, M.~J. (1978).
\newblock {Core condensation in heavy halos - A two-stage theory for galaxy
  formation and clustering}.
\newblock {\em \mnras}, 183:341--358.

\bibitem[{Zana} et~al., 2018a]{Zana2018A}
{Zana}, T., {Dotti}, M., {Capelo}, P.~R., {Bonoli}, S., {Haardt}, F., {Mayer},
  L., and {Spinoso}, D. (2018a).
\newblock {External versus internal triggers of bar formation in cosmological
  zoom-in simulations}.
\newblock {\em \mnras}, 473:2608--2621.

\bibitem[{Zana} et~al., 2018b]{Zana2018B}
{Zana}, T., {Dotti}, M., {Capelo}, P.~R., {Mayer}, L., {Haardt}, F., {Shen},
  S., and {Bonoli}, S. (2018b).
\newblock {Bar resilience to flybys in a cosmological framework}.
\newblock {\em \mnras}, 479:5214--5219.

\bibitem[{Zoldan} et~al., 2018]{Zoldan2018}
{Zoldan}, A., {De Lucia}, G., {Xie}, L., {Fontanot}, F., and {Hirschmann}, M.
  (2018).
\newblock {Structural and dynamical properties of galaxies in a hierarchical
  Universe: sizes and specific angular momenta}.
\newblock {\em \mnras}, 481:1376--1400.

\end{thebibliography}
\bibliographystyle{apalike}
\appendix

\section{Galactic sizes}\label{Appendix:Radius}
\subsubsection{Disk size}
One of the fundamental properties of galaxies is the radial size of their disks, as many evolutionary quantities depend on it \citep[such as the dynamical time of the disk and the galaxy SFR, for instance; for more details see e.g.][]{Henriques2015, Guo2011}. To compute the radii of both gaseous and stellar disk components, the model assumes exponential density profiles, so that both can be written as:
\begin{equation}
\Sigma^{\rm gas}(R) = \Sigma_{0}^{\rm gas} e^{-R_{\rm gas}/R_{\rm gas}^{sl}}    
\end{equation}
\vspace{-4mm}
\begin{equation}
    \Sigma^{\star}(R) = \Sigma_{0}^{\star} e^{-R_{\star}/R_{\star}^{sl}} \label{eq:Stellar_density_disk}
\end{equation}
\noindent
where $\Sigma_{0}^{\rm gas} = M_{\rm gas}/2\pi R_{\rm gas}^{sl}$ and $\Sigma_{0}^{\star} =  M_{\rm stellar}/2\pi R_{\star}^{sl}$ are the central surface densities of gaseous and stellar disks, respectively, while $R_{\rm gas}^{sl}$ and $R_{\star}^{sl}$ are their scale lengths. In particular, the two latter values are given by:

\begin{equation} \label{eq:Stellar_radii_scalelength}
R_{\rm gas}^{sl} = \frac{|\boldsymbol{J_{\rm gas}}|/M_{\rm gas}^{disk}}{2 V_{\rm max}}
\end{equation}
\begin{equation}
R_{\rm \star}^{sl} = \frac{|\boldsymbol{J_{\star}}|/M_{\rm stellar}^{\rm disk}}{2 V_{\rm max}}
\end{equation}

\noindent
are the total angular momentum values of the gaseous and stellar disk, respectively. It is clear, then, that it is crucial to model the time evolution of the disks angular momentum in order for the model to provide a good description of their radii.\\

During the evolution of galaxies in the model, both internal and external processes (such as cooling, star formation and SN feedback, or galaxy mergers respectively) induce modifications of the galactic disk's total angular momentum vector ($\boldsymbol{J}_{\rm gas}^{\rm Total}$). Following \cite{Guo2011}, the variation of $\boldsymbol{J}_{\rm gas}^{\rm Total}$ can be written as:  
\begin{equation}
\delta \boldsymbol{J}_{\rm gas}^{\rm Total} = \delta \boldsymbol{J}_{\rm gas,cooling} + \delta \boldsymbol{J}_{\rm gas,SF} + \delta \boldsymbol{J}_{\rm gas, merger}
\end{equation}
where each of the three components on the right-hand side is the variation of angular momentum induced on the gas disk by a specific process. 
An analogous relation can be written for the \textit{stellar} disk angular momentum, under the simplifying assumption that the only process contributing is star formation:
\begin{equation}
\delta \boldsymbol{J}_{\rm star}^{\rm Total} = \delta \boldsymbol{J}_{\rm gas,SF} 
\end{equation}


\subsubsection{Bulge size after a merger}

\begin{figure*} 
	\centering
    \includegraphics[width=1.\textwidth]{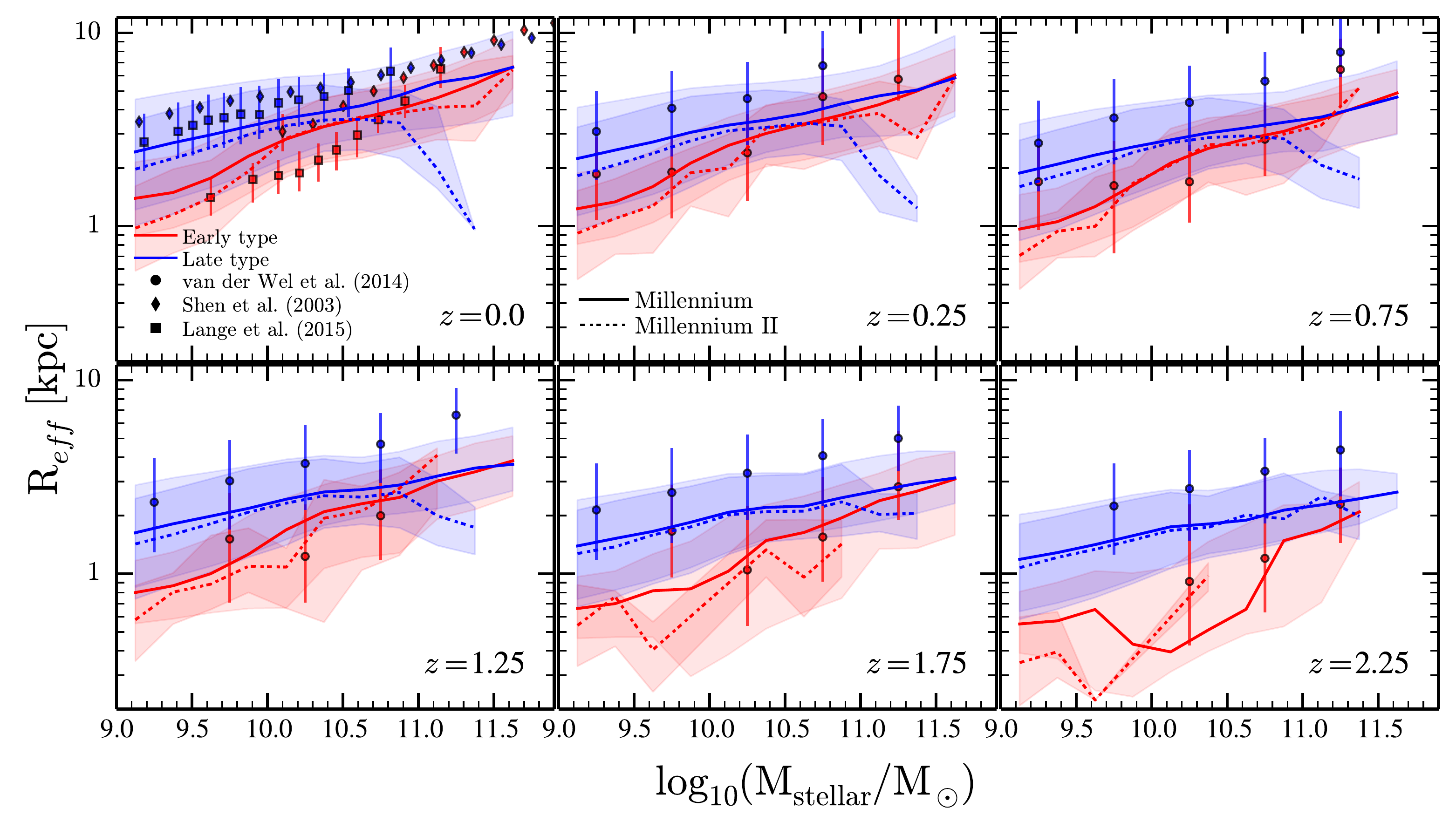}
	\caption{Effective radius of early and late type galaxies in the \texttt{Millennium} (solid lines) and \texttt{Millennium II} (dashed lines). Blue lines corresponds to late-type galaxies (\textit{disk-to-total} ratio $\rm D/T\,{>}\,0.8$) and red ones with early-type (\textit{bulge-to-total} ratio $\rm B/T\,{>}\,0.7$). Shaded areas corresponds to $1 \sigma$ dispersion. We compare the predictions with the data available from \protect\cite{Shen2003,vanderWel2014} and \protect\cite{Lange2015}.}
	\label{fig:Effective radii}
\end{figure*}

It has been shown that SAMs generate unrealistic bulge sizes if the dissipation of energy during gas rich mergers is not take into account \citep{Naab2006,Shankar2013,Zoldan2018}. Here we address this issue, implementing some analytic expressions presented in the literature. 
Following the standard picture, 
when a merger takes place the half-mass radii of the new bulge $\rm R_{new,bulge}$ is computed via energy conservation and virial theorem \citep{Covington2011,Guo2011,Porter2014,Tonini2016}: 
\begin{equation} \label{eq:Energies}
\begin{split}
\rm E_{f} = \sum_{i=1}^{i=2} E_{0}^i + E_{orbital} + E_{dissipation} = \\  \rm \sum_{i=1}^{i=2} E_{0}^i + E_{orbital} + C_{rad} \, \mathit{f}_{gas} \left( \sum_{i=1}^{i=2} E_{0}^i \right) = &  \rm \left( 1 + \mathit{f}_{gas} C_{rad}\right)\sum_{i=1}^{i=2} E_{0}^i + E_{orbital}
\end{split}
\end{equation}
where $\rm E_{f}$,  $\rm E_{0}^i$, $\rm E_{orbital}$ and  $\rm E_{dissipation}$ are, respectively, the total binding energy of the merger remnant, the self-binding energy of the $i$ progenitor, the orbital energy and the losses via dissipation by the gas component due to shocks. The latter can be written as: 
\begin{equation}
\rm E_{dissipation} = C_{rad} \, \mathit{f}_{gas} \left( \sum_{i=1}^{i=2} E_{0}^i\right)
\end{equation}
where $\rm C_{rad}$  is an efficiency parameter set to 2.75 \citep{Covington2011} and $\rm \mathit{f}_{\rm gas} = \rm \sum_{i=1}^{i=2} M_{gas}^{i}/(\sum_{i=1}^{i=2} M_{gas}^{i} + M_{stellar}^{i})$, the total merger gas fraction. Therefore, developing Eq.\eqref{eq:Energies} $\rm R_{new,bulge}$ can be computed from:
\begin{equation} \label{eq:bulgeformation_merger}
\begin{split}
\rm C\frac{G \, (M_{P1} + M_{P2})^2}{R_{new,bulge}} =   C  \left( 1 + \mathit{f}_{gas} C_{rad}\right) \left( \frac{G\,M^2_{P1}}{R_{P1}} + \rm \frac{G\,M^2_{P2}}{R_{P2}} \right) &  \\ + \;\rm  \mathit{f}_{orb} \frac{G\, M_{P1}M_{P2}}{R_{P1} + R_{P2}}
\end{split}
\end{equation}
where $f_{\rm orb}$ quantifies the orbital energy of the system and $\rm C$ is a structural parameter. From hereafter, we assume $C\,{=}\,0.5$ \citep{Guo2011} and $f_{\rm orb}\,{=}\,0$ \citep[see the discussion in][]{Shankar2013}. The values of $\rm M_{P1}$ and $\rm M_{P2}$ are respectively the total mass of the progenitor 1 and 2 involved in the bulge formation/growth after the merger and $\rm R_{P1}$, $\rm R_{P2}$ its respective half-mass radii.

Following the \textit{minor} merger formalism described in Section~\ref{sec:mergers_and_smooth_accretion} we assume that $\rm M_{P1}\,{=}\, M_{bulge}^{Central}$ and $\rm M_{P2} \,{=}\, M_{stellar}^{Satellite}$. 
Since during these processes the gas does not reach the bulge region, we assume no dissipation losses, i.e  $\rm E_{dissipation} = 0$ \citep[see more details in ][]{Shankar2013}. On the other hand, in \textit{major} mergers we allow dissipation effects as a consequence of the violent interaction setting $\rm M_{P1}$ and $\rm M_{P2}$ as the total baryonic mass (stellar and gas) of the central and satellite galaxy, respectively. To take into account that throughout this events the central parts of the dark matter halo is expected to behave with similar stellar dynamic, $\rm M_{P1}$ is changed by $\rm M_{P1} \,{=}\, M_{bar}^{P1} + \alpha \,M_{halo}(r<R_{P1})$ \citep[see][]{Shankar2013,Lagos2018}. Following \cite{Lagos2018} we set $\rm \alpha \,{=}\,2$. The value of $\rm M_{\rm halo}(r<R)$ is computed assuming a Navarro-Frenk-White profile \citep{NFW1996} with a concentration parameter computed using the fit presented in \cite{Dutton2014} for the Planck cosmology.


\subsubsection{Bulge size after a DI}\label{sec:DI_bulge_radi}

The bulge formation during this DI processes has to be modeled as well. The full analysis is described in \cite{Guo2011} but here we will summarize the main characteristics. When the galaxy does not host any bulge, the half mass radius of the newly formed bulge ($\rm R_{bulge}^{DI} $) can be found by solving: 
\begin{equation}\label{eq:Size_Bulge_DI}
\rm \Delta M_{\star}^{DI} = 2 \pi \Sigma_{0}^{\star} R_{\rm \star}^{sl} \left[ R_{\rm \star}^{sl} - \left( R_{bulge}^{DI} + R_{\rm \star}^{sl} \right)e^{-R_{bulge}^{DI}/R_{\rm \star}^{sl}}\right]
\end{equation}
where we had take into account the stellar surface density of Eq.\eqref{eq:Stellar_density_disk} and scale length of Eq.\eqref{eq:Stellar_radii_scalelength}. In the case of a pre-existing bulge, the DI bulge is generated as before but assuming that it mergers instantaneously with the old one in the same process explained in Eq.\eqref{eq:bulgeformation_merger} with $\rm M_{P1}\,{=}\,M_{Bulge}(t_{before}^{DI})$ and $\rm M_{P2}\,{=}\, \Delta M_{\star}^{DI}$ and the parameters $\rm C_{rad} \,{=}\, 0.0$, $\rm f_{orb} \,{=}\, 2.0$ to take into account the fact that the inner disc and the pre-existing bulge are concentric and have no relative motion.\\

In Fig~\ref{fig:Effective radii} we present the redshift evolution of galaxy sizes for MS and MSII (solid and dashed lines, respectively). The 3D galaxy effective radius ($\mathrm{R}_{eff}^{3D}$) has been computed as the mass weighted average of the bulge ($\rm R_{bulge}$) and stellar disc ($1.68R_{\star}^{sl}$) half-mass radii. To compare with data, we have converted the 3D into 2D projected half-mass radii ($\mathrm{R}_{eff}$) by using the conversion factor $0.68$ presented in \cite{Shankar2013}. Blue and red lines represents respectively the population of late-type (\textit{disk-to-total} ratio $\rm D/T\,{>}\,0.8$) and early-type ($\rm B/T\,{>}\,0.7$) galaxies. We have presented the predictions at 6 different redshift $z \,{=}\, 0,0.25,0.75,1.25,1.75$ and $2.25$. We compare them with the data available in \cite{Shen2003,vanderWel2014} and \cite{Lange2015}. As we can see, low massive galaxies in both early and late type population follow the observational trend. Nevertheless, for the most massive galaxies we predict slightly smaller radius. Probably, this is the consequence of the fact that we under-predict (overpredict) the elliptical (spiral) population in the most massive stellar mass bins (see Fig~\ref{fig:Morphology_DI_effect}). In spite of that, we are able to reproduce the redshift evolution of the stellar-size relation with a remarkable agreement with the observations.

\section{Convergence in the morphology for Millennium and Millennium II: A matter of major, minor mergers and smooth accretion}\label{Appendix:Smooth_accretion}

\begin{figure*}
	\centering
    \includegraphics[width=1\textwidth]{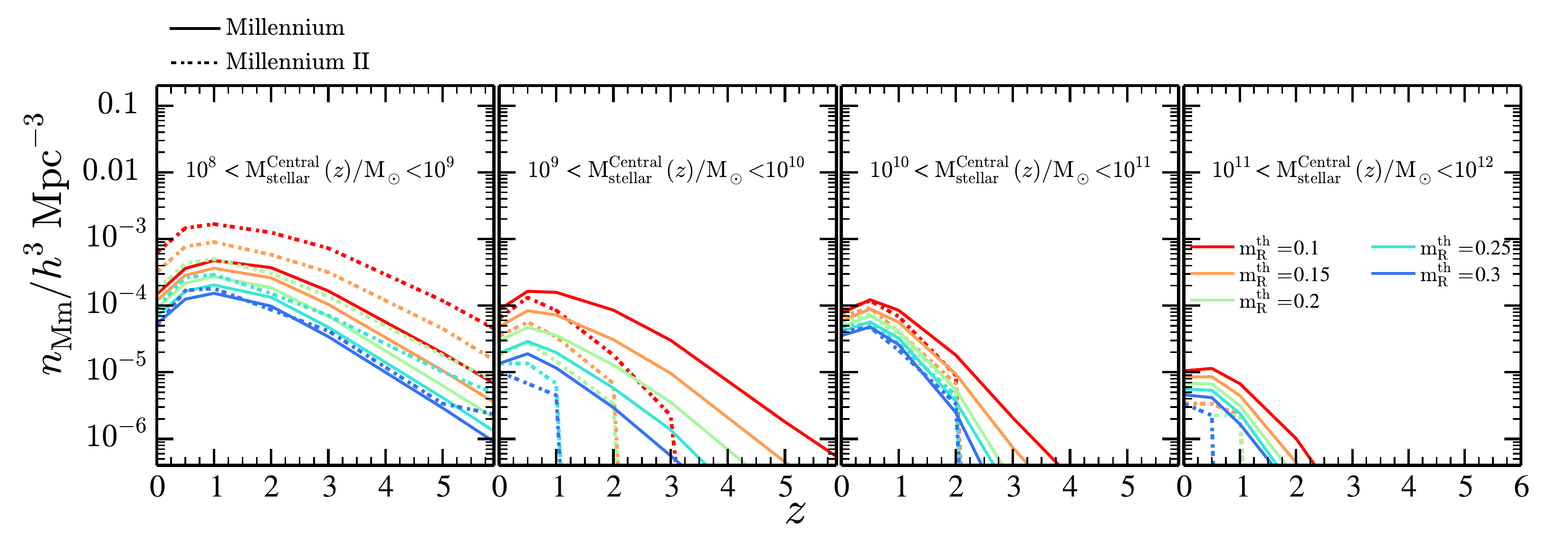}
    \includegraphics[width=1\textwidth]{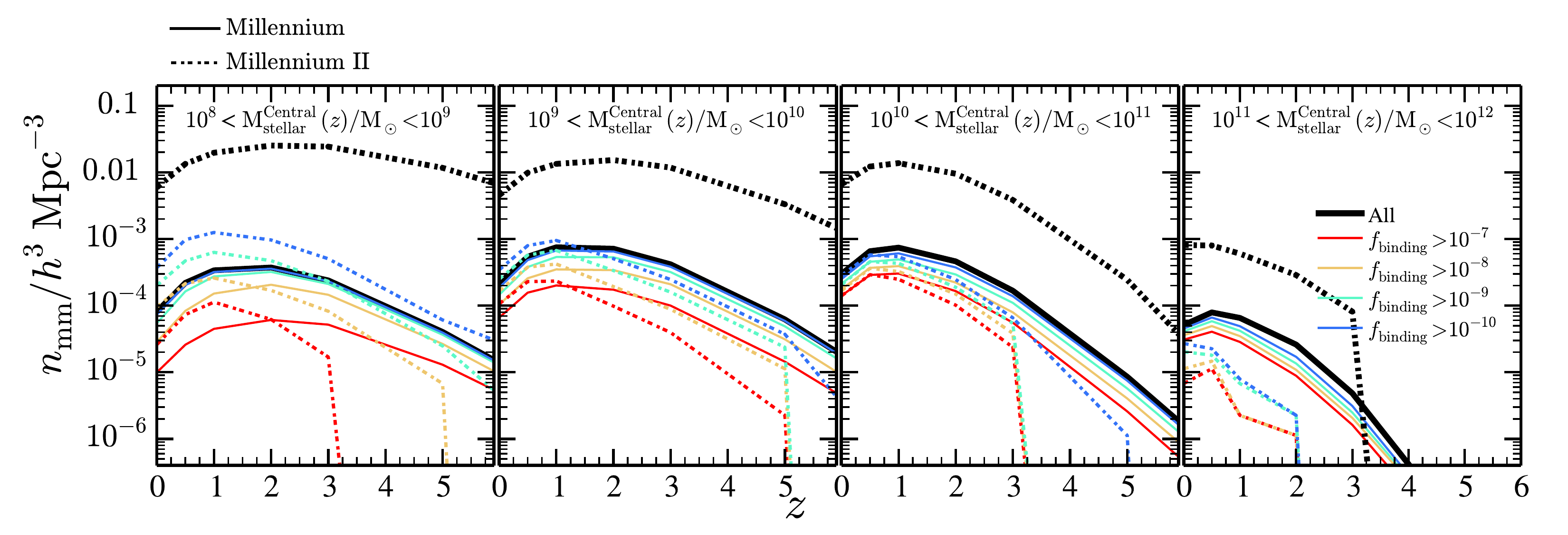}
	\caption{\textbf{Upper row}: Number density of major mergers $\mathit{n}_{\rm Mm}$. Different colors corresponds to different major/minor merger threshold ($\rm \rm m_R^{th}$). Solid and dashed lines represent respectively the predictions for MS and MSII merger trees. Each panel corresponds to different stellar masses of central galaxies at the moment of the merger. \textbf{Upper row}: Number density of minor mergers $\rm \mathit{n}_{mm}$. Each panel corresponds to different stellar masses of central galaxies at the moment of the merger. Here we have established $\rm m_{R}^{th}\,{=}\,0.2$. Different colors corresponds to different $\rm \mathit{f}_{binding}$ thresholds. Solid and dashed lines represent respectively the predictions for MS and MSII merger trees.}
\label{fig:number_density}
\end{figure*}

In this appendix we explore the convergence of the $z\,{=}\,0$ galaxy morphology at intermediate and low stellar masses between \texttt{Millennium} and \texttt{Millennium II}. By studying the mergers characteristics we find out that the responsible of the morphological disagreement between MS and MSII is a combination of major mergers and extreme minor mergers lead by small \textit{dwarf galaxies} ($\rm M_{stellar}\,{\lesssim}\, 10^{6}M_{\odot}$).\\

\begin{figure*}
	\centering
 \includegraphics[width=1.0\textwidth]{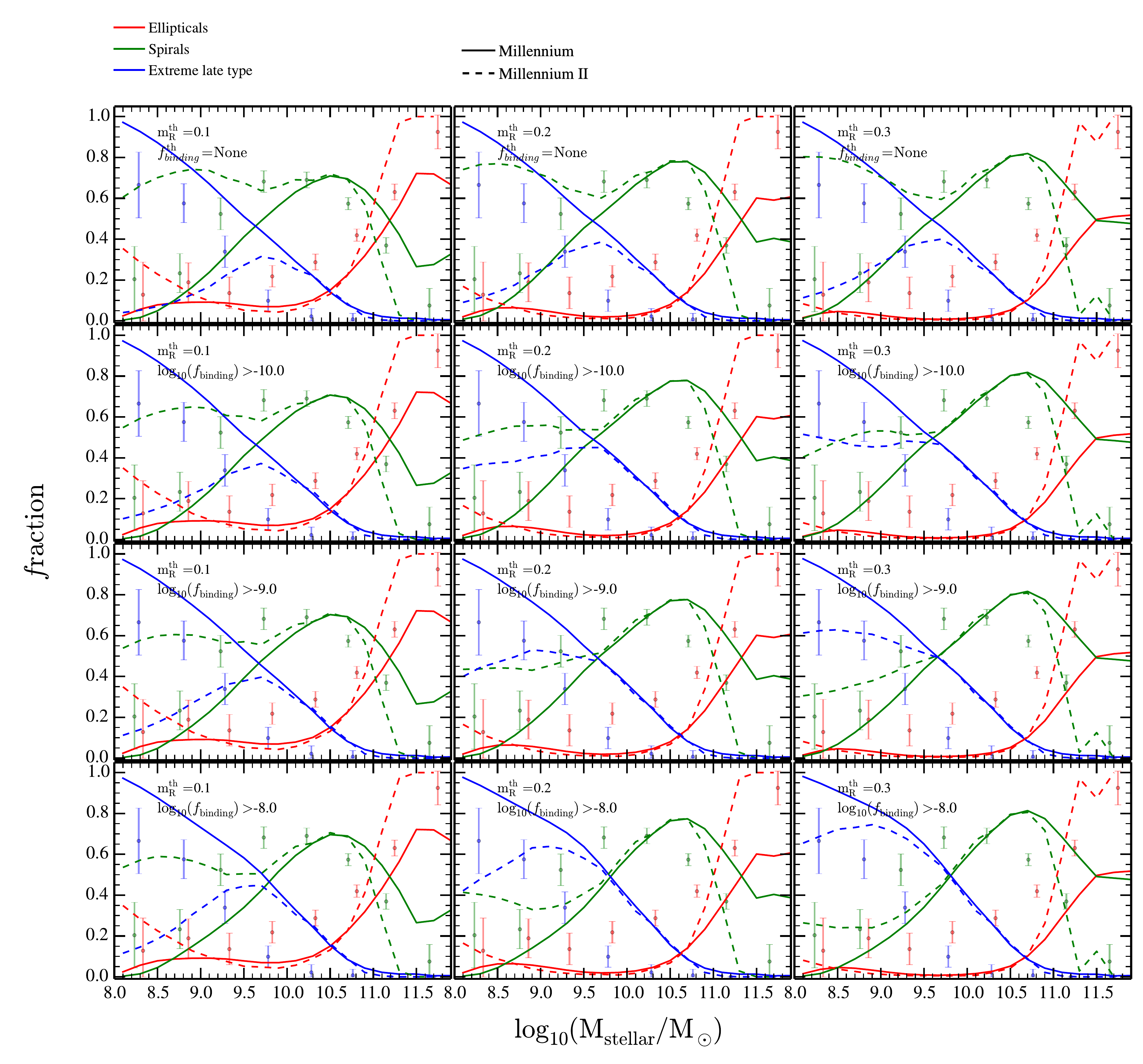}
	\caption{Evolution of morphology with different combinations of $f_{\rm binding}^{\rm th}$ and major/minor merger threshold $\rm m_{R}^{th}$. Each row and column correspond to a fix value of $f_{\rm binding}^{\rm th}$ and $\rm m_{R}^{th}$. Here we have presented the values $f_{\rm binding}^{\rm th}\,{=}\,10^{-10}, 10^{-9}, 10^{-8}$ and $\rm m_{R}^{th}\,{=}\,0.1,0.2,0.3$}
\label{fig:all_distribution_morph}
\end{figure*}

In order to investigate the $z\,{=}\,0$ morphology discrepancy between Millennium simulations, we started to explore the major mergers predictions. In Fig~\ref{fig:number_density} upper panel it is presented the redshift evolution of MS and MSII major mergers number density $n_{\rm Mm}$ at various $\rm m_R^{th}$ values. Each panel correspond to different central galaxy stellar mass at the moment of the merger, $\mathrm {M_{stellar}^{Central}(\mathit{z})}$. While galaxies with $\rm M_{stellar}^{Central}(\mathit{z})\,{>}\,10^9 \, M_{\odot}$ display similar $n_{\rm Mm}$ values in MS and MSII at any $\rm m_R^{th}$ threshold\footnote{The small disagreement is just due to MSII box size: massive galaxies are rare in a $\,{\sim}\, 100 \, [\mathrm{Mpc}/h]$ box side, especially at high-$z$.}, the $n_{\rm Mm}$ predictions for galaxies with $\rm M_{stellar}^{Central}(\mathit{z})\,{<}\,10^9 \, M_{\odot}$ diverge. At small thresholds ($\rm m_{R}^{th} <0.2$) the deviation between MSII and MS is almost one order of magnitude regardless of redshift. Increasing the $\rm m_R^{th}$ value ($\rm m_{R}^{th}\,{>}\,0.2$) the difference is reduced to a factor 3. Such disagreement can be easily understood by the fact that MSII is able to resolve smaller subhalos around central galaxies. Hence, small galaxies hosted in the \textit{friend-of-friend} central subhalo experience more frequently mergers with satellites galaxies of comparable \textit{baryonic} mass. As can be seen in Fig~\ref{fig:all_distribution_morph} first row the change of $\rm m_{R}^{th}$ has also an impact in the $z\,{=}\,0$ galaxy morphology. The high N-body resolution of MSII combined with low values of $\rm m_R^{th}$ (set to 0.1 in the \textit{standard model}) favors an increase of the elliptical galaxies at small stellar masses and overestimate the spiral population at $\rm M_{stellar}\,{\sim}\,10^{8-9.5} M_{\odot}$. Our analysis suggest that an improvement in the convergence of MS and MSII galaxy morphology and in the MSII disk-dominated galaxy population is achieved by imposing large $\rm m_{R}^{th}$ values. From hereafter we decide to use $\rm m_{R}^{th} = 0.2$ (closer to other thresholds imposed in others SAMs, see \cite{Somerville2001,Hatton2003,Lacey2016,Lagos2018}), based on the $n_{\rm Mm}$ number densities presented in Fig~\ref{fig:number_density}.

\begin{figure}
\centering
\includegraphics[width=0.9\columnwidth]{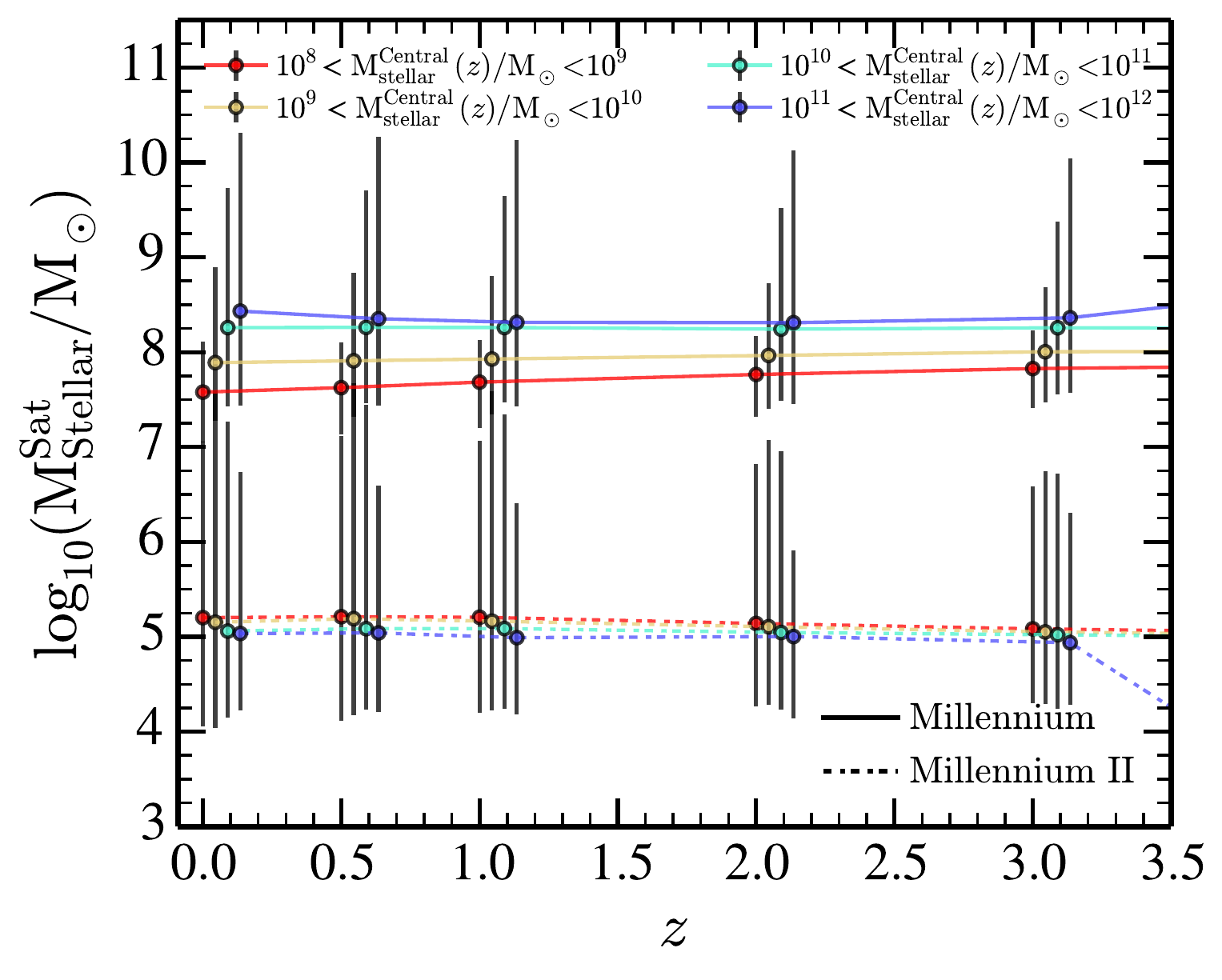}
\caption{Median stellar mass of the merging satellite galaxy ($\rm M_{Stellar}^{Sat}$). We have established a value of major/minor merger separation of $\rm m_{R}^{th} = 0.2$. Bars represents the $\rm 2\sigma$ value and colors different stellar mass bins of central galaxies at the moment of the merger ($\rm M_{Stellar}^{Central}(\mathit{z})$).}
\label{fig:median_stellar_mass}
\end{figure}

Nevertheless, as can be seen in Fig~\ref{fig:all_distribution_morph} first row the difference in $n_{\rm Mm}$ is not the unique cause of the MS and MSII morphology deviation. It is needed to explore the effect of the other type of galaxy mergers, i.e \textit{minor mergers}. In Fig~\ref{fig:number_density} lower panel we present, as we did with before, the number density of minor mergers $n_{\rm mm}$ as a function of redshift split in different central galaxy stellar mass. As we can see, the figure shows something expected: \LGalaxies run on top of MSII mergers trees predicts higher $n_{\rm mm}$ than run on the MS ones (see black lines). In particular the differences increases when we decrease the central stellar mass: form $\rm 1\,dex$ at masses $\rm M_{stellar}^{Central}(\mathit{z}) >10^{11} M_{\odot}$ up to $\rm 2\,dex$ at masses $ \rm M_{stellar}^{Central}(\mathit{z}) \, {\sim} 10^{8 - 9} M_{\odot}$. To explore the characteristics of the merging satellite galaxies, in Fig \ref{fig:median_stellar_mass} it is presented at different redshifts ($z<3$) their typical stellar mass. MS predicts a median merging satellite mass $\rm {\sim}\,10^{7.5} M_{\odot}$ with a small redshift evolution. Besides, more massive galaxies experience minor merger with slightly more massive galaxies. On contrary, MSII predicts smaller merging satellites $\rm {\sim}\,10^5 M_{\odot}$ (\textit{dwarf galaxies}) with not redshift evolution and dependence with the central galaxy stellar mass. In particular, we have found that these small mergers are the ones that lead the morphological change in galaxies with $\rm M_{stellar}\,{<}\,10^{9.5} M_{\odot}$. Such extreme interactions enlarge the bulges of the small central galaxies by incorporating the whole stellar mass of the satellites while their stellar disks are unable to increase in mass as the cold-gas content is no large enough to reach the critical mass imposed by the SAM to trigger an episode of star formation ($\rm M_{crit}\,{=}\,2.4{\times}10^{9} M_{\odot}$, see Eq.S14 of \cite{Henriques2015})\footnote{Note that a more accurate description of star formation might come by linking this process with the molecular gas component instead of the total cold gas \cite[see][]{Lagos2011}, as also discussed in \cite{Henriques2015}.}. Besides, merger induced bursts are not efficient either in this task. According to the merger ratios and the SAM efficiency parameters, less than the $\rm 0.2$\% of the total cold gas component is transformed in stars, i.e $\rm{\lesssim}\,10^5 M_{\odot}$ of new stars is added in the disk. Hence, the combination of the high number density of small interactions, the inefficient star formation and the simple minor merger recipe of \LGalaxies produces the unrealistic bust in the \textit{bulge-to-total} ratio in the low mass population of MSII, as can be seen in Fig~\ref{fig:all_distribution_morph}. All this points to the needed to update the minor mergers prescription implemented in \LGalaxies, as this appears to be not fully valid for treating interactions with extreme mass-ratios, particularly common in the MSII merger trees. We thus introduced a new set of prescriptions to include \textit{smooth accretion} as an additional channel for galaxy interactions \citep[see e.g.][]{Abadi2003,Peniarubia2006,Sales2007,Kazantzidis2008}.\\

\begin{figure}
	\centering
\includegraphics[width=\columnwidth]{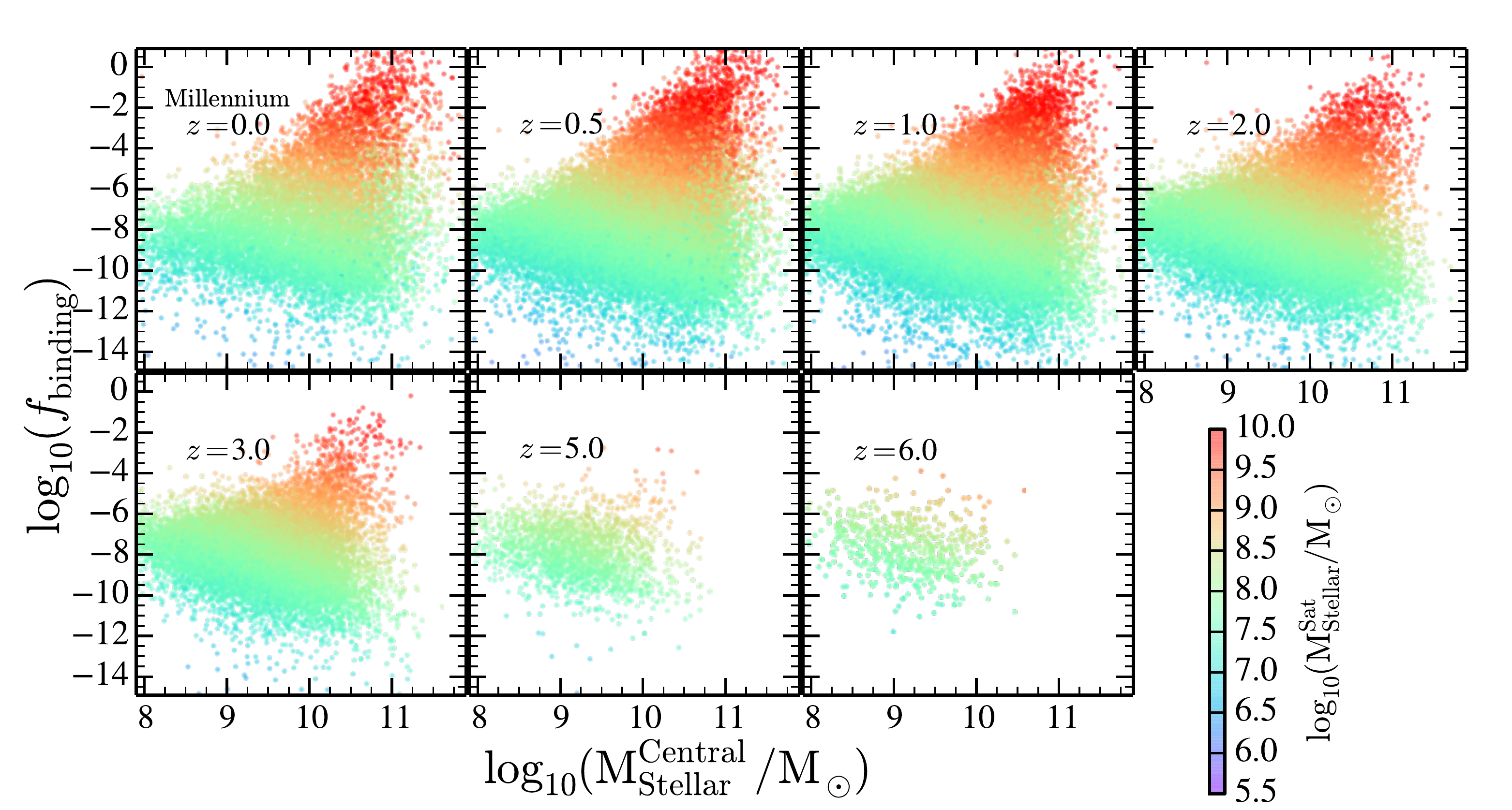}
    \includegraphics[width=\columnwidth]{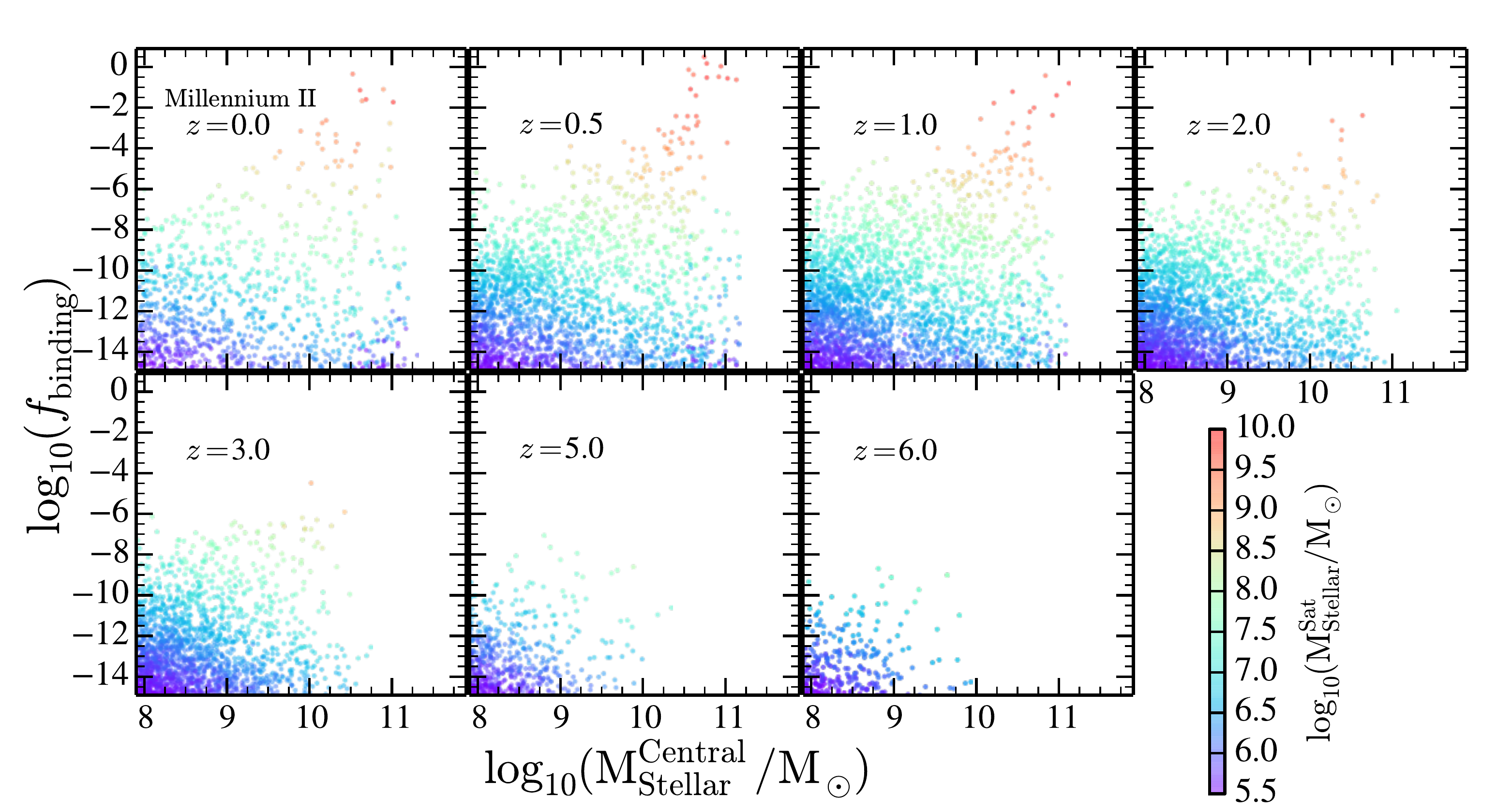}
	\caption{Plane $f_{\rm binding} - \rm M_{stellar}^{Central}$ for MS (top) and MSII (bottom) at different redshifts. The color codes the stellar mass of the merging satellite galaxy.} 
\label{fig:binding_energy}
\end{figure}

As we said, probably the crude minor merger recipe implemented in \LGalaxies is not completely valid in MSII. Its merger trees allow us to resolve the mergers of \textit{dwarf galaxies} whose merger interaction may not be completely address with the \LGalaxies standard recipe. In order to improve this scenario, we are going to allow another different minor interaction: \textit{smooth accretion} \citep{Abadi2003,Peniarubia2006,Sales2007,Kazantzidis2008}. While in  minor merges we follow the standard procedure presented \LGalaxies assuming that the whole stellar mass of the satellite galaxy is able to keep bound during the merger episode and reaches the central galaxy bulge, \textit{smooth accretions} are characterized by a deposit of the whole satellite stellar mass onto the central galaxy stellar disk. This scenario takes place when the stellar system (bulge and disk) of the satellite galaxy does not have the enough energy to keep it together and it is progressively diluted in the central galaxy stellar disk without the possibility of reaching the central galaxy center. We want to emphasize that \textit{smooth accretion} concept introduced here is not related with the already implemented tidal stripping events before the merger. \textit{Smooth accretion} goes beyond it and takes into account the redistribution of the satellite stellar mass that would happen during its interaction with the central galaxy disk throughout the galaxy-galaxy collision.\\

\begin{figure*}
\centering
\includegraphics[width=1.0\columnwidth]{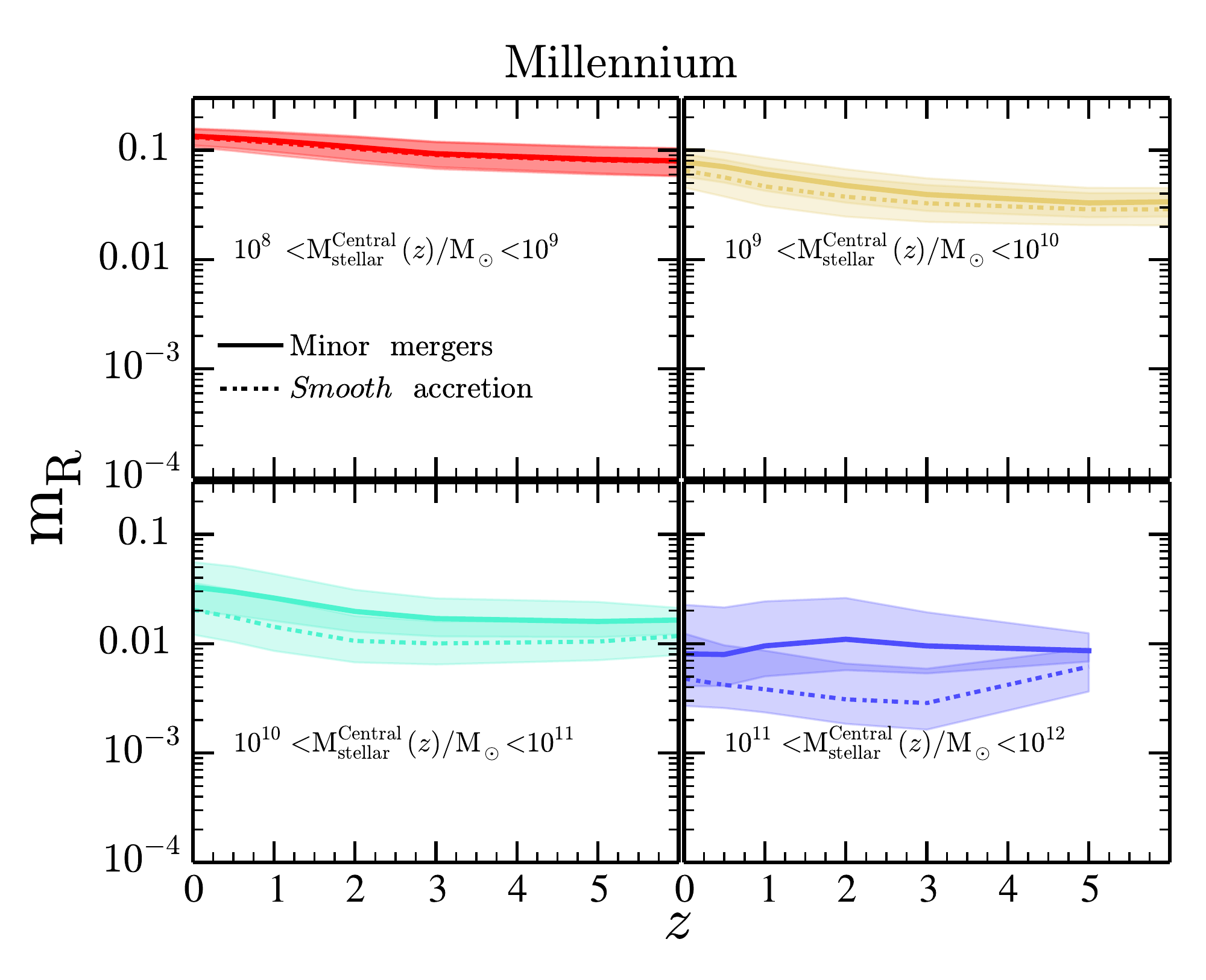}
\includegraphics[width=1.0\columnwidth]{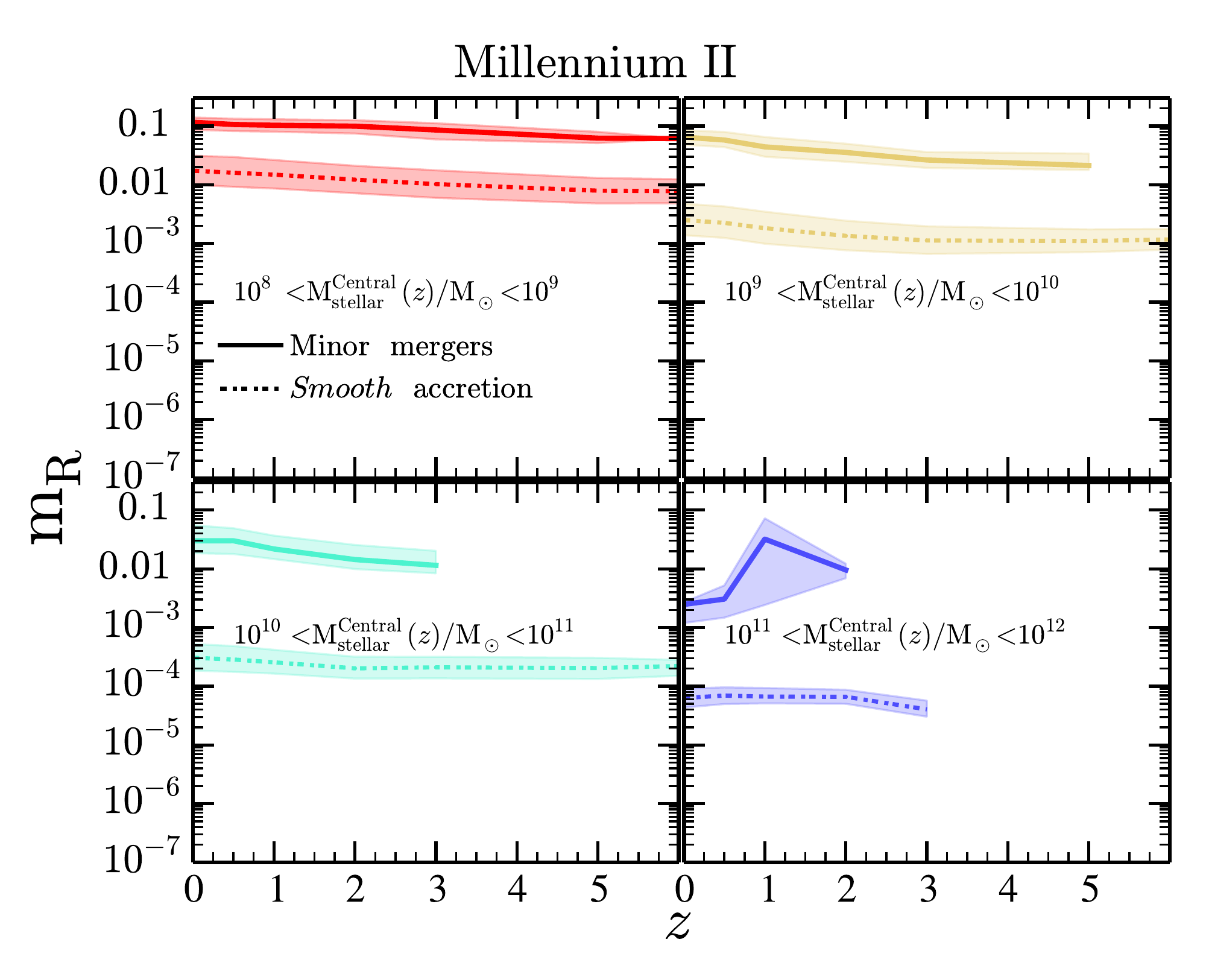}
\includegraphics[width=1.0\columnwidth]{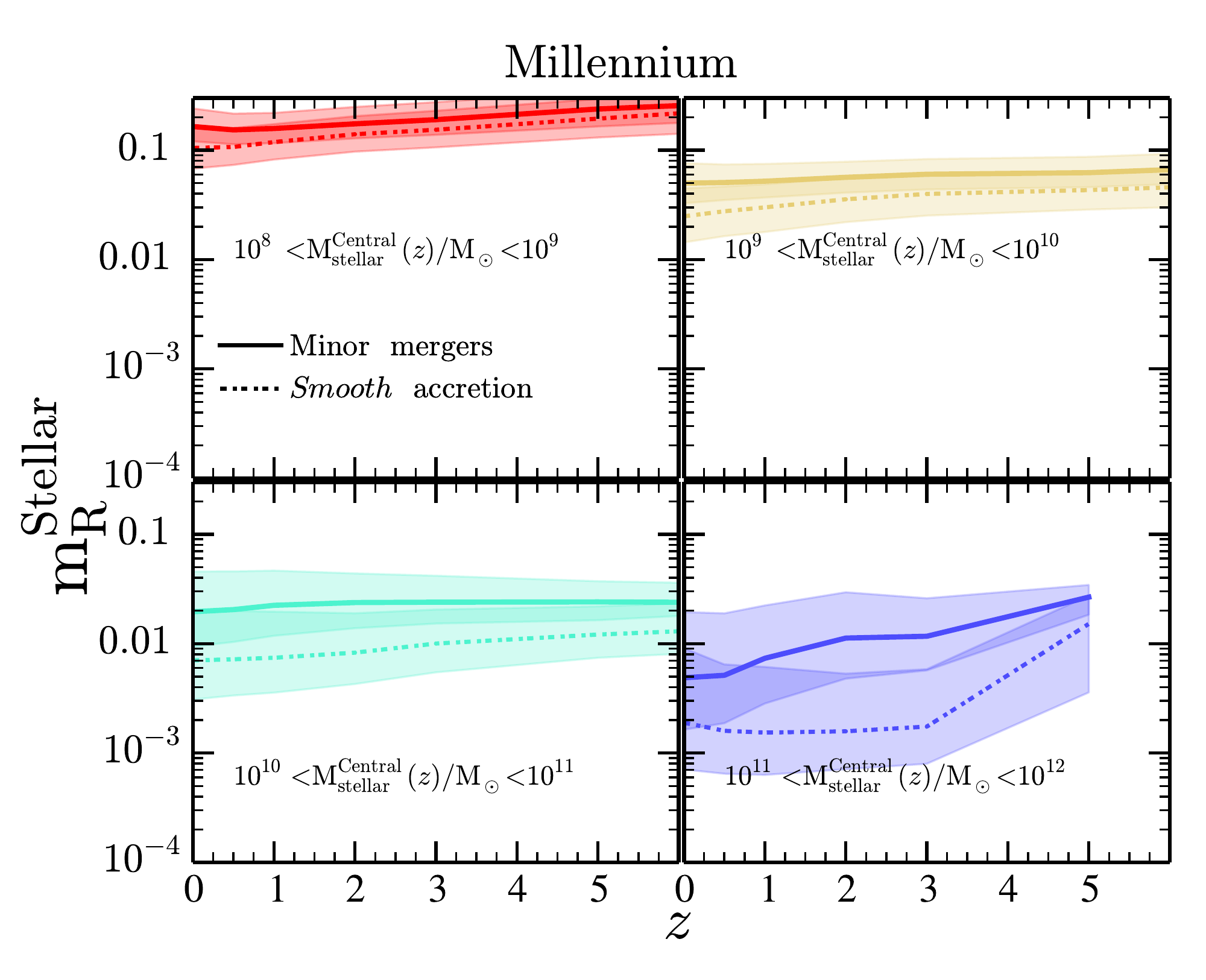}
\includegraphics[width=1.0\columnwidth]{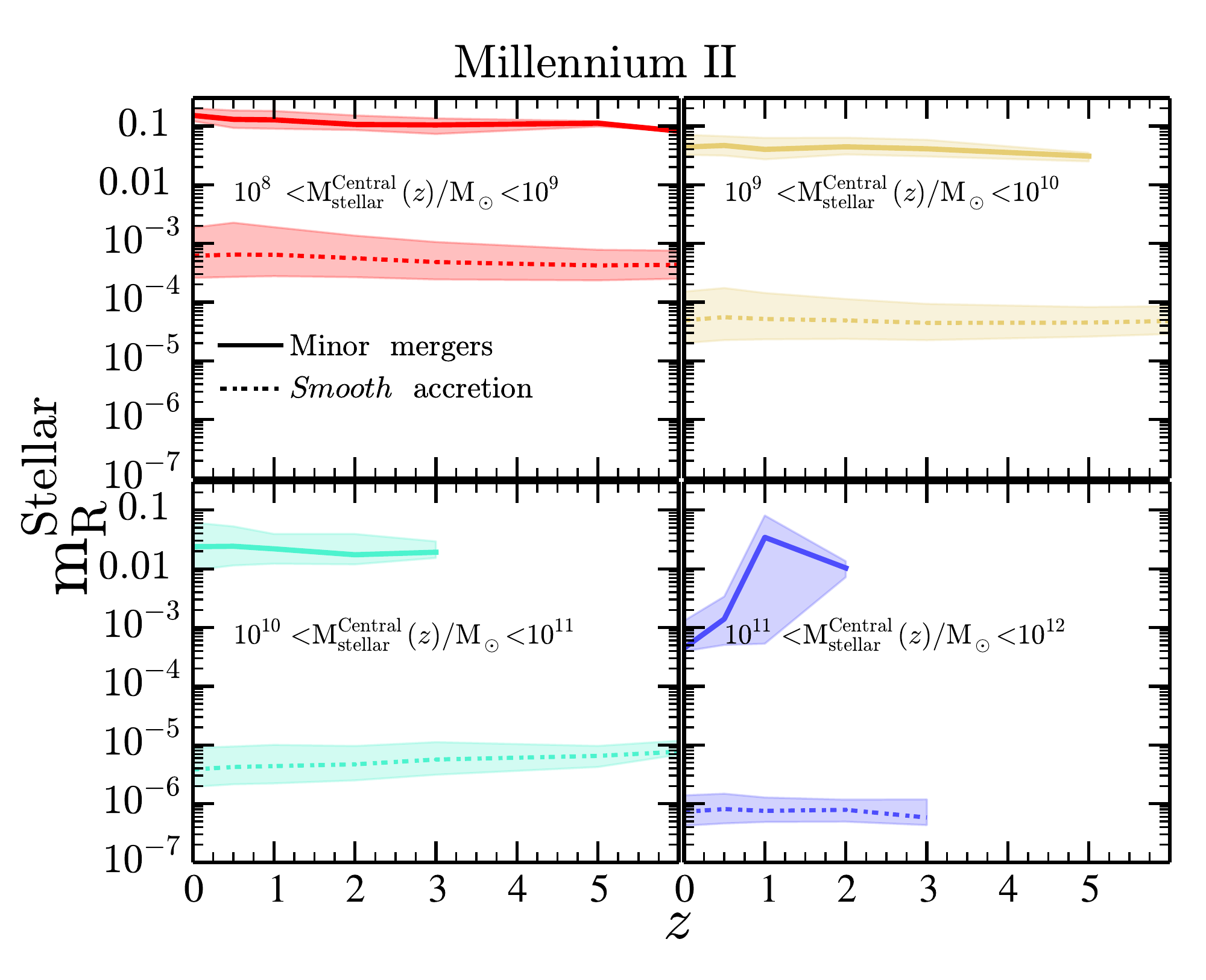}
\caption{\textbf{Upper panels}: Typical bayonic merger ratios, $\rm m_R$, for mergers with $\rm m_R < m_R^{th}$  ($\rm m_R^{th} = 0.2$) in the MS (left) and MSII (right) simulation. The episodes have been divided between minor mergers (solid lines) and \textit{smooth accretion} (dotted dashed lines). Each panel corresponds to a different bin of central galaxy stellar mass at the moment of the merger, $\rm M_{Stellar}^{Central}(\mathit{z})$. The shaded area represents the $1\sigma$ value. \textbf{Lower panels}: The same but for the stellar merger ratios.}
\label{fig:Merger_ratio_distribution_MR}
\end{figure*}

In order to establish in which systems the minor mergers or \textit{smooth accretions} take place, we are going to study the ratio of the two merging galaxies binding energies,  $f_{\rm binding}$. For the binding energy definition we only consider the interacting systems: for the satellite binding energy, $\rm E_{binding}^{Satellite}$, we only consider its stellar mass while for the central one, $\rm E_{binding}^{Central}$, we use the total mass in the disk (gas + stars). Therefore: 
\begin{equation}\label{eq:binding_ap}
f_{\rm binding} = \frac{\rm E_{binding}^{Satellite}}{\rm E_{binding}^{Central}} = \frac{\rm M^2_{Sat,Stellar}}{\rm M^2_{\rm Cent, disk}}\frac{\rm R_{disk}^{Central}}{\rm R_{Stellar}^{Sat}} 
\end{equation}
where $\rm R_{Stellar}^{Sat}$ is the mass-weighted average of the half-mass radii of the satellite bulge and the disc components and  $\rm R_{disk}^{Central}$ the same but using the cold and stellar disk of the central galaxy. Large values of $f_{\rm binding}$ means that the two interacting systems have similar binding energy so the satellite galaxy might survive the interaction inside of the central disk and reach the centre of its massive companion. On the contrary, low values of $f_{\rm binding}$ imply the the central galaxy can easily unbound the satellite stellar system inside its disk. In Fig~\ref{fig:binding_energy} we present the plane $f_{\rm binding} - \rm M_{stellar}^{Central}$ at different redshifts for MS and MSII. The color encodes the satellite stellar mass. The figure shows that $f_{\rm binding}$ is span in a wide range of values with a clear stellar mass trend and independence with redshift. On one hand, large $f_{\rm binding}$ are concentrated in the more massive galaxies ($\rm {\gtrsim} 10^{10}M_{\odot}$) as a natural consequence of the fact that they can experience minor interactions with massive galaxies harder to unbound. On the other hand, small central galaxies ($\rm {\lesssim}\,10^{10}M_{\odot}$) display smaller $f_{\rm binding}$ values during their minor interactions. This ones happen with satellites of low stellar mass compared with the central galaxy stellar and gas disk mass (as we will see in the stellar merger ratio of Fig~\ref{fig:Merger_ratio_distribution_MR}).\\


In Fig~\ref{fig:number_density} lower panel it is presented the evolution of $n_{mm}$ when we impose different $f_{\rm binding}$ values to set apart minor mergers from \textit{smooth accretion}. Different colors correspond to different thresholds in $f_{\rm binding}$. As we can see, smalls thresholds ($ f_{\rm binding}\,{>}\,10^{-10}$) have almost a null effect in MS but with dramatic consequences for MSII as they are able to reduce $\rm{\sim}\,2\,dex$ the value of MSII $n_{\rm mm}$. Among all the possibles values of $ f_{\rm binding}$ we find a reasonable $n_{\rm mm}$ convergence at $ f_{\rm binding} \,{\lesssim}\, 10^{8}\,{-}\,10^{9}$. Naturally, different $ f_{\rm binding}$ cuts have also a different repercussion in in the $z{=}0$ galaxy morphology. In order to study this, in Fig \ref{fig:all_distribution_morph} second column we present the morphology at different $f_{\rm binding}$ thresholds, fixing $\rm m_{R}^{th}\,{=}\,0.2$. Other columns display the same but varying $\rm m_{R}^{th}$ too. As we can see, despite $f_{\rm binding}^{\rm th}\,{<}\,10^{-10}$ values have an improvement in the low mass galaxies it is not enough to make MS and MSII converge. When we impose $f_{\rm binding}^{th}\,{>}\,10^{-9}$, the improvement in MSII is remarkable. Notice that different $f_{\rm binding}$ thresholds have a minimum effect in MS. We have found that the best threshold is $ f_{\rm binding}^{\rm th} = 10^{-8.5}$ in binding energy ratios to differentiate between \textit{smooth accretion} ($f_{\rm binding}\,{<}\,f_{\rm binding}^{\rm th}$) and minor merger ($f_{\rm binding}\,{>}\,f_{\rm binding}^{\rm th}$). Notice, that even though this procedure is a way to make converge the two simulations and make the MSII follow the observational data, one could try to implement others prescriptions like an smooth transition between major-minor merger like \cite{Hatton2003} and \cite{Somerville2008} do or try to implement the fact that during major mergers part of the galactic disc could survive (see \cite{Hopkins2009} and \cite{Hopkins_and_Somerville2009}). Nevertheless, given that our approx works and is the simplest thing we do not implement any of the previous cases but in futures works this topic should be address and take into account.\\

Finally, we have explore for minor mergers and \textit{smooth accretion} the typical baryonic and stellar merger ratio, $\rm m_R$ and $\rm m_R^{Stellar}$ respectively.  Fig~\ref{fig:Merger_ratio_distribution_MR} presents the results as a function of redshift and central galaxy stellar mass at the moment of the merger, $\rm M_{Stellar}^{Central}(\mathit{z})$. 
Regarding the baryonic merger ratio, we can see that both MS and MSII display the increasing trend of $\rm m_R$ value towards lower stellar masses. Interestingly, independently of redshift and central galaxy stellar mass both MS and MSII show that \textit{smooth accretion} display smaller merger ratios than minor merger, fact that is more evident in the MSII than in the MS (consequence of resolution effects). Concerning $\rm m_R^{Stellar}$, we see a similar behavior to the $\rm m_R$ one. However, in this case the difference between minor mergers and \textit{smooth accretion} is more extreme. While minor mergers $\rm m_R^{Stellar}$ values are between 0.1 - 0.01, \textit{smooth accretion} ones display values of $10^{-3} - 10^{-7}$. 

\bsp 
\label{lastpage}
\end{document}